\documentclass{aa}  

\usepackage[colorlinks=true, linkcolor=blue, citecolor=blue]{hyperref}
\usepackage{graphicx}
\usepackage{txfonts}
\usepackage{xcolor}
\usepackage{amsmath}	    
\usepackage{amssymb}	    
\usepackage{siunitx}        
\usepackage{nicefrac}       
\usepackage{mathtools}      
\usepackage{xcolor}         
\usepackage{textcomp}       

\graphicspath{{figures/}}

\usepackage{graphicx}	    
\usepackage{stfloats}  		

\newcommand{\orcid}[1]{\protect\href{https://orcid.org/#1}{\protect\includegraphics[width=8pt]{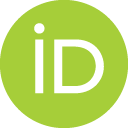}}}

\DeclareSIUnit\dex{dex}
\DeclareSIUnit\year{yr}
\DeclareSIUnit\fwhm{FWHM}
\DeclareSIUnit\ppt{ppt}
\DeclareSIUnit\ppm{ppm}
\DeclareSIUnit\ppmh{ppm\,h^{-\nicefrac{1}{2}}}
\DeclareSIUnit\arcsec{arcsec}
\DeclareSIUnit\pixel{pixel}
\DeclareSIUnit\electron{{e^-}}
\DeclareSIUnit\proton{{p^+}}
\DeclareSIUnit\electronvolt{{eV}}
\DeclareSIUnit\adu{ADU}
\DeclareSIUnit\dn{DN}
\DeclareSIUnit\hit{hits}
\DeclareSIUnit\events{events}
\DeclareSIUnit\photon{photons}
\DeclareSIUnit\volt{V}
\DeclareSIUnit\magnitude{mag}
\DeclareSIUnit\au{AU}
\DeclareSIUnit\pc{pc}
\DeclareSIUnit\exposure{exposure}
\DeclareSIUnit\Rsun{R_{\odot}}
\DeclareSIUnit\Msun{M_{\odot}}
\DeclareSIUnit\Lsun{L_{\odot}}

\newcommand{\paren}[1]{\left(#1\right)} 
\newcommand{\parenf}[1]{\left[#1\right]}

\newcommand{\metal}[0]{\text{[Fe/H]}}
\newcommand{\logg}[0]{\text{log}\,g}
\newcommand{\teff}[0]{T_{\text{eff}}}
\newcommand{\sole}[1]{{#1}_{\odot}}

\newcommand{\tx}[2]{{#1}_{\text{#2}}} 			   	

\newcommand{\gdor}{$\gamma\,\text{Dor}$}
\newcommand{\dsct}{$\delta\,\text{Sct}$}
\newcommand{\bcep}{$\beta\,\text{Cep}$}
\newcommand{\platosim}{\texttt{PlatoSim}}
\newcommand{\platonium}{\texttt{PLATOnium}}
\newcommand{\affogato}{\textsc{Affogato}}
\newcommand{\cortado}{\textsc{Cortado}}
\newcommand{\doppio}{\textsc{Doppio}}
\newcommand{\Pb}[0]{{\mathcal{P}}}

\makeatletter
\newcommand{\Hline}{%
    \noalign {\ifnum 0=`}\fi \hrule height 1pt
    \futurelet \reserved@a \@xhline
}
\newcolumntype{\Vline}{@{\hskip\tabcolsep\vrule width 1pt\hskip\tabcolsep}}
\makeatother

\begin{document} 

\title{MOCKA -- A PLATO mock asteroseismic catalogue: \\ 
Simulations for gravity-mode oscillators}

\author{
N.~Jannsen\inst{1}\orcid{0000-0003-4670-9616}\and
A.~Tkachenko\inst{1}\orcid{0000-0003-0842-2374}\and
P.~Royer\inst{1}\orcid{0000-0001-9341-2546}\and
J.~De~Ridder\inst{1}\orcid{0000-0001-6726-2863}\and
D.~Seynaeve\inst{1}\orcid{0000-0002-0731-8893}\and
C.~Aerts\inst{1,2,3}\orcid{0000-0003-1822-7126}\and
S.~Aigrain\inst{4}\orcid{0000-0003-1453-0574}\and
E.~Plachy\inst{5,6}\orcid{0000-0002-5481-3352}\and
A.~Bodi\inst{5,7}\orcid{0000-0002-8585-4544}\and
M.~Uzundag\inst{1}\orcid{0000-0002-6603-994X}\and
D.~M.~Bowman\inst{8,1}\orcid{0000-0001-7402-3852}\and
D.~J.~Fritzewski\inst{1}\orcid{0000-0002-2275-3877}\and
L.~W.~IJspeert\inst{1}\orcid{0000-0002-9241-3894}\and
G.~Li\inst{1}\orcid{0000-0001-9313-251X}\and
M.~G.~Pedersen\inst{9}\orcid{0000-0002-7950-0061}\and
M.~Vanrespaille\inst{1}\orcid{0000-0002-0420-2473}\and
T.~Van~Reeth\inst{1}\orcid{0000-0003-2771-1745}
}

\institute{
Institute for Astronomy, KU Leuven, Celestijnenlaan 200D bus 2401, 3001 Leuven, Belgium \\ \email{nicholas.jannsen@kuleuven.be} \and
Department of Astrophysics, IMAPP, Radboud University Nijmegen, PO Box 9010, 6500 GL Nijmegen, The Netherlands \and
Max Planck Institute for Astronomy, Koenigstuhl 17, 69117 Heidelberg, Germany \and
Sub-department of Astrophysics, Department of Physics, University of Oxford, Oxford OX1 3RH, United Kingdom \and
HUN-REN CSFK Konkoly Observatory, MTA Centre of Excellence, Konkoly Thege Mikl\'os \'ut 15-17, H-1121 Budapest, Hungary \and
ELTE E\"otv\"os Lor\'and University, Institute of Physics and Astronomy, 1117, P\'azm\'any P\'eter s\'et\'any 1/A, Budapest, Hungary \and
Department of Astrophysical Sciences, Princeton University, 4 Ivy Lane, Princeton, NJ 08544, USA \and
School of Mathematics, Statistics and Physics, Newcastle University, Newcastle upon Tyne, NE1 7RU, United Kingdom \and
Sydney Institute for Astronomy, School of Physics, University of Sydney, Sydney, NSW 2006, Australia
}

\date{\today}
 
\abstract
{With ESA’s PLATO space mission set for launch in December 2026, a new photometric legacy and a future of new scientific discoveries await. By exploring scientific topics distinct from the core science program, the PLATO complementary science program (PLATO-CS) provides a unique opportunity to maximise the scientific yield of the mission.}
{In this work we investigate PLATO’s potential for observing pulsating stars across the Hertzsprung--Russell diagram, distinct from the core science program. Specifically, a PLATO mock asteroseismic catalogue (MOCKA) of intermediate to massive stars is presented as a benchmark to highlight the asteroseismic yield of PLATO-CS in a quantitative way. MOCKA includes simulations of $\beta$~Cephei, slowly pulsating B (SPB), $\delta$~Scuti, $\gamma$~Doradus, RR Lyrae, Cepheid, hot subdwarf, and white dwarf stars. In particular, main-sequence gravity (g) mode pulsators are of interest as some of these stars form an important foundation for the scientific calibration of PLATO. Their pulsation modes primarily probe the radiative region near the convective core boundary, thus making them unique stellar laboratories to study the deep internal structure of stars.}
{MOCKA is based on a magnitude limited ($G\lesssim17$) \textit{Gaia} catalogue and is a product of realistic end-to-end \texttt{PlatoSim} simulations of stars for the first PLATO pointing field in the Southern hemisphere, which will be observed for a minimally 2-yr duration. Being a state-of-the-art hare-and-hound detection exercise, the simulations of this project explore the impact of spacecraft systematics and stellar contamination for on-board PLATO light curves.}
{We demonstrate for the first time PLATO's ability to detect and recover the oscillation modes for main-sequence g-mode pulsators. We show that an abundant spectrum of frequencies is achievable across a wide range of magnitudes and co-pointing PLATO cameras. Within the magnitude limited regimes simulated ($G \lesssim 14$ for $\gamma$~Doradus stars and $G \lesssim 16$ for SPB stars) the dominant g-mode frequency is recovered in more than 95\% of the cases. Furthermore, an increased spacecraft noise budget impacts the recovery of g modes more than stellar contamination by variable stars.}
{MOCKA help us to understand the limits of the PLATO mission as well as highlight the opportunities to push astrophysics beyond current stellar models. All data products of this paper are made available to the community for further exploration. The key data products of MOCKA are the magnitude limited \textit{Gaia} catalogue of the first PLATO pointing field, together with fully reduced light curves from multi-camera observations for each pulsation class.}


\maketitle

\section{Introduction}\label{sec:introduction}


PLAnetary Transits and Oscillation of stars \citep[PLATO;][]{rauer2014plato, rauer2024plato} is the next medium-class ESA mission dedicated to space photometry. With a payload utilising a multi-telescopic design \citep[covering a sky area of \SI{2132}{\deg\squared};][]{pertenais2021unique}, PLATO will monitor about a quarter of a million bright stars ($V<\SI{15}{\magnitude}$) over its nominal 4-yr mission duration. The primary aim of PLATO's core science program is to discover and characterise Earth-like planets orbiting in the habitable zone of Sun-like stars.


Space-based missions like PLATO deliver exquisite photometric quality over long baselines valuable for studying a zoo of variable phenomena. Indeed, experience from dedicated space photometers like MOST \citep{walker2003most}, CoRoT \citep{auvergne2009corot}, \textit{Kepler} \citep{borucki2010kepler}, and TESS \citep{ricker2015tess} have shown that, even with a minimal observational budget, the scientific outcome of these missions extends well beyond the primary science goals. To exploit the full potential of the PLATO mission for scientific topics that are distinct from the core science program, the PLATO complementary science program (PLATO-CS) has been designed \citep{tkachenko2024eas, conny2024tkas}. With 8\% of PLATO's telemetry budget being offered to the Guest Observer (GO) program through open competitive calls to the community \citep{heras2024eas}, PLATO-CS will play a crucial role in preparing these calls. 


To help explore the potential of the different variable phenomena of PLATO-CS (grouped into dedicated mission work packages)%
\footnote{\tiny{\url{https://fys.kuleuven.be/ster/research-projects/plato-cs}}}, %
simulations provide essential diagnostic quantities before mission launch. Common for all PLATO's scientific disciplines, the underlying internal and external noise sources will dictate how well the astrophysical signal can be preserved. The impact of spacecraft systematics and data post-processing strongly depends on the photometric signature \citep[e.g. see highlights from previous photometric space missions by][]{garcia2011preparation, handberg2014automated, vanderburg2014technique, aigrain2015precise, lund2015k2p2, aigrain2016k2sc, luger2016everest, handberg2021tess, lund2021tess, maxted2022analysis}. Consequently, to validate the success of the PLATO-CS we treat the following questions:
\begin{itemize}
\itemsep0cm
\item Which instrument (normal versus fast cameras), data products (pixel imagettes versus light curves), and observing cadence (\SI{2.5}{\second}, \SI{25}{\second}, \SI{50}{\second}, and \SI{600}{\second}) are needed?
\item How many cameras are minimally needed, and how does this comply with the number density of star with different spectral type across the field-of-view (FOV)?
\item Given that the PLATO passband is designed for solar-type dwarf and subgiant stars, what is the typical limiting magnitude for detecting variability of more massive and/or evolved stars?
\item What is the effect of spacecraft systematics, and how does this depend on, for example the intra-pixel position of a target star on the detector, the radial distance away from the optical axis of the pupil, or ageing effects of the cameras?
\item What is the effect of stellar contamination, and how does this interplay with instrumental systematics?
\end{itemize}
Ultimately, answering these (and many more) questions with the usage of mock simulations gives a strong indication of the expected outcome and, more importantly, the full potential of PLATO-CS. 


As an integrated part of the workforce providing diagnostics for the `pulsating stars' work package of PLATO-CS, we present the PLATO mock asteroseismic catalogue (MOCKA). Since the success of asteroseismic studies crucially depends on the detection and identification of as many stellar pulsation modes as possible \citep[e.g. see reviews by][]{cunha2007asteroseismology, aerts2010asteroseismology, chaplin2013asteroseismology, hekker2017giant, aerts2019angular, bowman2020asteroseismology, aerts2021probing}, MOCKA is the first simulated catalogue to benchmark the asteroseismic potential of the mission. As we aim to provide a PLATO mock asteroseismic catalogue for the PLATO-CS, the stars in question are all more massive and/or more evolved than the Sun. With MOCKA, we target eight classes of stellar pulsators with synthetic light curves that are well suited to address questions related to the photometric precision needed for seismic modelling. Table~\ref{tab:variables} shows the different pulsators of MOCKA and their generic asteroseismic characteristics. 


This work focusses on $\gamma$ Doradus (\gdor{}) and slowly pulsating B-type (SPB) stars as they together form a critical benchmark sample of gravity (g) mode scientific calibrators for the PLATO mission \citep{rauer2024plato}. Being late-A to early-F spectral types, \gdor{} stars are located on and near the main sequence with masses between \SIrange{1.3}{2.0}{\Msun}, whereas SPB stars are typically found on the main sequence with masses between \SIrange{2}{8}{\Msun} \citep{mombarg2024two}. Together with the pressure (p) mode $\delta$ Scuti (\dsct{}) pulsators, these mass regimes probe the transition phases between convective versus radiative outer envelope ($\sim$\SI{2}{\Msun}), which allows asteroseismic inferences of the underlying physics \citep{aerts2021probing}. 

The g modes of \gdor{} and SPB stars primarily probe the radiative region near the convective core boundary. In the asymptotic regime of low-frequency high-order, consecutive modes are equally spaced in period and exhibit a characteristic period $\Pi_0$ \citep{shibahashi1979modal, tassoul1980asymtotic}. The quasi regularity of the modes in period allows the construction of a so-called period-spacing pattern, which is a key diagnostic tool for unravelling the physics of the stellar interior, such as the gradient in chemical composition \citep[e.g.][]{miglio2008probing, degroote2010deviations, moravveji2016subinertial, mombarg2019asteroseismic}, the near-core rotation profile \citep[e.g.][]{vanreeth2016interior, ouazzani2017new, papics2017signatures}, the transport of angular momentum \citep[e.g.][]{kurtz2014asteroseismic, saio2015asteroseismic, vanreeth2018sensitivity, ouazzani2019gamma, pedersen2022stellar}, and recently stellar magnetism \citep[e.g.][]{vanbeeck2020detecting, loi2020effect, mathis2021probing, roi2023asteroseismology}.


In this paper, we provide a small general overview of MOCKA in Sect.~\ref{sec:mocka}. We explain the generation of the stellar catalogue in Sect.~\ref{sec:catalogue} and the models of variability in Sect.~\ref{sec:variability}, which are used as input for the simulations. Section~\ref{sec:simulations} provides a detailed explanation of the setup and the execution of the mock simulations, together with the post-processing pipeline developed to provide the final data products (being light curves and pulsation modes). Next, we present and discuss the results of recovering pulsation modes for g-mode pulsators in Sect.~\ref{sec:results}, and finally, in Sect.~\ref{sec:conclusions} we conclude our findings in the context of the future prospects for PLATO(-CS). 

\section{Overview of MOCKA}\label{sec:mocka}

\begin{table*}
\caption[]{MOCKA's arsenal of pressure (p) and gravity (g) mode pulsators. The first four columns display generic information about each pulsating class. The last three columns show the number of stars ($N_{\star}$), the limiting magnitude ($\Pb_{\rm max}$), and the final cadence ($\delta t$) of each sample simulated.}
\begin{center}
\begin{tabular}{lllr|ccr}
\hline\hline
Variability & Spectral 	& Dominant 			& Typical & $N_{\star}$ & $\Pb_{\rm max}$ 	& $\delta t$ 	\\
class	& type		& mode character	& periods &				& [mag]				& [s]			\\	
\hline
Cepheid	(Ceph)					& F6-K2 & Low-order, radial p modes	 		& \SIrange{1}{300}{\day} 	& 2000 				& 17 & 600	\\
RR Lyrae (RR Lyr)				& F0-G9 & Low-order, radial p modes 		& \SIrange{0.3}{1.0}{\day}	& 2000 				& 17 & 600	\\
Slowly pulsating B (SPB)	& B3-B9 & High-order, non-radial g modes	& \SIrange{0.2}{3.2}{\day}	& 4000 				& 16 & 600	\\
$\gamma$ Doradus (\gdor)	& F4-F0 & High-order, non-radial g modes 	& \SIrange{0.2}{3.0}{\day} 	& 4000 				& 14 & 600	\\
$\beta$ Cephei (\bcep)		& O8-B6 & Low-order, p- and g modes 		& \SIrange{2.5}{8.0}{\hour}	& 2000 				& 16 & 50 	\\
$\delta$ Scuti (\dsct)		& A0-F5 & Low-order, non-radial p modes 	& \SIrange{0.2}{8.0}{\hour} & 4000 				& 16 & 50	\\
sdBV (V361\,Hya, V1093\,Her) & B0-B9 & Non-radial p- and g modes 		& \SIrange{1}{180}{\min} 	& $17 \times 10$ 	& 17 & 25	\\
White dwarf (WD: DA, DB, DO) & D(OBA) & High-order, non-radial g modes 	& \SIrange{0.5}{25}{\min} 	& $30 \times 10$ 	& 17 & 25	\\
\hline
\end{tabular}
\label{tab:variables}
\end{center}
\end{table*}

The MOCKA catalogue has been generated using the PLATO camera simulator, \platosim{}%
\footnote{\tiny{\url{https://github.com/IvS-KULeuven/PlatoSim3}}} %
\citep{jannsen2024platosim}. Specifically, the \platosim{} software package provides a toolkit dubbed \platonium{}, which transforms \platosim{} (being a camera simulator) into a mission simulator, meaning that realistic instrumental systematics for the entire payload are easily configured in a coherent way in accordance to the number of co-pointing cameras and the mission requirements. With \platosim{} being an extremely feature-rich pixel-based simulator, \platonium{} greatly alleviates the manual process of setting up a new \platosim{} project, meanwhile being designed for parallel computing. Aside from instrumental systematics, the toolkit provides scripts to generate custom stellar sky catalogues (either using the PLATO input catalogues \citep[PIC;][]{montalto2021all} or the \textit{Gaia} database \citep{gaia2016mission}, and realistic models of stellar variability (c.f. Sect.~\ref{sec:variability}).


Compared to the PIC, which focusses on FGKM dwarf and subgiant stars, the stellar catalogue used in this work is queried from the third data release \citep[DR3;][]{gaia2023dr3} of the \textit{Gaia} mission. Since the selection of the long-duration observational phases \citep[LOPs;][]{nascimbeni2021plato} a slight modification has been made to the LOP south (LOPS1 $\rightarrow$ LOPS2; Nascimbeni et al. subm.), which is the first definite pointing for PLATO. The choice of using the \textit{Gaia} catalogue (as compared to a synthetic one) is based on: i) a more realistic distribution of relative pixel positions and brightnesses of target stars with respect to stellar contaminants, and ii) a realistic crowding metric \citep[e.g. more massive pulsators are found in crowded regions within the galactic thin disc;][]{bowman2022cubespec}. In practice, the final LOPS2--\textit{Gaia} DR3 catalogue was generated from full-frame CCD images of \platosim{} while injecting stars with $2 \lesssim G \lesssim 17$ (c.f. Appendix~\ref{app:catalogue}).


\begin{figure}[t!]
\center
\includegraphics[width=\columnwidth]{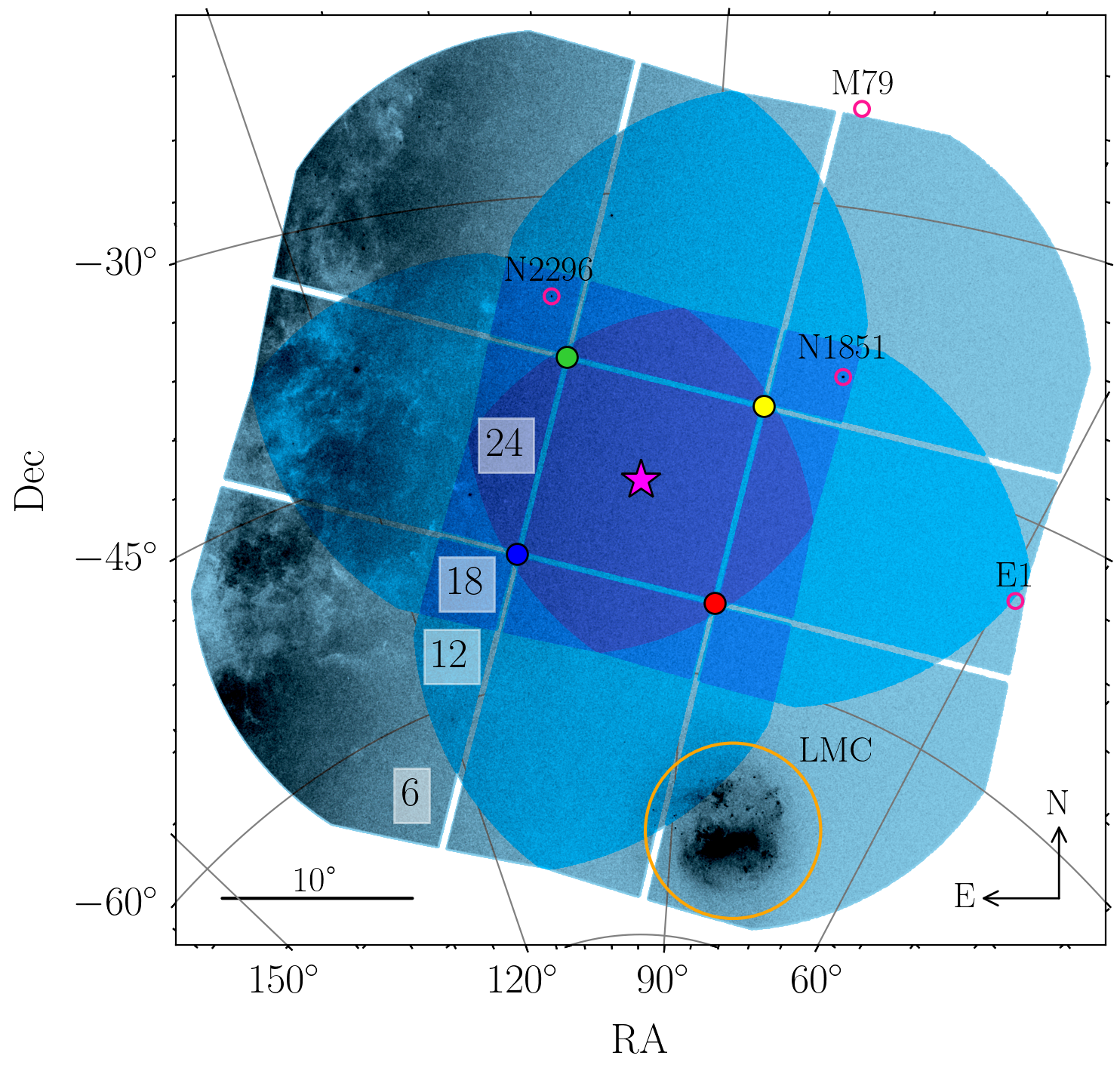}
\caption[]
{Illustration of the first PLATO pointing field called LOPS2. The platform pointing (being parallel to the pointing of the two F-CAMs; see magenta star) is centred at the equatorial coordinate $(\alpha, \delta) = (\SI{95.31043}{\degree}, \SI{-47.88693}{\degree})$, with zero rotation with respect to the Galactic equator. The N-CAM overlap of $\tx{n}{CAM}\in\{6, 12, 18, 24\}$ is illustrated with an increasing darker shade of blue (also indicated in the white boxes), and the blue, green, yellow, and red dot show the pointing of N-CAM group one, two, three, and four, respectively. The black transparent map highlights dense sky regions such as the location of the Milky Way plane, the Large Magellanic Cloud (LMC, encircled in orange), and a few globular clusters \citep[pink circles, from][]{harris1996catalog}.} 
\label{fig:LOPS2}
\end{figure}

\begin{figure}[t!]
\center
\includegraphics[width=\columnwidth]{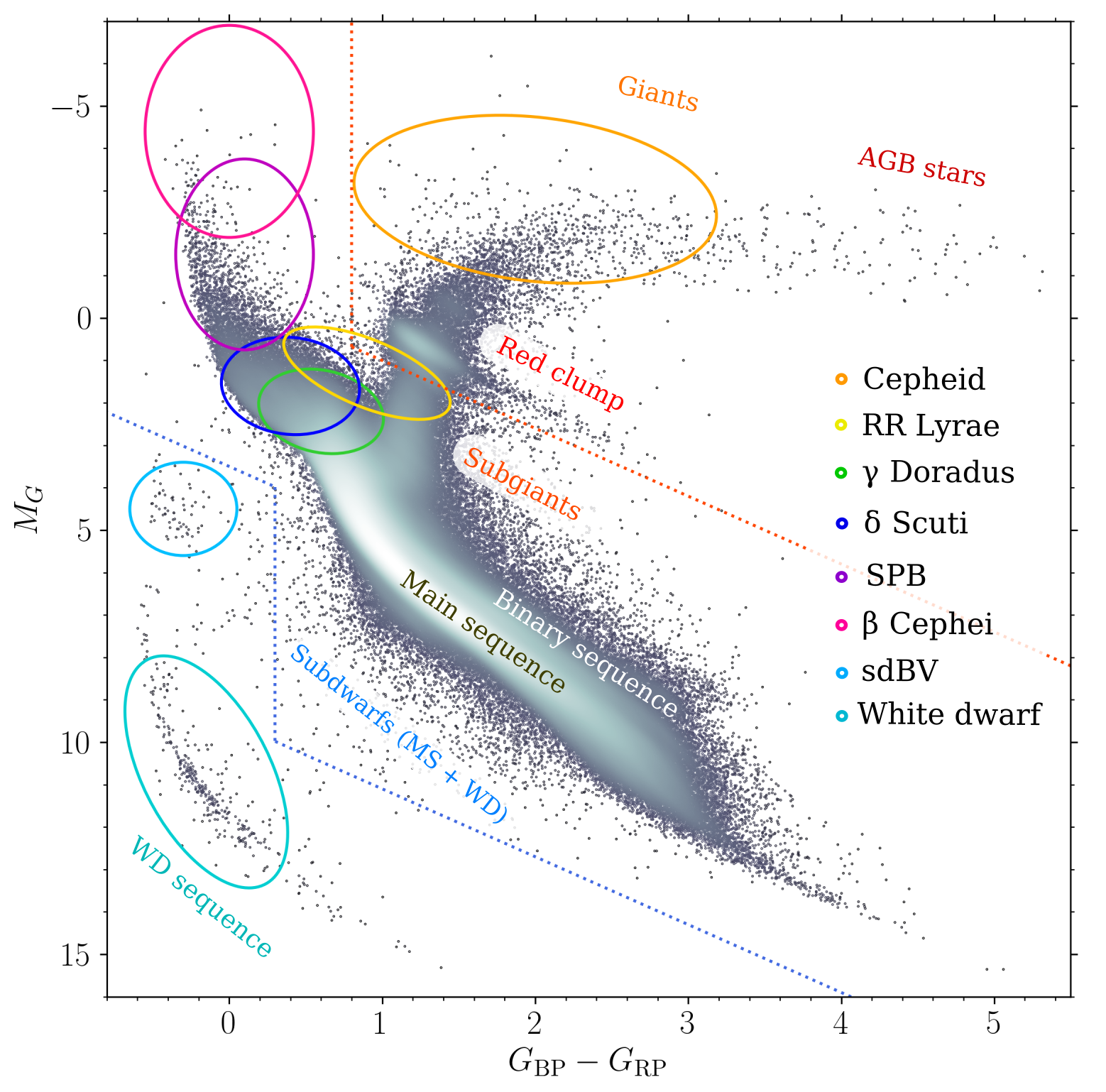}
\caption[]
{The colour absolute magnitude diagram (CaMD) of stars from the LOPS2 (see Fig.~\ref{fig:LOPS2}) within \SI{1}{\kilo\pc} from the Sun. Here $G$, $\tx{G}{BP}$, and $\tx{G}{RP}$ refer to the \textit{Gaia} full-bandwidth, blue, and red passband, respectively. Commonly known CaMD features are highlighted with coloured text. Each of the coloured circles indicates the approximate region of a stellar pulsation class that is a part of MOCKA.}
\label{fig:CaMD}
\end{figure}

Figure~\ref{fig:LOPS2} shows the LOPS2 pointing in equatorial coordinates with the characteristic geometric feature of six co-pointing cameras, situated in four camera groups each with an opening angle of \SI{9.2}{\degree} relative to the platform pointing. The result is an overlap of the FOV for the so-called normal cameras (N-CAMs) counting $\tx{n}{CAM}\in\{6, 12, 18, 24\}$. Figure~\ref{fig:CaMD} shows the colour absolute magnitude diagram (CaMD) of stars from the LOPS2 within a distance of \SI{1}{\kilo\pc} from the Sun. The ellipses represent the approximate regions of where each pulsating class is expected to be found. 

Our catalogue contains three batches of simulated data for each stellar sample: 
\begin{enumerate}
\itemsep0cm
\item Simulation batch (called \affogato{}) is a best-case scenario where the instrumental systematics are `as expected' and stellar variability from nearby contaminant stars is excluded;
\item Simulation batch (called \cortado{}) is a worst-case scenario regarding instrumental systematics but with `quiet' stellar contaminants, and;
\item Simulation batch (called \doppio{}) is a worst-case scenario regarding variable stellar contaminants (with all contaminants being variable) but with instrumental systematics `as expected' by the mission.
\end{enumerate}
 
\section{Stellar input catalogue}\label{sec:catalogue}

The LOPS2--\textit{Gaia} DR3 catalogue shown in Fig.~\ref{fig:LOPS2} contains 7,757,180 stars in the magnitude range $2 \lesssim G \lesssim 17$ and forms the base from which we construct a target catalogue for each asteroseismic sample. Like any other survey, \textit{Gaia} is magnitude limited and suffers from several observational biases \citep[e.g.][]{schonrich2019distances, rybizki2022classifier}. To mitigate the inclusion of artefacts in our stellar samples, we start from the \textit{Gaia} CaMD, shown in Fig.~\ref{fig:CaMD}.


We first account for the extinction as measured in the \textit{Gaia} passband, $\tx{A}{G}$ (\texttt{GSP-PHOT}). Our catalogue contains 6,602,912 stars with an extinction measurement. We consider these as potential target stars (c.f. Sect.~\ref{sec:cat_samples}). The remaining million or so (typically faint) stars  without an extinction value may be a stellar contaminant. With an uncertain location in the CaMD, we assume that all of these stars have a zero extinction and acknowledge the small bias this may enforce in the target-to-contaminant magnitude distribution of our simulations (c.f. Sect.~\ref{sec:sims_subfield}).


Next we convert the stellar magnitudes from the \textit{Gaia} $G$ passband to the PLATO N-CAM passband, $\Pb$ \citep[for a passband comparison, see Fig.~6 of][]{marchiori2019flight}. The passband conversion is derived from atlases of stellar atmospheric models for dwarfs and giants (c.f. Appendix~\ref{app:passband}, using the \textit{Gaia} colour, $\tx{G}{BP}-\tx{G}{RP}$). As shown in Fig.~\ref{fig:CaMD}, we separate dwarfs from giants using an upper CaMD boundary (with giants above the dotted red line) and dwarfs from compact objects using a lower CaMD boundary (with compact objects below the blue dotted line). Determined from a linear fit to the red clump, the gradient of the boundary definitions is parallel to the direction of interstellar reddening, such that $\tx{M}{G} \propto 1.6 \, (\tx{G}{BP}-\tx{G}{RP})$.

\subsection{Target star sky catalogue}\label{sec:cat_samples}


Before we start to refine each asteroseismic sample, we first remove bright stars that will saturate the PLATO detectors. \cite{jannsen2024platosim} showed that for stars of $\Pb \sim \SI{7.5}{\magnitude}$ charge bleeding results in a non-conservative photometric measurement. Hence, we reject 3175 bright stars with $\Pb < \SI{7.5}{\magnitude}$. To keep the analysis simple, we only intend to simulate targets as if they are all single stars. Following \cite{penoyre2020binary}, we exclude binaries by rejecting stars with a reduced unit weight error defined by \textit{Gaia} in excess of 1.2. Moreover, we remove all stars without a measurement of either $\{G,\,\tx{G}{BP}-\tx{G}{RP},\,\tx{A}{G},\,\varpi\}$ and stars with extremely large relative parallax uncertainties (i.e. $\sigma_{\varpi}/\varpi > 1$). After all these cuts, we are left with 5,683,986 potential target stars.


The base of each asteroseismic sample (except for compact pulsators) is constructed from a query of all stars defined by the ellipses of Fig.~\ref{fig:CaMD}. Each ellipse was define based on known pulsators from \citet[][Fig.~3]{gaia2018dr3varability}. Since these CaMD instability regions comply approximately with the mass range of intermediate to massive stars, the on-sky crowding metric per pulsation class is well conserved using this methodology. Next, each sample is trimmed using the respective gross spectral type ranges given in Table~\ref{tab:variables}, except for \gdor{} and \dsct{} stars for which further physical cuts are made (as we discuss in Sect.~\ref{sec:cat_parameters}). 


For each asteroseismic sample, we homogeneously assembled $N_{\star}$ stars as indicated in the last column of Table~\ref{tab:variables}. The choice of $N_{\star}$ was set by the number of potential targets available in each class after all cuts. We required that each sample has an equal observability count of $\tx{n}{CAM}\in\{6, 12, 18, 24\}$ while having an approximately uniform star count over the range of apparent magnitudes $7.5 < \Pb < 17$. The upper magnitude limit ($\tx{\Pb}{max}$, Table~\ref{tab:variables}) was adjusted given the detectability of the maximum amplitude for each type of pulsator. As an example, even though both are gravito-inertial pulsators along the main sequence, the \gdor{} stars have lower amplitudes than the SPB stars \citep{vanreeth2015detecting, pedersen2021internal}, implying that we simulated the former class only up to $\Pb = 14$ while the latter up to $\Pb = 16$. Each detection-magnitude threshold was set by PLATO's noise budget derived from a separate set of simulations (c.f. Appendix~\ref{app:NSR}). To conform to these requirements, we use a stellar magnitude histogram, corresponding to the four camera visibilities, as the inverse weight to randomly draw $N_{\star}/4$ stars in each $\tx{n}{CAM}$-bin of each asteroseismic sample. We note that due to the effective number of a available targets in our limited \textit{Gaia} catalogue, this methodology does not provide a perfect equal, but sufficient, count over each $\tx{n}{CAM}$--$\Pb$ range (see Appendix~\ref{app:parameters}).

\subsection{Target star parameters}\label{sec:cat_parameters}

We define a stellar parameter space of each (target) star in order to apply an instrumental amplitude correction (c.f. Sect.~\ref{sec:variability}). For this we need an estimate of the stellar mass, $M$, radius, $R$, luminosity, $L$, effective temperature, $\tx{T}{eff}$, surface gravity, $\text{log} g$, and metallicity, \metal{}. In the following we provide a short summary how the stellar parameter space of each asteroseismic sample was created, while leaving an in depth description to Appendix~\ref{app:parameters} for the interested reader. 
 
\begin{figure*}[t!]
\center
\includegraphics[width=\columnwidth]{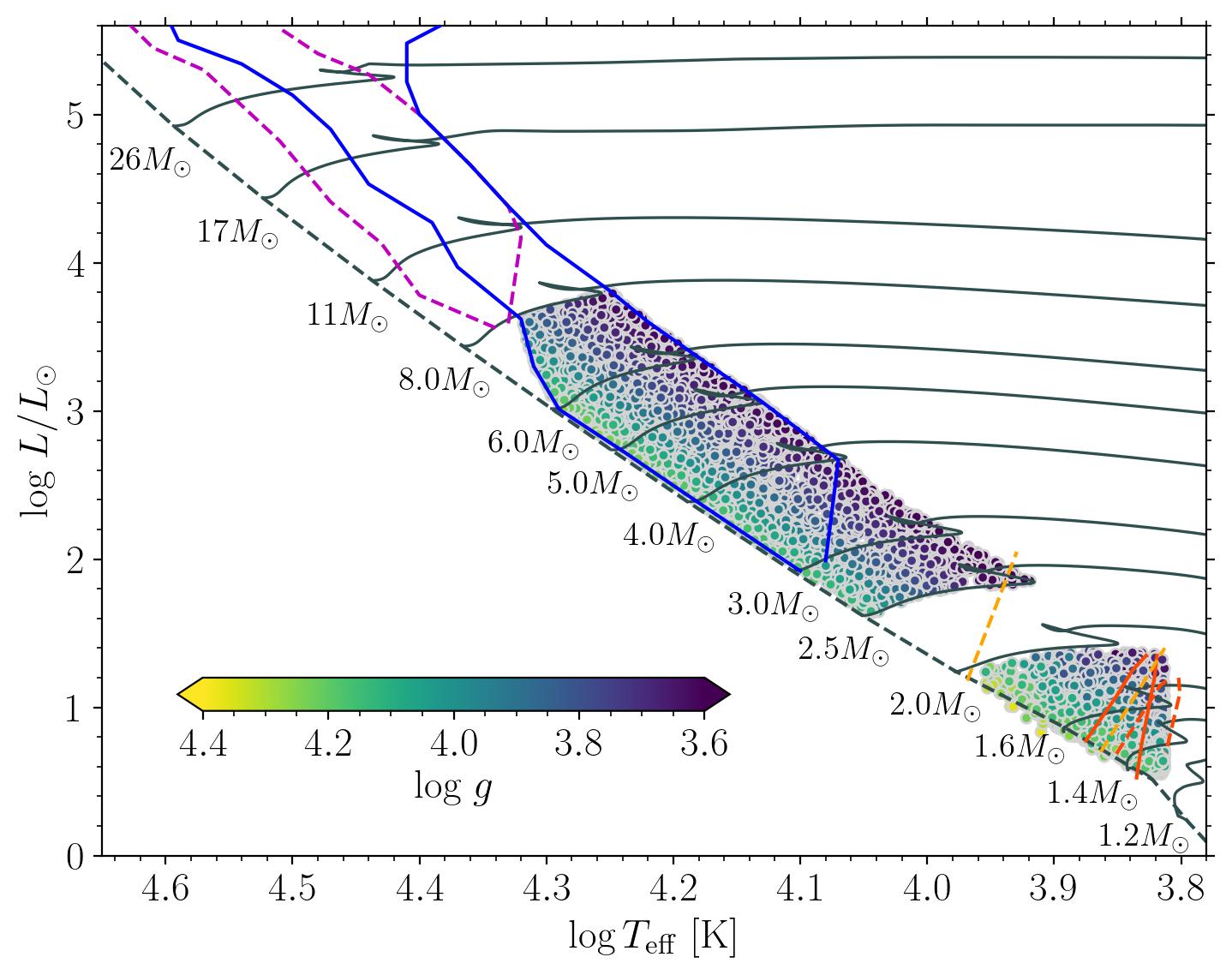}
\includegraphics[width=\columnwidth]{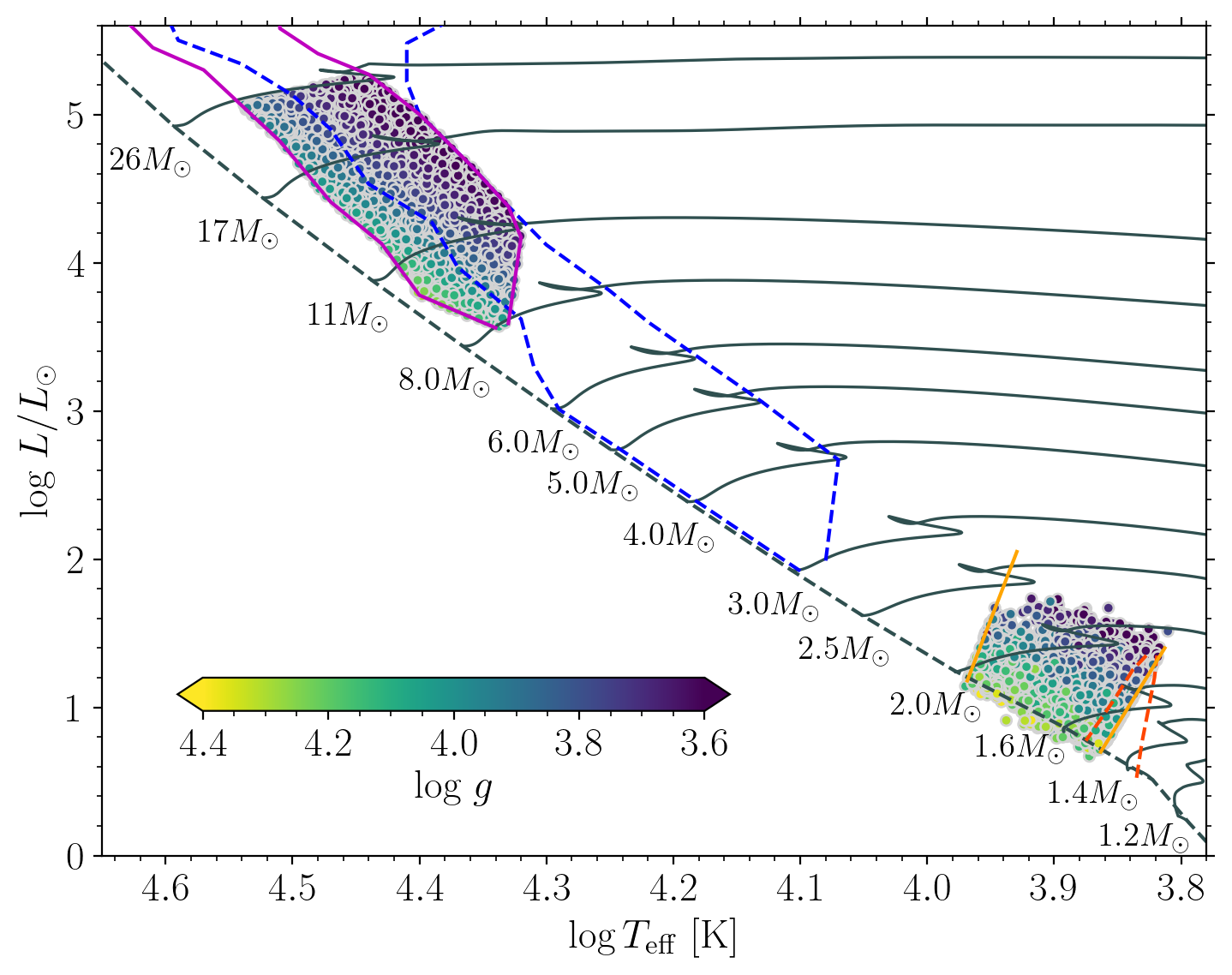}
\caption[]
{HRD of the \gdor{} and SPB (left panel), and \dsct{} and \bcep{} (right panel) stellar samples. The colour gradient indicates the stellar surface gravity, and over-plotted are MIST evolutionary tracks \citep{choi2016mesa}. Solid lines indicate instability strips for gravity modes in the left panel and pressure modes in the right panel. The red lines show the edges of the theoretical \gdor{} instability strip from \cite{dupret2005convection} (corresponding to a mixing length of respectively $\alpha_{\rm MLT}=2$ (solid lines) and $\alpha_{\rm MLT}=1.5$ (dashed lines) in the left-hand plot). The orange lines represent the observational \dsct{} instability strip from \cite{murhpy2019gaia}. The blue and purple lines are the theoretical instability regions of SPB and \bcep{} stars, respectively, calculated by \cite{burssens2020variability}.}
\label{fig:HRD_samples}
\end{figure*}

At this stage, the majority of our target stars have \{$M$, $R$, $\teff$, $\logg$, \metal{}\} defined by \textit{Gaia}. Due to the strong model biases from the \textit{Gaia} pipeline for these parameters regarding massive and evolved stars \citep[and to some degree for lower mass dwarfs, e.g.][]{fritzewski2024age}, we only rely the \textit{Gaia} \texttt{FLAME} pipeline%
\footnote{\tiny{\url{https://gea.esac.esa.int/archive/documentation/GDR3/}}} %
for the AF-type dwarfs, namely the \gdor{} and \dsct{} samples. While the selection of \gdor{} stars was guided by the theoretical instability strip of \citet[][red lines of Fig.~\ref{fig:HRD_samples}]{dupret2005convection}, the \dsct{} stars were established using the observational instability strip of \citet[][orange lines of Fig.~\ref{fig:HRD_samples}]{murhpy2019gaia}. The parameter space of the SPB and \bcep{} sample was artificially generated by defining a HRD polygon using evolutionary tracks and instability strips of \citet[][blue and purple lines of Fig.~\ref{fig:HRD_samples}, respectively]{burssens2020variability}.

The parameter space of evolved stars was likewise artificially created. For the sample of RR Lyrae and Cepheid pulsators, each physical quantity was generated using a simple normal distribution while drawing the mean and standard deviations using literature results (see Table~\ref{tab:rrlyrae} and \ref{tab:cepheids}, respectively). For the hot subdwarf B (sdB) and white dwarf (WD) pulsators, we also use literature values wherever possible (see reference of Sect.~\ref{sec:var_tar_compact}). Otherwise the parameter space was assigned using a best knowledge estimate given the spectroscopic and/or photometric information of each compact pulsator.

\section{Models of variability}\label{sec:variability}

A huge number of stellar variables all across the CaMD have been discovered by ground-based surveys, such as OGLE \citep{udalski1992ogle}, and space-based missions, primarily by CoRoT, \textit{Kepler}, \textit{K2}, TESS, and \textit{Gaia}. In particular, \cite{gaia2018dr3varability} cross matched several of the above surveys with the Gaia DR2 for the identification of variable sources. To avoid introducing biases from sparse sampling and changing data quality from one survey to another, we generate each variable model using either a database of high-precision space-based photometric measurements or a theoretical framework.

Following the former option means finding a suitable sample for which pulsation modes (frequency, amplitude, and phase) can be used more broadly in a statistical sense. Then, starting from an equidistant time series, $t$ (with a sampling of $\delta t = \SI{25}{\second}$, and a duration of $\Delta t = \SI{2}{\year}$), we use a slightly modified formalism from \cite{vanreeth2015detecting} to create each light curve:
\begin{equation}\label{eq:F(t)}
F(t) = A \parenf{\frac{f(t)}{\max[f(t)]} - \Biggl\langle\frac{f(t)}{\max[f(t)]}\Biggr\rangle} \,,
\end{equation}
with
\begin{equation}\label{eq:f(t)}
f(t) = \paren{1 + \frac{1}{A}\sum_{i=1}^N a_i \sin\paren{2\pi\,\nu_i \, t + \phi_i}}^{\gamma} \,.
\end{equation}
Here every $i^{\rm th}$ pulsation mode has a cyclic frequency, $\nu$ (together with an angular frequency $\omega=2\pi\,\nu$ and period $P = 2\pi/\omega = 1/\nu$), an amplitude, $a$, and a phase, $\phi$. $A$ is the maximum peak-to-peak amplitude of the light curve, and the index, $\gamma$, describes the asymmetry of the pulsations in the time domain. We use $\gamma=2.2$ for \gdor{} and SPB stars \citep{vanreeth2015detecting} and otherwise a power of unity.%
\footnote{Due to the direct usage of pulsation modes of RR Lyrae and Cepheid (c.f. Sect.~\ref{sec:var_tar_rrlyr} and \ref{sec:var_tar_ceph}), the light curve asymmetry is preserved. While some high amplitude \dsct{} and \bcep{} stars also show strongly asymmetric light curves \citep[e.g.][]{bowman2016amplitude, stankow2005catalog}, we do not model these stellar minorities in this work.} %
Furthermore, throughout Sect.~\ref{sec:var_targets}, a log-normal fit is used to describe the distribution of mode amplitudes, $a_i$, from the stellar sample in question. In this work, we restrict ourselves to simulations based on stable frequencies, $\nu_i$.

While this work uses observations from different instruments, a correction is needed to rescale the pulsation amplitudes observed in a given passband to that of the PLATO passband. This effect is particularly important for the current analysis of intermediate to massive stars, as the amplitude ratios of OBAF-type stars decrease with increasing central passband wavelength \citep[e.g.][]{heynderickx1994photometric, de2004asteroseismology, hey2024confronting, fritzewski2024mode}. The passband correction (PC) is modelled%
\footnote{The PC can also be determined in a model independent way using concomitant light curves from multiple instruments \citep[e.g.][]{bowman2016amplitude}.} %
as the quotient of the effective passband flux of a given star, which in turn is the product of the total spectral response, $S_{\lambda}$, and the spectral energy density, $F_{\lambda}$, from a synthetic spectrum:
\begin{equation}\label{eq:PC}
\text{PC}_{Y \rightarrow X} = \frac{\int_{\lambda_{X_1}}^{\lambda_{X_2}} F_{\lambda} \, S_X \, \text{d}\lambda}{\int_{\lambda_{Y_1}}^{\lambda_{Y_2}} F_{\lambda} \, S_Y \, \text{d}\lambda} \,,
\end{equation}
where $S_X$ is the response function for passband $X$ (i.e. the PLATO passband) and $Y$ is the response function for passband $S_Y$ (i.e. the reference passband of the photometric data). When possible, we use the PHOENIX %
library \citep{husser2013new} of high resolution spectral models as it covers a dense grid of stellar bulk parameters \{$M$, $R$, $\teff$, $\logg$, \metal{}\}. Since the PHOENIX grid only extends to $\tx{T}{eff}=\SI{12.2}{\kilo\kelvin}$, we use the ATLAS9 %
library \citep{castelli2003modelling} for earlier spectral type stars (extending to $\tx{T}{eff}=\SI{50}{\kilo\kelvin}$).

Additionally, the finite data sampling of astronomical instruments also alters the observed amplitudes. This phenomenon is described by the amplitude visibility function \citep[or `apodization' c.f.][]{hekker2017giant} which is a strong function of frequency \citep[e.g.][]{bowman2016amplitude, bowman2018characterizing}. Expressed in terms of the normalised sinc function, the apodization correction (AC) factor is given by:
\begin{equation}\label{eq:AC}
\text{AC}_{a_1 \rightarrow a_0} = \frac{a_0}{a_1} = \sqrt{\text{sinc}^{-2}\paren{\frac{\nu}{\tx{\nu}{Ny}}}} \,,
\end{equation}
where $a_0$ is the true signal amplitude and $a_1$ is the observed signal amplitude. It is evident from Eq.~\eqref{eq:AC} that the strongest suppression are near integer multiplets of the Nyquist frequency, $\tx{\nu}{Nq}=0.5/\delta t$.

\subsection{Target variables}\label{sec:var_targets}

To simulate the target variables, we primarily use \textit{Kepler} observations, as this mission shares a similar photometric precision, a similar observing strategy, and instrumental systematics common for a spacecraft located in the second Lagrange point (L2), as planned for PLATO. Hereto, the 4-yr \textit{Kepler} light curves serve as an excellent benchmark for creating the asteroseismic signals of our targets. For the remaining cases, light curves from TESS and OGLE were used due to their typical much larger sample size of variable stars.

\subsubsection{$\gamma$ Doradus stars}\label{sec:var_tar_gdor}

\gdor{} stars are high-order non-radial g-mode pulsators whose oscillations are excited by the flux blocking mechanism at the base of their convective envelopes \citep{guzik2000driving, dupret2005convection}. They are intermediate mass ($\sim$\SIrange{1.2}{2}{\Msun}) Population-I main-sequence stars with oscillation periods between $\sim$\SIrange{0.2}{3}{\day}.

To generate the oscillation modes for each mock object, we use the sample of 611 \textit{Kepler} \gdor{} stars from \cite{gang2020gravity} as benchmark. From this sample, three classes of oscillation modes dominate: dipole sectoral prograde g modes $(l,m)=(1,1)$, quadrupole sectoral prograde g modes $(l,m)=(2,2)$, and retrograde dipole Rossby (r) modes $(k,m)=(-2,-1)$ with an occurrence rate of around 62\%, 19\%, and 12\%, respectively. For simplicity, we generate synthetic light curves including only dipole sectoral g modes. Assuming that the period spacing changes linearly with period (imitating a smooth chemical gradient at the convective core boundary), we construct the mode periods, $P_i$, using the formalism by \citet[][typically used to construct a period-spacing échellogram]{gang2020gravity}:
\begin{equation}\label{eq:P_i}
P_i = \Delta P_0 \frac{(1+\Sigma)^i - 1}{\Sigma} + P_0 = \Delta P_0 \paren{n_i + \epsilon} \,,
\end{equation}
where $P_0$ and $\Delta P_0$ are respectively the first period and first period-spacing in the pattern, $\Sigma$ is the gradient of the period-spacing pattern, and $n_i \equiv [(1+\Sigma)^i - 1]/\Sigma$ is the normalised index, with $\epsilon = P_0/\Delta P_0$. A gradient in the period spacing pattern is a result of stellar rotation shifting the pulsation frequencies \citep{bouabid2013effects}. Prograde modes, considered in this project, show a downward gradient as seen by an observer in an inertial frame of reference (\citealt{vanreeth2015detecting}; Aerts et al. subm.).

\begin{figure}[t!]
\center
\includegraphics[width=\columnwidth]{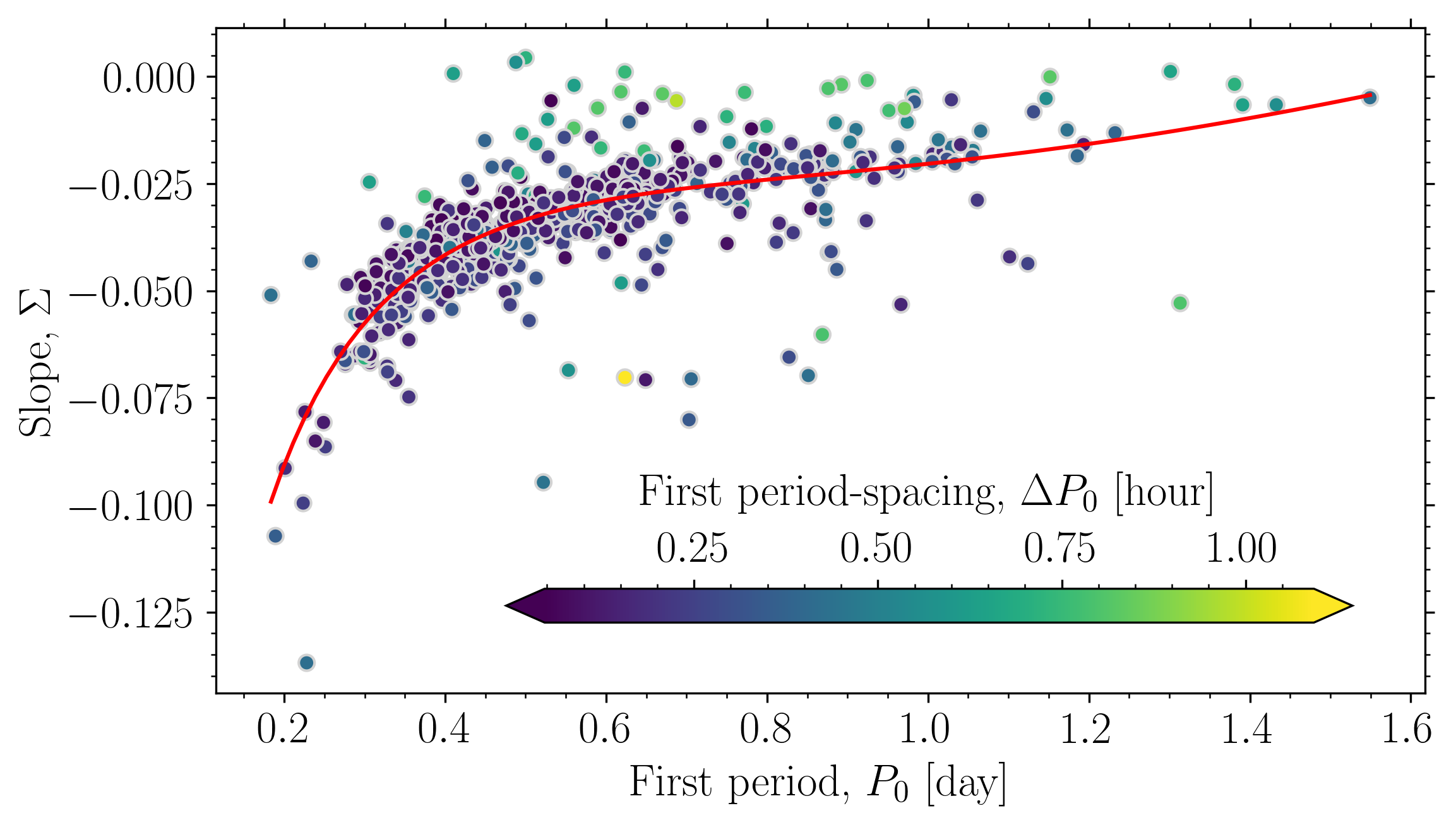}
\caption[]
{Model fit to the gradient--period relation for dipole sectoral prograde g modes from \cite{gang2020gravity}. Typically this diagram features the mean period, but here we correlate the first mode period in the period spacing pattern with the gradient. The colour scaling shows the value of the corresponding first period spacing.}
\label{fig:gdor_slope_vs_period}
\end{figure}

Since the gradient and the mean period spacing are correlated (as a consequence of the stellar rotation rate), we model the gradient from a (arbitrary) best fit model to the dipole mode periods of lowest radial order:
\begin{equation}\label{Sigma}
\Sigma(P_0) = c_1 \cdot e^{-c_2 \cdot P_0} + c_3 \cdot \log(c_4 \cdot P_0) + c_5 \,,
\end{equation}
where $\{c_1,\,c_2,\,c_3,\,c_4,\,c_5\}$ are model coefficients. Figure~\ref{fig:gdor_slope_vs_period} shows the model fit. From the \cite{gang2020gravity} sample, we construct a kernel density estimation (KDE) for $P_0$, $\Delta P_0$, and the total number of modes in the period-spacing pattern. A weighted extraction of these parameters was made, allowing also $\Sigma$ be computed with Eq.~\eqref{Sigma}.

While generating the mode periods with Eq.~\eqref{eq:P_i}, care was taken to avoid unphysical period-spacing patterns that would imply a rotational velocity above the critical value (i.e. if $\nu_i > \SI{3.3}{\per\day}$ the model generation was reset by drawing $P_0$ again). Moreover, the peak of maximum amplitude may be either the first or last mode period, which is typically not observed \citep{vanreeth2015detecting, gang2020gravity}. Hence the number index of this mode in the pattern is used to swap with the central mode period in the pattern. The number index is allowed a random uniform offset of $\tx{n}{offset} \in [-5, 5]$. Lastly, using the model generation described above, a synthetic light curve for each star was iteratively generated using Eq.~\eqref{eq:F(t)}.

\subsubsection{Slowly pulsating B stars}\label{sec:var_tar_SPB}

SPB stars are are high-order non-radial g-mode pulsators, whose oscillations are excited by the opacity ($\kappa$) mechanism operating in the partial ionisation zone of iron-group elements \citep{dziembowski1993opacity}. These stars are intermediate mass ($\sim$\SIrange{2.5}{8}{\Msun}) Population-I main-sequence stars with oscillation periods between $\sim$\SIrange{0.2}{3}{\day}.

\begin{figure}[t!]
\center
\includegraphics[width=\columnwidth]{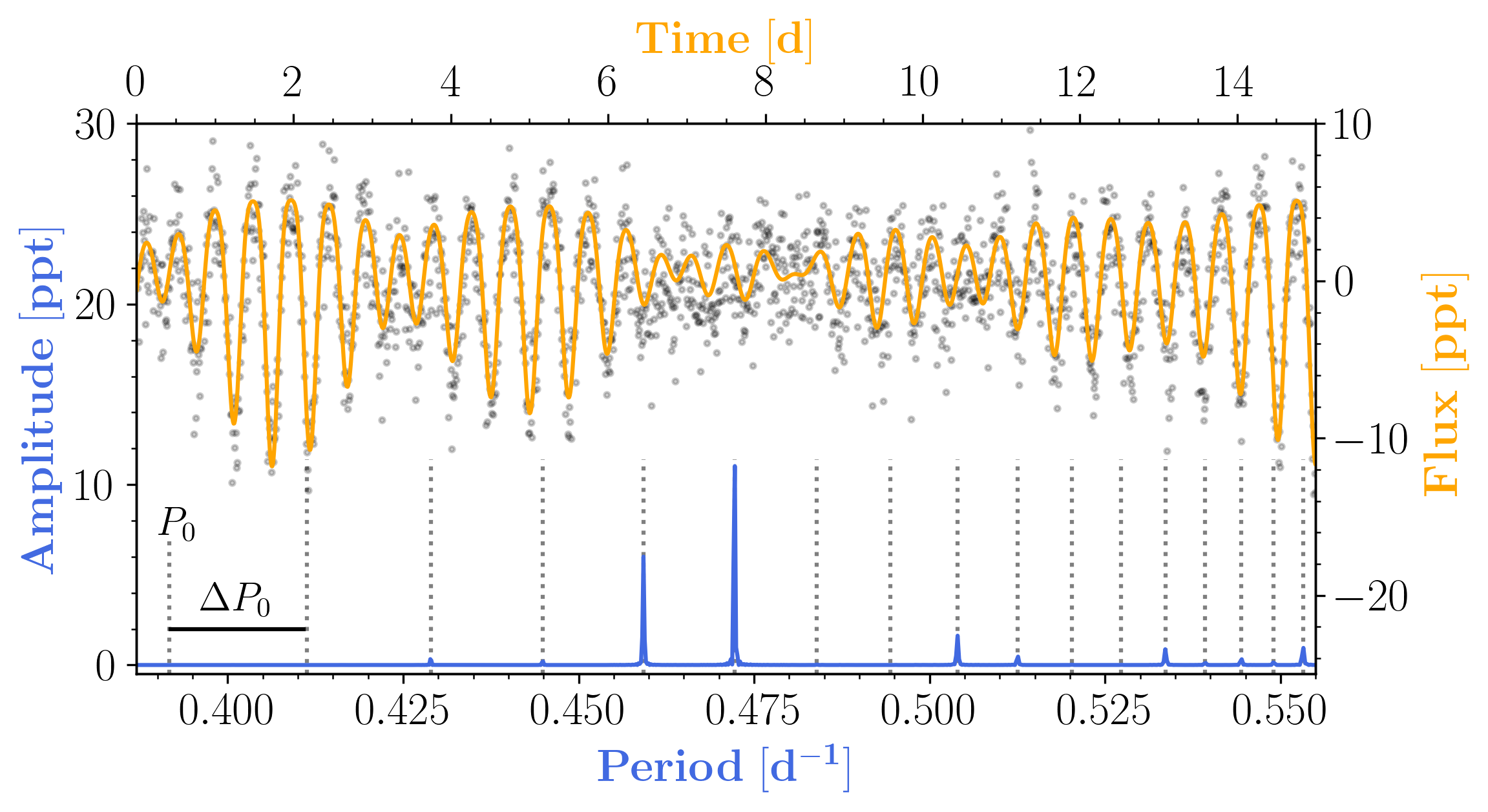}
\caption[]
{Example of a simulated SPB star. The lower/left axes belong to the amplitude spectrum (blue line), where we have highlighted the mode frequencies (dotted lines), the first mode period ($P_0$), and the first period spacing ($\Delta P_0$). The upper/right axes belong to the corresponding light curve (shown for the first 15 days) in its noise-less form (orange line) and simulated form (black points, representing a $\Pb\approx 10.3$ star observed with $\tx{n}{CAM}=6$).}
\label{fig:spb_ligthcurve}
\end{figure}

The \textit{Kepler} sample of 26 SPB stars from \cite{pedersen2021internal} was used to construct a model of the oscillation modes. Each noise-less light curve was computed following the same methodology as for \gdor{} stars of Sect.~\ref{sec:var_tar_gdor}, with the exception of $\Sigma$, which was drawn from its KDE distribution due to the low number statistics being disadvantageous for fitting it to a real \textit{Kepler} sample. Figure~\ref{fig:spb_ligthcurve} shows an example of a typical g-mode pulsation model of a SPB star.

\subsubsection{$\delta$ Scuti stars}\label{sec:var_tar_dsct}

\dsct{} stars are radial or low-order non-radial p-mode pulsators, whose oscillations are excited by the $\kappa$ mechanism acting in the partial ionisation zone of He II \citep{aerts2010asteroseismology}. Moreover, some \dsct{} stars have moderate radial order p modes excited by turbulent pressure \citep{antoci2014role, antoci2019first}. They are intermediate mass ($\sim$\SIrange{1.5}{2.5}{\Msun}) Population-I stars with pulsation periods between $\sim$\SIrange{0.2}{8}{\hour}. While these pulsators are found on the sub-giant branch, in this project we only consider main-sequence p-mode \dsct{} stars.


We use the \textit{Kepler} sample of 334 \dsct{} stars from \cite{bowman2018characterizing} to generate a database of synthetic light curves. Amongst these multi-periodic pulsators, 187 stars were selected to confidently avoid residual instrumental systematics present in the short cadence data. Due to the highly complex oscillation patterns amongst \dsct{} stars, we do not attempt to reproduce a physical model similar to that of the \gdor{} and SPB stars. For this work, we simply draw the number of modes, frequencies, and amplitudes directly from the KDE distribution established from the \textit{Kepler} sample. 


\subsubsection{$\beta$ Cephei stars}\label{sec:var_tar_bcep}

\bcep{} are low-radial order p- or g-mode pulsators whose oscillations are excited by the $\kappa$ mechanism acting in the partial ionisation zone of iron elements \citep{dziembowski1993opacity, gautschy1993on}. The \bcep{} stars are high mass ($\sim$\SIrange{8}{25}{\Msun}) Population-I stars, and their oscillation periods range between a few hours to several days. Most of them are main-sequence stars, but a significant fraction are (sub-)giants \citep[see][]{burssens2020variability}. 

As benchmark for the generation of oscillation modes, we use a sample of 196 \bcep{} stars initially assembled by \cite{de2023gaia} using Gaia DR3 observations and further refined and validated by \cite{hey2024confronting} and \cite{fritzewski2024mode} using TESS observations. Among this sample, 93 objects were already catalogued by \cite{stankow2005catalog}. The model generation of \bcep{} mock stars follows that of the \dsct{} sample (see Sect.~\ref{sec:var_tar_dsct}). 

\subsubsection{RR Lyrae stars}\label{sec:var_tar_rrlyr}

RR Lyrae stars are often called `classical radial pulsators' whose p modes are driven by the $\kappa$ mechanism acting in the partial ionisation zone of He\,I \citep{stellingwerf1984convection}. They are metal-poor, low-mass ($\sim$\SIrange{0.5}{0.9}{\Msun}) Population-II horizontal branch (i.e. He-shell burning) stars, where the majority pulsate in a dominant radial mode with period from $\sim$\SIrange{0.2}{1}{\day}. Originally from Bailey's classification scheme, today RR Lyrae are classified into RRab, RRc, and RRd types based on the amplitude and skewness in the light curves \citep[e.g.][]{mcnamara2014rrlyr, bono2020on}. A significant fraction of the RRab and RRc stars shows long-term amplitude and phase modulations, also known as the Tseraskaya-Blazko effect \citep{blazko1907mitteilung}.  

While partly being based on the work of \cite{molnar2022first}, we use a sample of 1538 synthetic RR Lyrae templates derived from TESS observations. Among these, we have 1105 RRab (99 being Tseraskaya-Blazko) stars, 382 RRc (17 being Tseraskaya-Blazko) stars, and 56 RRd stars. With such an abundant database of variable models we generate each noise-less light curve by simply drawing a source with replacement from the sample and introduce a small perturbation to the mode frequencies and amplitudes. We allow a constant (multiplicative) shift of all amplitudes of $\pm10\%$ compared to the dominant mode amplitude (resulting in a variety of light curve shapes seen for real RR Lyrae stars). Next, the first mode frequency was perturbed by $\pm10\%$ and the subsequent frequencies were adjusted proportionally, based on the original frequency ratios.

\subsubsection{Cepheid stars}\label{sec:var_tar_ceph}

Cepheid variables also belong to the class of `classical radial pulsators' excited by the $\kappa$ mechanism. While evolved stars of different masses may cross the classical instability strip at various evolutionary stages, the classification of Cepheids is quite diverse, forming three groups: classical Cepheids, anomalous Cepheids, and type II Cepheids \citep[see][for a review]{aerts2010asteroseismology}. 

While partly being based on the work of \cite{plachy2021tess}, we use a database of 2703 variable templates derived from TESS observations. Amongst 1339 classical Cepheids, our database contains 902 fundamental-mode, 366 overtone, and 71 double-mode pulsators. Amongst 264 anomalous Cepheids, our database contains 170 fundamental-mode and 94 first-overtone pulsators. Amongst 1078 type II Cepheids, our database contains 512 BL Herculis, 503 W Virginis star, and 63 RV Tauri stars. We follow the same methodology as used for the RR Lyrae sample to generate each mock object (see Sect.~\ref{sec:var_tar_rrlyr}).

\subsubsection{Compact pulsating stars}\label{sec:var_tar_compact}

sdB stars are evolved low- to intermediate-mass stars that have survived core helium (flash) ignition, and now populate the extreme blue end of the horizontal branch \citep[][see Fig.~\ref{fig:CaMD}]{heber2009hot, heber2016hot}. Pulsating sdB (sdBV) stars come in two flavours: V361 Hya and V1093 Her variables, both with modes excited by the $\kappa$ mechanism acting in the partial ionisation zone of iron elements \citep{charpinet1997potential, fontaine2012preliminary}. V361 Hya stars are slightly hot sdB stars with short-period ($\sim$\SIrange{1}{13}{\minute}), non-radial, low-order p modes. V1093 Her stars are on the other hand slightly cooler pulsating sdB stars with long-period ($\sim$\SIrange{1}{3}{\hour}), non-radial, low-order g modes. From the catalogue of 256 pulsating sdBV stars compiled by \cite{uzundag2024comprehensive}, we use the 17 richest pulsators observed by \textit{Kepler} suited for asteroseismic analysis. Among these, two are V361 Hya stars and 15 are V1093 Her stars.

Being the end product of stellar evolution for most (low- to intermediate mass) stars, WDs may enter one of three main instability regions as they slowly cool (i.e. DOV, DBV, and DAV). Which instability region they (potentially) enter depends on their atmospheric composition \citep[see the reviews of][]{fontaine2008pulsating, winget2008pulsating, corsico2019pulsating}. While \textit{Kepler}/K2 provided high-precision photometry for 27 DAV stars \citep[e.g.][Table~1]{hermes2017white}, only three DBV stars \citep{oestensen2011at, duan2021epic, zhang2024asteroseismology} and one DOV star \citep{hermes2017deep} were observed. In contrast, the on-going TESS survey has so far provided an abundant catalogue of WD pulsators, especially for H-deficient WDs. Considering the photometric precision and sample coverage, we simulate in total 30 WD pulsators with the highest number of detected modes, which includes 10 DAV \textit{Kepler}/K2 stars \citep{mukadam2004thirty, voss2006discovery, gianninas2006mapping, greiss2014asteroseismology, greiss2016search}, 10 DAV TESS stars \citep[c.f.][]{bognar2020tess, romero2022discovery, uzundag2023asteroseismological, romero2023asteroseismology}, 5 DBV TESS stars \citep[c.f.][]{bell2019tess, corsico2022pulsatingIII, corsico2022pulsatingV}, and 5 DOV TESS stars \citep[c.f.][]{corsico2021asteroseismology, uzundag2021pulsating, uzundag2022pulsating, oliveira2022pulsating, calcaferro2024pulsating}.

\subsection{Contaminating variables}\label{sec:var_contaminants}

With FGKM-type dwarf stars being the major occupant of our stellar (contaminant) catalogue, we additionally need to model their variability. Solar-like (FGK dwarfs) stars can exhibit convection-driven stochastic oscillations and various phenomenas of stellar activity (e.g. granulation and star spots), whereas M dwarfs are known to be magnetically active leading to frequent events of energetic flares. In the following we describe our adapted model for variable contaminants, including eclipsing binaries and a group of miscellaneous variables. 

\begin{figure}[t!]
\centering
\includegraphics[width=\columnwidth]{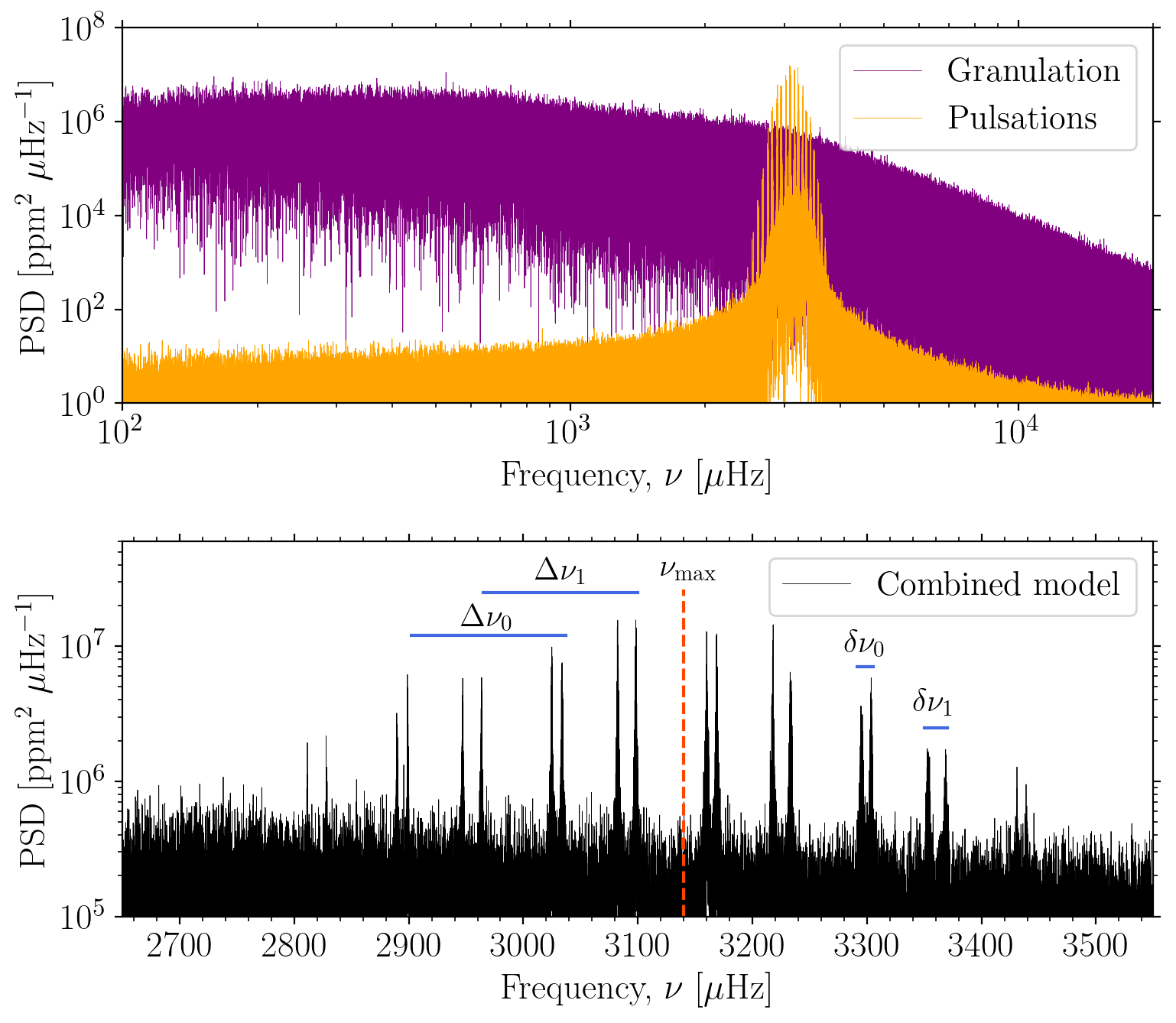}
\caption[]
{Power spectral density (PSD) diagrams showing granulation and pulsation signals caused by large convective envelopes in low-mass dwarfs. The PSD was computed for a 2-yr noise-less light curve. The bottom panel is a zoom-in on the combined model illustrating the standard envelope of excited modes (used to estimate the frequency of maximum power, $\tx{\nu}{max}$) and the small ($\delta\nu$) and large ($\Delta\nu$) frequency separations.}
\label{fig:oscillations}
\end{figure} 

\subsubsection{Solar-like oscillations}\label{sec:var_con_osc}

To generate convection-driven variability of solar-like stars in PLATO passband, we follow \cite{jannsen2024platosim}. In essence the stochastic oscillations are generated using 96 distinct pulsation mode frequencies of the Sun from the observational network BiSON %
\citep{chaplin1996bison,davies2014low,hale2016performance} and using asteroseismic scaling relations from \cite{kjeldsen1994amplitudes}. Granulation is modelled with two super-Lorentzian functions \citep[c.f.][]{kallinger2014connection} while scaling amplitudes using the methodology of \cite{corsaro2013bayesian}. The oscillation and granulation signals are modelled in the time domain directly using the formalism of \cite{de2006modelling}. Lastly, we use high resolution PHOENIX spectra to calculate the bolometric correction and scale the amplitude spectrum to what is expected in the PLATO passband. Figure~\ref{fig:oscillations} shows a model example of a solar-type star. 

\subsubsection{Star spots}\label{sec:models_con_spots}

\begin{figure}[t!]
\centering
\includegraphics[width=\columnwidth]{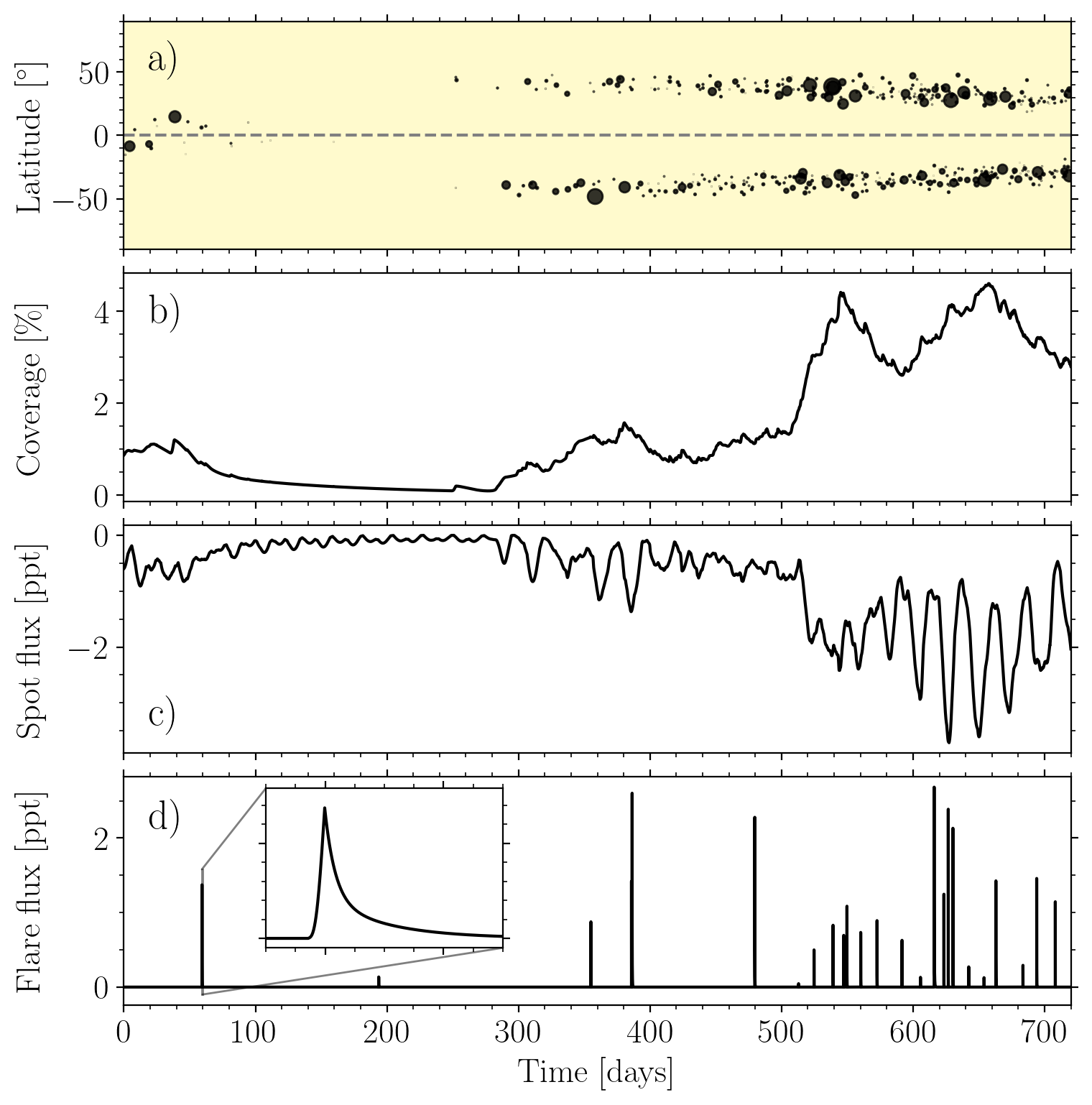}
\caption[]
{Illustration of a stellar on-axis activity model for a solar-like star simulated for two years. (a) Spot emergence diagram with the black dots being their latitudinal location and relative sizes at emergence with respect to the stellar surface. (b) Stellar surface area covered by spots (in percentage). (c) Relative flux darkening by the spot coverage (in ppt) as function time. (d) Relative flux brightening by stellar flares (in ppt) as function of time.}
\label{fig:starspot}
\end{figure} 

Stellar activity of solar-type stars is an important and non-negligible photometric noise component for PLATO photometry. From an observational point of view, stellar activity generally imprints itself as cyclic rotational modulations due to the presence of dark optical spots and bright faculae (and networks of the latter, so-called prominences), and transient events such as flares. In the effort of modelling spot modulations of solar-type stars in a broader scope for the PLATO core science program, we developed the code \texttt{pyspot}. We describe the functionality of the software in Appendix~\ref{app:spot}. In short, most parameter are being derived or selected at random from distributions based on \citet[][and references therein]{meunier2019activity}. Figure~\ref{fig:starspot} shows an example of the code's output, with panel a, b, and c showing the spot emergence diagram, the spot coverage diagram, and the relative flux diagram as function of time, respectively.

\subsubsection{Stellar flares}\label{sec:var_con_flares}

Flares are high energetic phenomena caused by reconnection of magnetic fields in the stellar atmosphere \citep{sun2015extreme}. Their dramatic increase of brightness, lasting on the order of minutes to hours, challenge the post-processing of space-based photometry, hence making them important to include in our simulations. To be representative of real stellar flares, we use the analytic model of \cite{davenport2014Akepler} which is based on a sample of more than 6100 single flare events of the M dwarf star GJ1243 that was observed by \textit{Kepler}. This model describes a fast polynomial rise and a two-phased exponential decay (c.f. Eq.~1-3). \cite{mendoza2022llamaradas} improved the model of \cite{davenport2014Akepler} using a more robust subsample of flares from GJ1243 and modern statistical methods. This model is not yet implemented in \platosim{}. With only slight differences between the two flare templates, the improved model is expected to have a minimal impact on the asteroseismic yields of this project, but a greater impact is expected for extrinsic variability research (such as eclipse signatures).

Given the difference in internal structure and magnetic topology of FGKM dwarf stars, we use the spectral type dependent distributions of \cite{doorsselaere2017stellar} to draw the amplitudes (Fig.~4) and the occurrence rates (Fig.~6) for each star. Using the occurrence rate, the total number of injected flares was scaled by the activity rate (compared to solar) determined by the software \texttt{pyspot}. Each flare duration is drawn from a uniform distribution between \SIrange{1}{200}{\min}, and following scaled to the flare amplitude such that more energetic events last longer. Finally, the spot coverage (also provided by \texttt{pyspot}) is used as weight while drawing the times of peak flaring flux. Hence, the flare occurrences realistically follow the the stellar magnetic cycle, as illustrated in Fig.~\ref{fig:starspot}d for an active solar-like star.

\subsubsection{Eclipsing binaries}\label{sec:var_con_EB}

The multiplicity fraction is a rapidly increasing function with mass, meaning almost all OB-type stars are found in multiple star systems \citep[see][for an overview]{offner2022origin}. Especially, eclipsing binary (EB) systems are overly presented in all-sky surveys \citep{ijspeert2024automated}. For simplicity, we only model EBs (ignoring higher orders of binarity, cataclysmic binaries, attached systems, etc.) with an intermediate to massive star primary. Moreover, exoplanet transits are excluded in this project as they would have a minimal effect on our analysis. 

We use a database of 3155 EB candidates observed by TESS from \cite{ijspeert2021allsky}, out of which 2946 systems were confidently selected as genuine EB systems. This catalogue contains detached main-sequence EBs with an OBA-type dwarf primary.

\subsubsection{Miscellaneous variables}\label{sec:var_con_mis}

All other stellar objects, not catagorised photometrically in previous subsections (acknowledging our enforced pigeonholing)%
\footnote{To make the set of models manageable, we exclude pseudo-periodic and non-periodic signals from Be stars, young stars with discs, and phenomena related to accretion, novae, and supernovae.}, %
are placed into a class of miscellaneous variables. We divide this class into long-period variables (LPVs) and short-period variables (SPVs). Generally, LPVs are red giant stars that populate the reddest and brightest regions of the CaMD and consist in terms of oscillatory behavior of Mira stars, semi-regular variables, slow irregular variables, and small-amplitude red giants. Due to the custom detrending approach employed in this work (which removes any trend longer than $\sim$\SI{30}{\day}; see Sect.~\ref{sec:sim_photometry}), LPVs are not considered as target stars but only as stellar contaminants. We use the extensive library of LPVs from OGLE \citep{udanski2008ogle, udalski2015ogleiv} for the model generation.

Being part of PLATO's calibration strategy, short-period chemically peculiar ApBp stars show extremely stable photometric signatures of chemcial spots that cause rotational modulation. From current studies we expect around 10\% of all B5-A5 stars to possess a fossil magnetic field and therefore exhibit stable rotational modulations caused by chemical stratification (i.e. spots) in such stars \citep{renson2009catalogue, chojnowski2019discovery, mathys2020long}. Hence, a significant fraction of these stars are expected in the PLATO FOV. We simulate their noise-less light curves by drawing the rotational frequencies from a uniform distribution covering the usual $[1,\,3]\,\si{\per\day}$. The light curve shapes of these stars is modelled with a simple sinusoid with occasional contribution of the second harmonic. While the amplitude modulations of these stars observed by \textit{Kepler} \citep{holdsworth2021roap} and TESS \citep{holdsworth2021tess, holdsworth2024tess} are typically below \SI{30}{\milli\magnitude}, we assume that all are high amplitude BpAp stars with an amplitude distribution of \SIrange{10}{30}{\milli\magnitude} in the PLATO passband.

\section{Simulations}\label{sec:simulations}

\subsection{Spacecraft systematics}\label{sec:sim_spacecraft}

Instrumental systematics are crucial to include in any realistic set of simulations \citep[e.g.][]{borner2024plato}. \platosim{} provides a wealth of options and features to configure the PLATO payload in this regard \citep{jannsen2024platosim}. In this project we configure \platosim{} with settings from the PLATO mission parameter database (MPDB)%
\footnote{\tiny{\url{https://indico.esa.int/event/407/contributions/7401/attachments/4799/7321/1220-Abstract-PLATO Mission Parameter Database.pdf}}}. %
Common for all simulations we assume that the properties of each camera (i.e. the telescope optical unit and the optics; \citealt{munari2022plato}; the detectors and the front-end electronics; \citealt{koncz2022plato}) are identical and that their noise sources are uncorrelated. All simulations include cosmic rays with a constant hit rate of \SI{10}{\events\per\centi\meter\squared\per\second}. Moreover, an analytic model of the point spread function (PSF; mimicking the expected variation in PSF morphology across the focal plane) was used and charge diffusion was activated using a Gaussian kernel of \SI{0.2}{\pixel}.

From a computational point of view, the most important systematics to consider are pointing error sources that may impact the shape and stability of the PSF, meanwhile introducing a pixel displacement of the PSF barycenter in the CCD focal plane. In particular, two systematic errors that alter the location of the PSF on the CCD are camera misalignments \citep[see][]{royer2022alignment} and imperfect pointing repeatability between consecutive mission quarters. We account for these by randomly drawing the camera pointing and the platform pointing from a uniform distribution with a $3\sigma$ confidence that the error is below that of the mission requirements. Furthermore, we apply differential kinematic aberration (DKA) using a realistic L2 orbit.

We differentiate the configuration of random and systematic noise sources between the three simulation batches: \affogato{}, \cortado{}, and \doppio{}. Drawing all parameters from the PLATO mission database, \affogato{} and \doppio{} complies to the MPDB `as expected' instrument design, whereas \cortado{} comply to the MPDB `as required' instrument design. In particular, the only difference between \affogato{}/\doppio{} and \cortado{} regarding spacecraft systematics are the differences in settings for detector noise, the attitude and orbit control system (AOCS) jitter, and the thermo-elastic distortion (TED) of the payload. Table~\ref{tab:systematics} shows the noise parameters related to the PLATO detectors given their requirement; a linear fit between beginning-of-life (BOL) and end-of-life (EOL) values is used to properly account for the time dependence of the CCD/FEE read noise and the dark current, which is updated at the beginning of each mission quarter. 

\begin{table}[t!]
\caption[]{Configuration parameters for the PLATO detectors. The CCD/FEE readout noise and the dark current are here tabulated at BOL, and EOL within the rounded brackets.}
\begin{center}
\begin{tabular}{lcc}
\hline\hline
Parameter & Expected & Required \\
\hline
FEE gain F/E side [\si{\adu\per\micro\volt}] & 0.0186 & 0.0186 \\
CCD gain F/E side [\si{\micro\volt}/\si{\electron}] & [2.04, 2.28] & [1.8, 2.5]\\
FEE read noise [\si{\electron\per\pixel}] & 32.8 (37.7) & 32.8 (37.7) \\
CCD read noise [\si{\electron\per\pixel}] & 23.2 (25.0)	& 24.5 (28.0) \\
Dark current [\si{\electron\per\second}]  & 0.54 (4.0)	& 1.2 (4.5)		\\
Dark current stability [\%]	              & 0.7			& 5.0 \\
\hline
\end{tabular}
\label{tab:systematics}
\end{center}
\end{table}

A common AOCS red noise jitter model sampled at \SI{0.1}{\hertz} was used, but the root mean square (rms) values of the Euler angles \{yaw, pitch, roll\} are configured differently: \affogato{} and \doppio{} use \{0.036, 0.034, 0.040\}~arcsec, and \cortado{} uses \{0.144, 0.136, 0.160\}~arcsec. The former values correspond to the rms values extracted from the simulations of a dynamical AOCS model representative for the mission requirements `as expected'. The TED for each camera and mission quarter is included using a second order polynomial model whilst uniformly drawing the model coefficients under the restriction that the amplitude in yaw, pitch, and roll cannot exceed \SI{9}{\arcsec} and \SI{2.25}{\arcsec} (being equivalent to \SI{0.6}{\pixel} and \SI{0.15}{\pixel})%
\footnote{The e2v CCD270 plate scale is $\SI{15}{\arcsec\per\pixel}$.} %
in three months, respectively. We further use a dynamical TED model to realistically add the repeating pattern of momentum dumps of the reaction wheels with a typical amplitude of $\sim$\SI{30}{\ppm} (which will happen every three days for PLATO). Next, a TED time series is created for each camera group via small perturbations in order to simulate a more realistic distribution of correlated noise across the platform. In summary, the AOCS jitter and TED amplitudes for \cortado{} are four times larger than those of \affogato{} and \doppio{}.

\subsection{The pixel subfield}\label{sec:sims_subfield}

For each simulation (i.e. star/camera/quarter configuration), only contaminant stars within a radial distance of \SI{45}{\arcsec} (equivalent to three pixels in the CCD focal plane) away from their corresponding target star are simulated. Although our stellar catalogue is complete up to $\Pb \sim 17$, to avoid a negligible contribution from faint stellar contaminants for the brighter targets, a relative target-to-contaminant magnitude threshold of $\Delta \Pb = 5$ was used (being equivalent to an intensity ratio of 100) to limit the number of contaminants in a single subfield.

As part of keeping the PLATO telemetry budget at a minimum, small $6\times\SI{6}{\pixel}$ imagettes, will be placed around each (non-saturated) target star. The usage of the imagettes in this project is however undesirable since long-term drifts of a star's barycentric position may move it outside the subfield and thus cause flux anomalies. By design, the subfield is always placed around a target star such that the target is within the central pixel of the subfield (i.e. maximally half a pixel offset from the central intra-pixel position). From the inclusion threshold of stellar contaminants, this means that all contaminants, before considering barycentric drift, are located within a central circular aperture of \SI{6.5}{\pixel} in diameter. In principle the maximum amplitude for the DKA can be as high as \SI{0.8}{\pixel} per mission quarter \citep{samadi2019plato}. However, for the vast majority of cases the DKA amplitude is much lower. Hence, the combined (TED and DKA) pixel displacement in a single mission quarter rarely exceeds \SI{1}{\pixel}. Thus, even for a worst case scenario of having the long-term drift purely in a horizontal or vertical direction, a $8\times\SI{8}{\pixel}$ subfield is an optimal trade-off regarding computational speed and preventing flux anomalies.

\subsection{Variability injection for contaminants}\label{sec:sims_contaminants}

For \platosim{} the variable injection happens at pixel-level and at run time. Thus, the database of noise-less light curves (c.f. Sect.~\ref{sec:catalogue}) are considered. For \affogato{} and \cortado{}, only the target star is variable and the injection follows Sect.~\ref{sec:variability}. For \doppio{} we also inject stellar variability into the contaminants, for which we first check for binaries (c.f. Sect.~\ref{sec:cat_samples}). If a contaminant star belongs to one of the 853,869 sources in our catalogue classified as a binary system, we randomly draw an EB model from our database (c.f. Sect.~\ref{sec:var_con_EB}). If not a binary system the classification continues to the next stage for single stars.

\begin{figure}[t!]
\center
\includegraphics[width=\columnwidth]{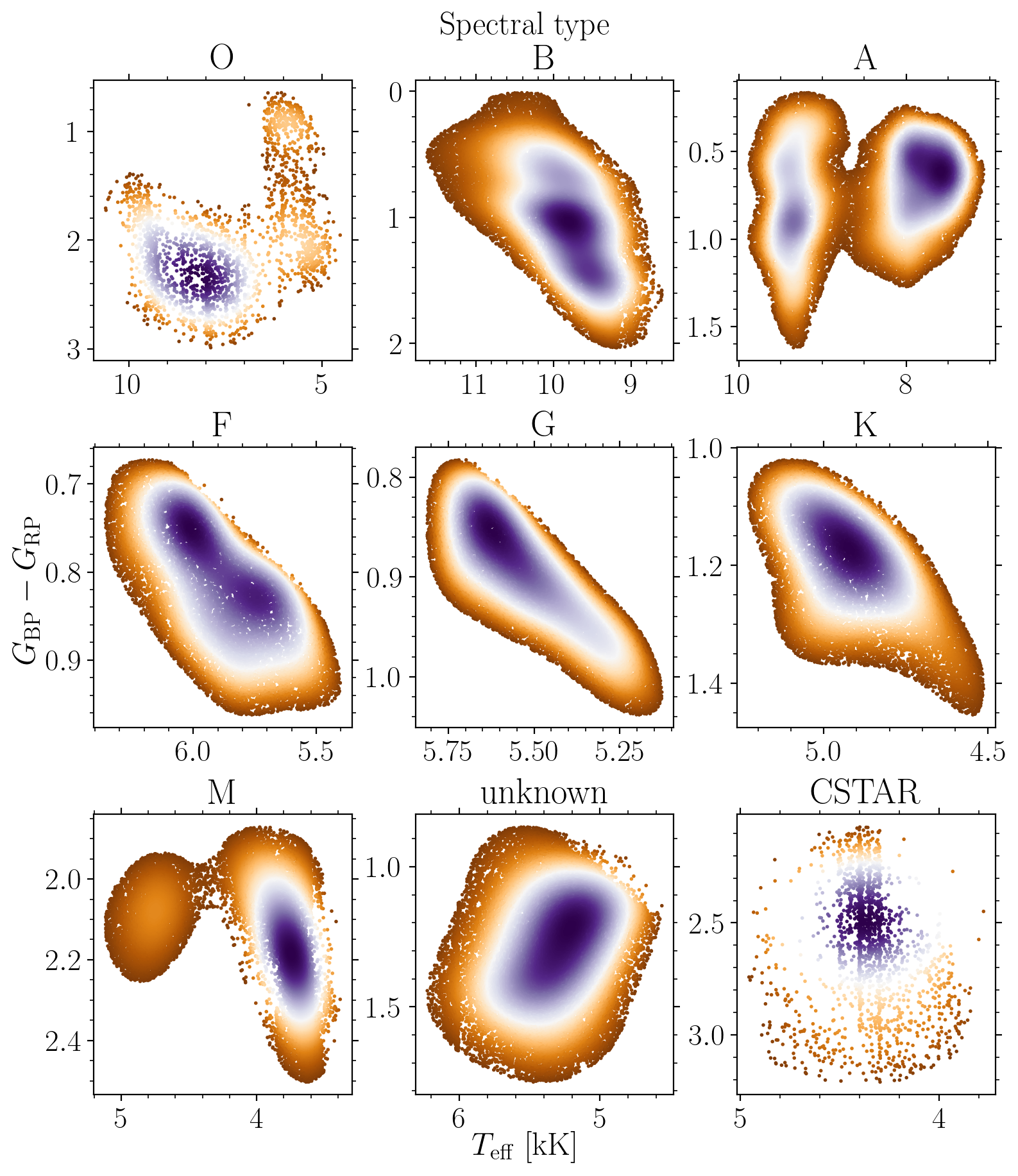}
\caption[]
{Gaussian KDE distribution per spectral type of the LOPS2--\textit{Gaia} DR3 single stars with valid stellar bulk parameters. The normalised density is scaled from low to high with brown to purple colour, respectively. A cut in the density region, discarding the lowest 25--50\%, has been made to each spectral type to better confine each region.} 
\label{fig:distributions_spectype}
\end{figure}

In the single star stage we first check that the stellar parameters \{$M$, $R$, $\teff$, $\logg$, \metal{}\} are defined for all stars within the subfield. Since \textit{Gaia} does not always provide these bulk parameters, we draw them from empirically estimated KDE distributions as in Fig.~\ref{fig:distributions_spectype}, assembled into spectral types provided by \textit{Gaia}. Each spectral type KDE is created from the subset of stars with valid stellar parameters, meanwhile an arbitrary 25-50\% density cut is used to remove potentially spurious solutions. It is clear from the unphysical bimodal structures (i.e. for O, A, and M-type), that the \textit{Gaia} pipeline suffers from systematic biases. The spectral type tag \texttt{CSTAR} are candidate carbon stars (whose complex morphology we ignore in this project) and the \texttt{unknown} spectral types are typically non-stellar or extragalactic objects. Furthermore, a smaller group of stars do not have any spectral classification (in total 27,981 objects, i.e 0.4\% of our catalogue), hence we merged these into the group of \texttt{unknown} stars. For each star we retrieve the stellar parameters using three cases based on Fig.~\ref{fig:distributions_spectype}: i) If the effective temperature and the colour are known, we select a star (and its parameters) from a closest match to these parameters of the respective KDE distribution; ii) if a star only has colour information, this is used to draw a star along a horizontal slice of the KDE for a given spectral type; iii) if no colour information is available and only the spectral type of a star is known, a direct weighted draw from the KDE distribution is performed. 

With each contaminant having a complete set of stellar parameters, we first check if a contaminant belongs to any of the pulsation classes of MOCKA. If true, we create its model as explained in the respective subsection of Sect.~\ref{sec:var_targets}. If false, we consider variability of lower mass dwarf and giant stars. As outlined in Sect.~\ref{sec:var_contaminants}, depending on the stellar parameters and magnitude cuts, we model stars later than spectral type F5 using one or more of the components: granulation, stochastic oscillations, spots, and flares. We introduce cuts in the CaMD using a linear relation of the form: $\tx{M}{G}=a\,(\tx{G}{BP}-\tx{G}{RP})+b$, with model coefficients $\{a\,,b\}$. 

\begin{figure*}[]
\center
\includegraphics[width=2\columnwidth]{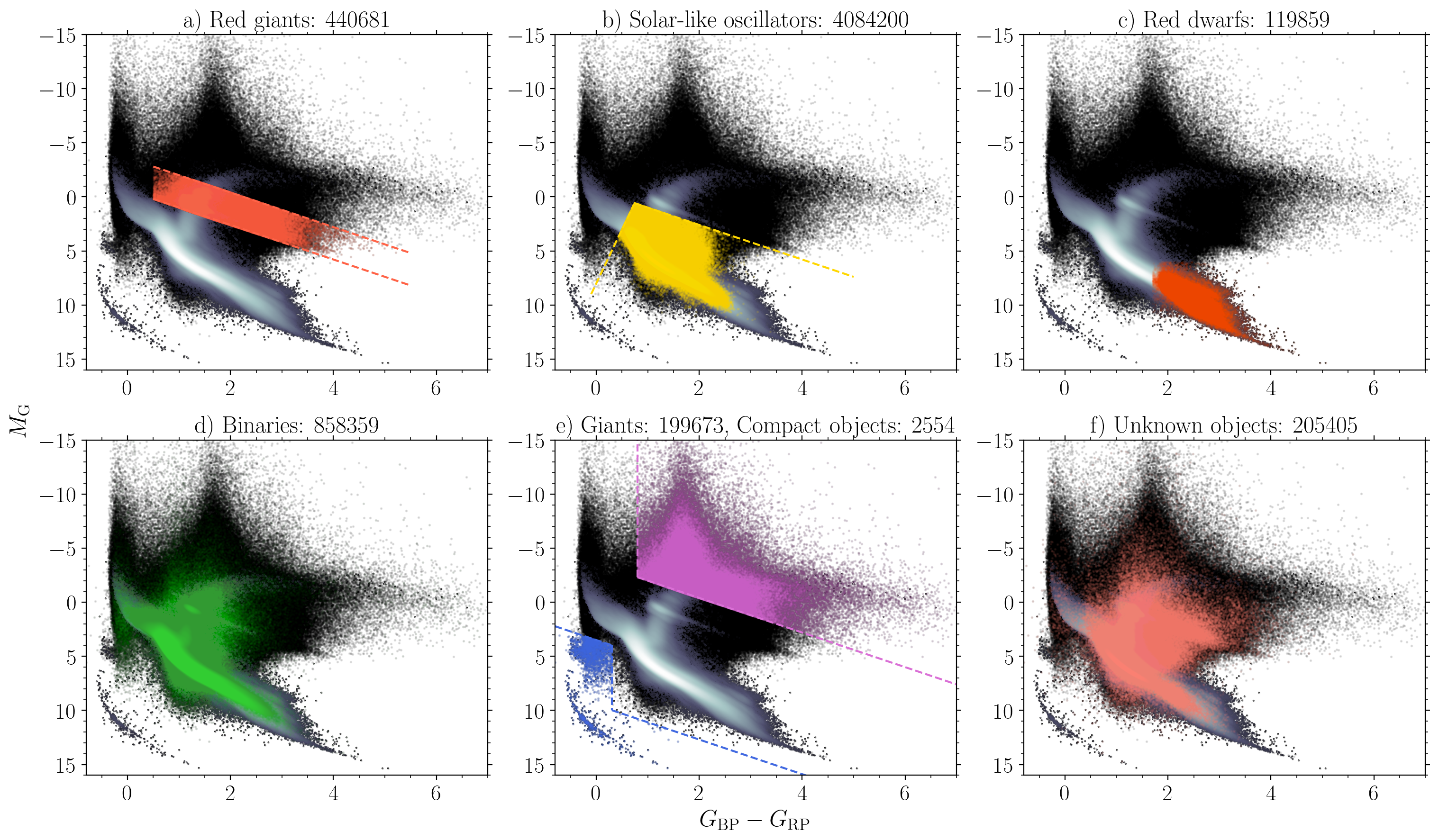}
\caption[]
{CaMD of different stellar classes used to assign a variable signal to stellar contaminants. The respective class is written in the title header together with the number count of the class. The upper panels show the confinement of all stars that feature solar-like variability: a) red giants, b) solar-like dwarfs and subgiants, and c) red dwarfs. The lower panels show the confinement of d) binaries, e) giants and compact objects (above the purple line and below the blue lines, respectively), and f) all objects with an \texttt{unknown} spectral type. The black points are all stars in the LOPS2 ($G \lesssim 17$), whereas the metallic contour defines stars within \SI{1}{\kilo\pc} from the Sun.} 
\label{fig:variable_contaminants}
\end{figure*}

Since dwarf, subgiant, and giant stars of spectral type F5-K7 have a convective envelope, the expectation is that all of them are solar-like oscillators. We first confine red giant stars in the CaMD within an upper boundary using $(a,\,b)=(1.6,\,-3.6)$ and a lower boundary using $(a,\,b)=(1.6,\,-0.6)$, respectively. Further we require that $\tx{G}{BP}-\tx{G}{RP}>0.5$ and $\log g < 3.8$ (Fig.~\ref{fig:variable_contaminants}a). We confine solar-type (main-sequence and subgiant) stars using the lower red giant star boundary in addition to a CaMD cut using $(a,\,b)=(-10,\,8)$, while further requiring that $\log g > 4.2$ (Fig.~\ref{fig:variable_contaminants}b). We furthermore confine rotational spotted variables in CaMD using the observational dearth of these stars below $\tx{G}{BP}-\tx{G}{RP} < 0.4$ \citep[e.g. see][]{gaia2018dr3varability}, which theoretically corresponds to the transition region between stars having radiative versus convective outer envelopes \citep{kippenhahn2013stellar}. To increase the variety among the solar-like stars, each star is assigned a stellar flare probability based on its spectral type being $p\rm(F, G, K, M)=(20\%, 80\%, 90\%, 100\%)$. Lastly, we confine active M dwarfs using the cuts of $\tx{G}{BP}-\tx{G}{RP}>1.7$ and $\tx{M}{G} > 6$. Shown in Fig.~\ref{fig:variable_contaminants}c, these red dwarf stars were modelled to have spots and flares.

If not assigned as a binary (see Fig.~\ref{fig:variable_contaminants}d), a solar-like oscillator, or a red dwarf, the classification simply assumes that the stellar contaminant is a miscellaneous variable. As mentioned in Sect.~\ref{sec:var_con_mis}, we model short and long period miscellaneous variables. Shown in Fig.~\ref{fig:variable_contaminants}e, we confine LPVs in the CaMD as those above the upper red giant boundary and with $\tx{G}{BP}-\tx{G}{RP}>0.8$. Lastly, any star with an \texttt{unknown} spectral classification (see Fig.~\ref{fig:variable_contaminants}f) is simply classified as a SPV. Hence, we artificially increase the noise budget by assuming that all extra-galactic objects are SPVs. 

\subsection{Photometric pipeline}\label{sec:sim_photometry}

The GO programmes of PLATO, demanding on-board photometry, will rely on an optimal aperture mask algorithm by \cite{marchiori2019flight}. This algorithm is implemented in \platosim{} and constructs a pixel mask based on a simple trade-off between lowering the noise-to-signal ratio (NSR) while mitigating stellar contamination. In particular, the algorithm defines a stellar pollution ratio (SPR; an important parameter for the forthcoming analysis) measuring how much contaminating flux leaks into the aperture mask. As part of the mission strategy \citep{rauer2024plato}, the aperture mask of each star may be updated periodically during a mission quarter given that a lower NSR can be achieved. The mask-update frequency is currently undecided. Hence, to be conservative, we choose an update every \SI{30.5}{\day} (which may or may not enforce a trigger twice per quarter). 

As illustrated by \citet[][Fig.~18]{jannsen2024platosim}, the current best estimate of a single PLATO camera light curve shows several levels of instrumental systematics. The three dominant instrumental features are the long-term trends caused by displacement of the PSF barycenter (due to the effect of TED, KDA, etc.), large jumps in the mean flux level between consecutive mission quarters (due to changes of optical throughput), and a slow continuous flux decrease (due to ageing effects). Additionally, relevant for the on-board photometry, mask-updates likewise introduce jumps in the mean flux level.     
     
While the official post-processing pipeline is still under construction, we have implemented a simple reduction pipeline needed to perform our analysis. Several sub-modules of the official post-processing pipeline exist \citep[e.g. the \texttt{REPUBLIC} detrending algorithm;][]{barragan2024republic}. Hence, as a better performance is expected from the official pipeline, the employed pipeline should be a conservative approach with respect to the removal of systematics (potentially counteracting the increased complexity of reality versus simulations). Appendix~\ref{app:pipeline} shows each of the reduction steps presented in the following.

To correct for the long-term systematic trends in the light curves, we use a polynomial model for detrending the signal of each individual camera and mission quarter segment. If a mask-update was triggered, each sub-segment was detrended individually. To account for the varying segment lengths (i.e. roughly $\SI{30}{\day}$, $\SI{60}{\day}$, and $\SI{90}{\day}$) we perform a model comparison between a polynomial fit of first to fourth degree by obtaining the Bayesian information criterion (BIC) probability: 
\begin{equation}
p_i(\text{BIC}) = \frac{e^{-\frac{1}{2}\Delta\text{BIC}_i}}{\sum_{j=1}^J e^{-\frac{1}{2}\Delta\text{BIC}_j}} \,,
\end{equation}
where $\Delta \text{BIC} \equiv \text{BIC} - \text{BIC}_{\text{min}}$ and BIC being determined from each ordinary least squares fit. The model with the highest BIC probability was selected for the fit.

Outliers were removed only after the detrending to avoid sharp features in the light curve (e.g. flux jumps due to mask-updates) to be flagged as outliers. Closely resembling PLATO's on-board outlier rejection algorithm, we employ the software \texttt{Wōtan}%
\footnote{\tiny{\url{https://github.com/hippke/wotan}}} %
\citep{hippke2019wotan} and use an upper and lower threshold in multiples of the median absolute deviation (MAD) with the following magnitude dependence:
\begin{equation}
\text{MAD}(\Pb) =
\begin{cases}
5,   & \text{for} \quad \Pb \leq 10 \\
4.5, & \text{for} \quad 10 < \Pb < 11 \\
4,   & \text{for} \quad \Pb \geq 11 \,.
\end{cases}
\end{equation}
From empirical tests, the length of the filter window was set to half a day, and the middle point in each window was calculated using the median.

For each star, the multi-camera and multi-quarter light curves were combined to a single dataset. Next, a second iteration of the outlier removal was applied following the procedure described above. Data points sharing the exact same timings (i.e. camera observations from the same camera group) were averaged. Depending on the pulsation class, the final light curve per star was sampled at a cadence of $\delta t = \{25,\,50,\,600\}\,\text{s}$ (c.f. Table~\ref{tab:variables}), corresponding to a Nyquist frequency of $\tx{\nu}{Nq} = \{1728,\, 864,\, 72\}\,\text{d}^{-1}$, respectively.%
\footnote{We note that due to the \SI{6.5}{\second} time offset between the four CCDs, the final cadence of the delivered photometry is in fact a time average.}
 
The uncertainty on the flux measurements was computed as the internal scatter in the time series to account for all additional sources of error introduced by previous reduction steps. From a computational point of view, $\sigma$ was calculated using a carbox filter of size $N$. Hence, for each time stamp $i$, we equated the fractional uncertainty by calculating a short-term filter $\sigma_i(N=\SI{1}{\hour})$. To robustly estimate the uncertainties in the presence of intrinsic stellar variability, a long-term filter $\mu_i(N=\SI{10}{\day})$ was applied to the filtered signal and used as the mean flux error.  

\begin{figure}[]
\center
\includegraphics[width=\columnwidth]{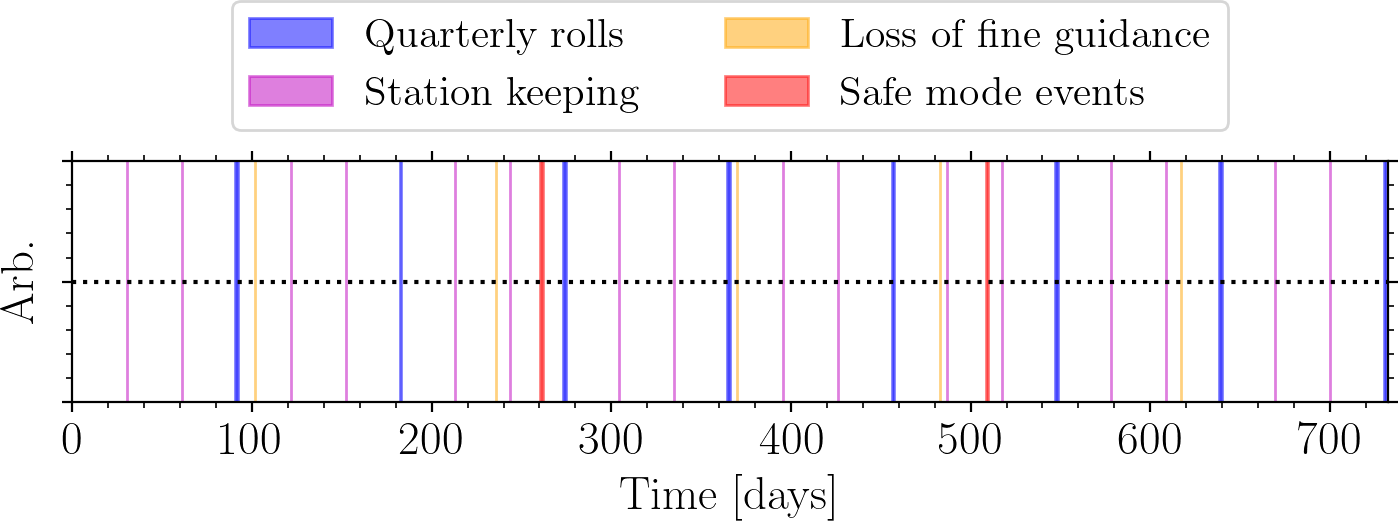}
\caption[]
{Assembly of data gaps included prior to the frequency analysis. Besides the quarterly rotational realignment of the spacecraft (blue lines), we also consider downtime due to station keeping manoeuvres (purple lines), loss of fine guidance (orange lines), and safe mode events (red lines).} 
\label{fig:data_gaps}
\end{figure}

The final light curves of MOCKA span two years and include quarterly data gaps of exactly one day (allowing a custom mission downtime distribution to be applied). As shown in Fig.~\ref{fig:data_gaps}, a realistic distribution of data gaps was introduced for further analysis in this project. This was done to properly account for the impact on the asteroseismic analysis from a degraded window function \citep[e.g.][]{garcia2014impact, pascualgranado2018impact}.  

In line with the latest performance study, we apply inter-quarter data gaps (approximately every \SI{91}{\day}) with a duration of $\Delta T_i = 1.5 + \mathcal{U}(-0.5, 0.5)\,\si{\day}$, where a duration anomaly (drawn from a uniform distribution) was included. To maintain PLATO's nominal Lissajous orbit around L2, a station-keeping manoeuvre is planned every \SI{30}{\day}. As the duration of these manoeuvres is still unknown, we conservatively estimate it to be $\Delta T_i = 2.5 + \mathcal{U}(-0.5, 0.5)\,\si{\hour}$. Lastly, using statistics from the 4-year \textit{Kepler} mission, we further introduced data gaps from coarse pointing events (i.e. temporary loss of fine guidance resulting in an increased photometric scatter) and safe mode events (i.e. temporarily operation shut-offs due to unexpected events). The duration of these two events was computed with $\Delta T_i = 30 + \mathcal{U}(-15, 15)\,\si{\min}$ and $\Delta T_i = 1 + \mathcal{U}(-0.5, 0.5)\,\si{\day}$, respectively.%
\footnote{Related to data gaps, in principle any large movement of the spacecraft (e.g. quarterly rotations, safe mode events, etc.) will introduce small thermal changes across the payload (affecting mounts, optics, and detectors), which gradually relaxes typically on time scales from hours to days. Seen as exponential decaying structures in the extracted light curve, these so-called `thermal transients' are caused by an increased detector gain or/and general PSF breathing. Due to PLATO's aperture mask strategy this may have a significant effect on the on-board photometry, hence, we acknowledge their importance for future simulations.} 

\subsection{Frequency extraction}\label{sec:sim_frequency}

With the delivery of high-quality space-based data, the recovery of mode frequencies increasingly depends on the iterative prewhitening strategy employed \citep[e.g. see][for a comparison between five different methods for g modes in SPB stars]{vanbeeck2021detection}. The frequency extraction in this work was performed with the software \texttt{STAR\,SHADOW}%
\footnote{\tiny{\url{https://github.com/LucIJspeert/star_shadow}}} %
\citep{ijspeert2024automated}. Although developed for the detection of eclipsing binaries, this software contains a robust iterative prewhitening procedure: i) it extracts the frequency of the highest amplitude one by one, directly from the Lomb-Scargle periodogram \citep{lomb1976least, scargle1982studies} as long as it reduces the BIC of the model by $\Delta\text{BIC}<2$, and; ii) a multi-sinusoid non-linear least-squares optimisation is performed, using groups of frequencies to limit the number of free parameters. 

We emphasise that the prewhitening strategy used by \texttt{STAR\,SHADOW} is conceptually different from a more standardised procedure of invoking a stopping criterion based on signal-to-noise ratio (SNR). Thus, as an alternative to the $\Delta\text{BIC}<2$ stopping criterion, we extended the iterative prewhitening module of \texttt{STAR\,SHADOW} implementing a SNR stopping criterion. In our analysis we recorded all extracted frequencies until a certain SNR threshold. A commonly used criterion is $\text{SNR} = 4$, which was empirically determined from ground-based observations of p-mode oscillators by \cite{berger1993nonradial}. The validity of this criterion in comparison to space-based surveys is however questionable \citep[e.g. see][]{baran2015detection, zong2016amplitude, baran2021detection, bowman2021towards}. In fact, the premise of a standardised SNR criterion relies on the assumption of Gaussian noise, in which only the cadence, $\delta t$, and number of data points, $N$, are model parameters \citep{baran2021detection}. 

\begin{figure}[t!]
\center
\includegraphics[width=\columnwidth]{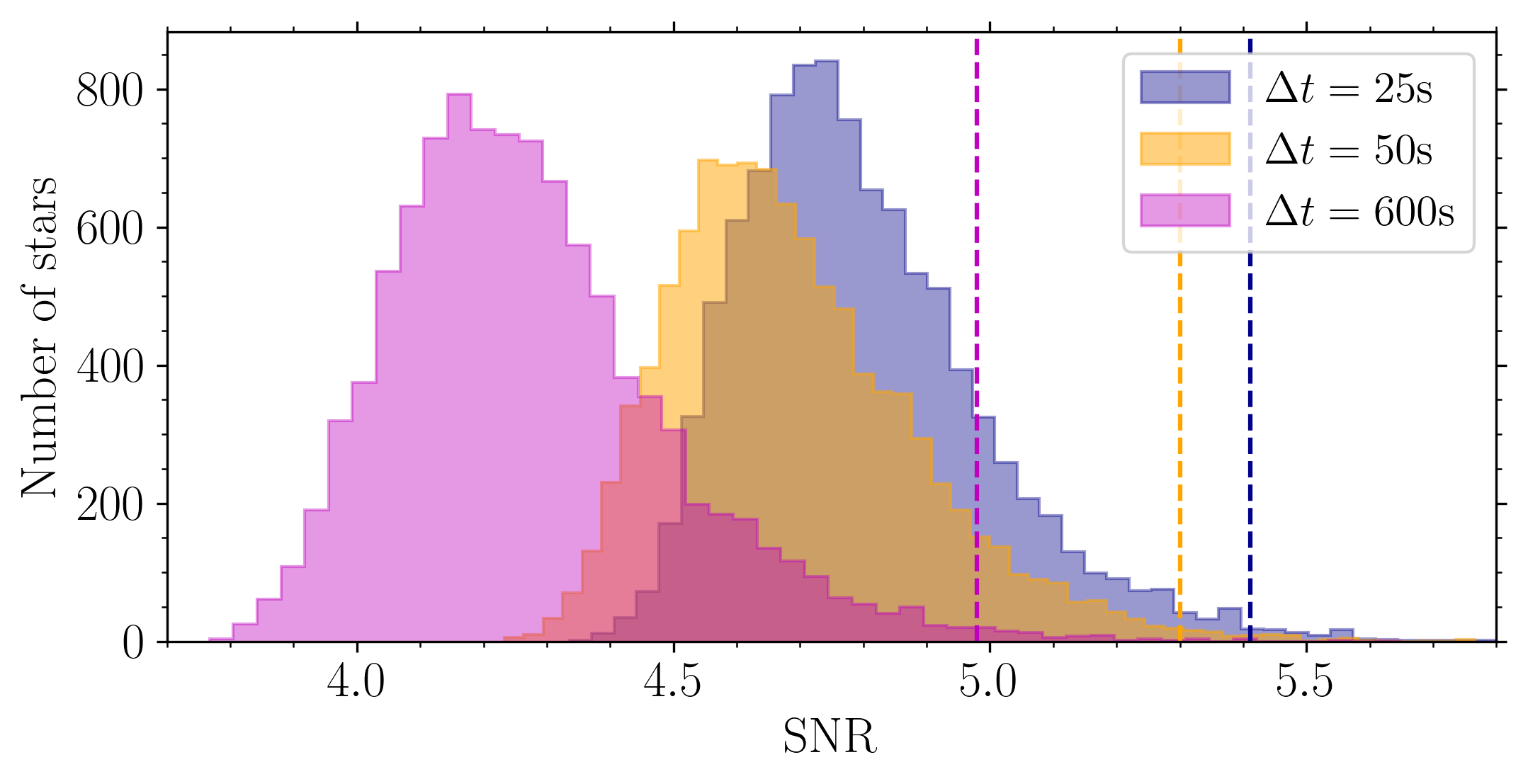}
\caption[]{Histogram of the SNR of the highest amplitude frequency extracted from 10\,000 synthetic white noise PLATO light curves. This example shows the optimal SNR criterion (dashed lines) estimated as the $\text{FAP}=1\%$ from time series with a duration of eight mission quarters.} 
\label{fig:snr_criterion}
\end{figure}

We investigated what SNR criterion is optimal for PLATO light curves as function of sampling $\delta t \in \{25,\,50,\,600\}\,\text{s}$ and number of mission quarters $\tx{n}{Q} \in [1,\,8]$. We assume that two days are lost due to downtime (approximately agreeing with Fig.~\ref{fig:data_gaps}), meaning that a mission quarter of $\sim\SI{89.31}{\day}$ contains $N=\{304128,\,152064,\,4224\}$ data points for the nominal, twice nominal, and long N-CAM cadence, respectively. We generated 10\,000 synthetic light curves, including only white noise for each duration and sampling, and used, for consistency, \texttt{STAR\,SHADOW} to compute the Lomb-Scargle periodogram of each light curve up to the Nyquist frequency. Following \cite{baran2021detection}, the amplitude spectrum was standardised using its median to get rid of the dependence of the rms average noise level. The SNR value of the highest amplitude peak was computed as the ratio between the peak-amplitude and the local noise level using a \SI{1}{\per\day} frequency window in the residual periodogram. Lastly, we calculate the false alarm probability (FAP) from the resultant histogram. Figure~\ref{fig:snr_criterion} shows an example of the computed criterion for $\tx{n}{Q}=8$ and $\text{FAP}=1\%$. 

\begin{figure}[t!]
\center
\includegraphics[width=\columnwidth]{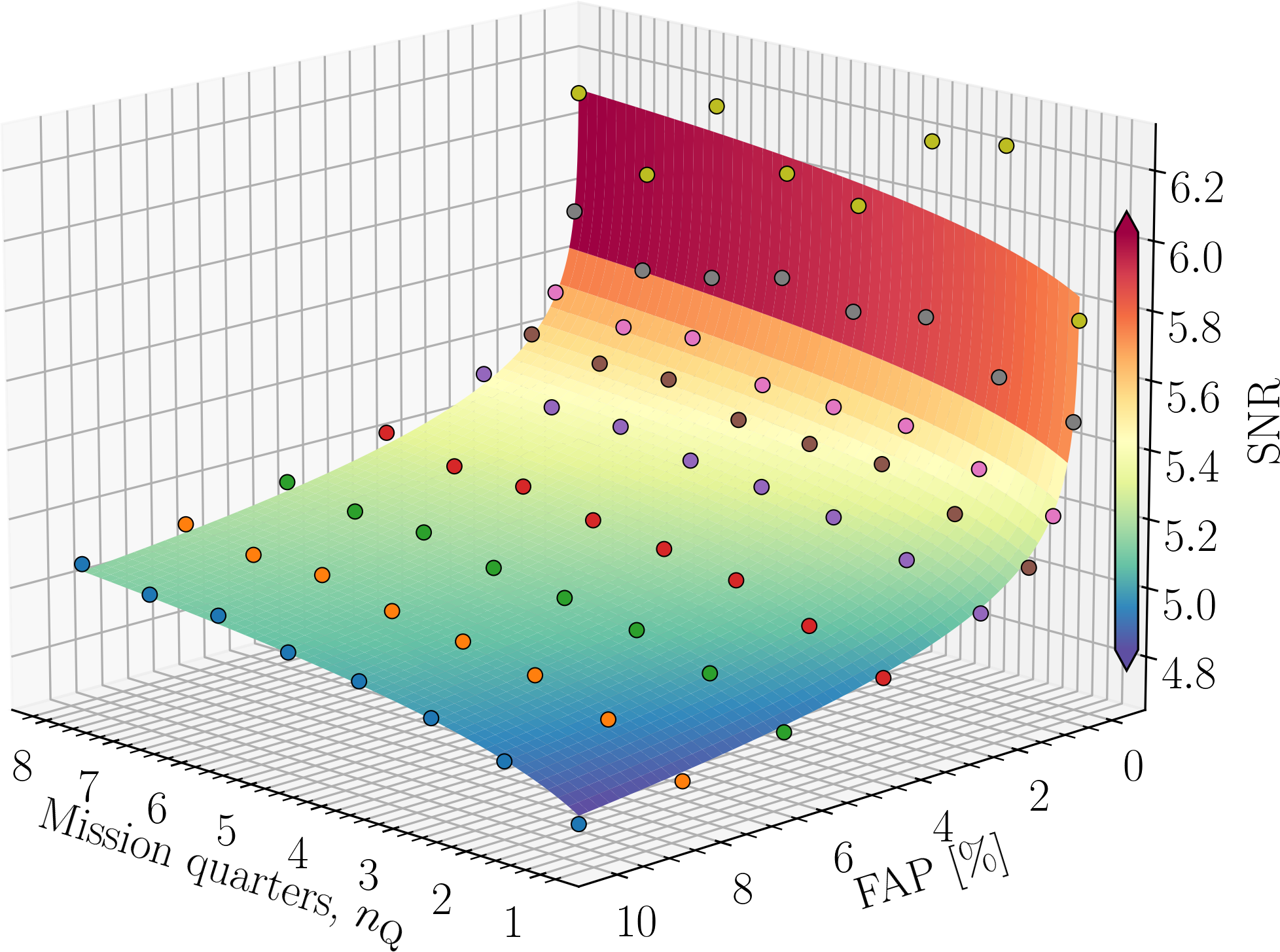}
\caption[]{Optimum significance surface for PLATO light curves with a cadence of $\delta t = \SI{25}{\second}$. The surface fit was performed with Eq.~\eqref{SNR_cadence} to the grid points (sorted in colour after the FAP, with semi-regular grid points of \{10, 8, 6, 4, 2, 1, 0.5, 0.1, 0.01\}\%).} 
\label{fig:snr_surface}
\end{figure}

\cite{baran2021detection} showed that the $\text{SNR}(\tx{n}{Q})$ can be described by a simple logarithmic relation. This is also true for $\text{SNR}(\text{FAP})$ due to the log-normal behaviour of the SNR histograms (see Fig.~\ref{fig:snr_criterion}). Performing the above exercise for each cadence, mission quarter duration, and over a semi-regular grid of $\text{FAP} \in [0.01, 10]\%$, results in a so-called significance criterion surface for each cadence:
\begin{equation}\label{SNR_cadence}
\text{SNR}_{\delta t}(\tx{n}{Q},\,\text{FAP}) = d_1\,\ln\,\tx{n}{Q} + d_2\,\ln\,\text{FAP} + d_3 \,,
\end{equation}
where the best fit coefficients $\{d_1,\,d_2,\,d_3\}$ are listed in Table~\ref{tab:snr_coefficients}. Eq.~\eqref{SNR_cadence} states that the SNR criterion is increasing: i) the shorter the cadence; ii) the longer the duration of the time series, and; iii) the lower the FAP. The latter two interferences are clearly illustrated in Fig.~\ref{fig:snr_surface} showing the best fit SNR surface to the simulations with a \SI{25}{\second} cadence.

\begin{table}[t!]
\caption[]{Best fit coefficients to Eq.~\eqref{SNR_cadence}, describing the optimal SNR surface, for the three standardised PLATO cadences.}
\begin{center}
\begin{tabular}{cccc}
\hline\hline
Cadence [s] & $d_1$ & $d_2$ & $d_3$ \\
\hline
25 		& 0.13217934 	& -0.15429918  	& 5.12996448 \\
50 		& 0.10930811 	& -0.14909435  	& 5.05757044 \\
600 	& 0.15992845 	& -0.16496221  	& 4.62691121 \\ 
\hline 
\end{tabular}
\end{center}
\label{tab:snr_coefficients}
\end{table}

It is worth noticing that for the most commonly adopted thresholds in the literature of $\text{FAP}=1\%$ and $\text{FAP}=0.1\%$, the optimal SNR is always larger than four, independent of the duration of the light curve. This highlights that the traditional threshold is too optimistic for continuous space-based data \citep[agreeing with previous studies, e.g.][]{degroote2010deviations, baran2015detection}. Furthermore, since PLATO light curves consist of co-added observations of multiple cameras, residual instrumental systematics are difficult to avoid. Instrumental systematics typically show excess power at low frequencies in amplitude spectra, which increases the probability of detecting higher amplitude noise peaks. Thus, any SNR value determined using Eq.~\eqref{SNR_cadence} should be seen as a lower boundary when dealing with long-period variable signals. In the forthcoming analysis, the SNR threshold computed for $\tx{n}{Q}=8$ and $\text{FAP}=0.1\%$ were selected, which for $\delta t \in \{25,\,50,\,600\}\,\text{s}$ are $\text{SNR} \in \{5.76,\,5.63,\,5.34\}$.

As part of the final post-processing step, we computed the frequency and amplitude precision of the pulsation modes from the amplitude spectrum (see Fig.~\ref{fig:pipeline_model_comparison} for the workflow), rather than from analytical formulae as in \cite{montgomery1999derivation} or an adaptation of it as in \cite{zong2021oscillation}. \texttt{STAR\,SHADOW} uses more advanced numerical techniques to compute the frequency and amplitude precisions \citep[see][for details]{ijspeert2024automated}. From the list of observed pulsation modes (extracted using the BIC criterion), we queried and matched each mode one by one to the list of injected modes. First, this was done by searching for all possible modes within a \SI{0.1}{\per\day} frequency window from the injected mode frequency. If multiple mode frequency candidates were found, a closest match to the input frequency and amplitude was computed by minimising $\chi^2 = (a_i/\sigma_{a_i})^2 + (\nu_i/\sigma_{\nu_i})^2$, with $\sigma_{a_i}$ and $\sigma_{\nu_i}$ being the relative difference between the observed and injected amplitude and frequency, respectively. Next, a residual diagram in frequency and amplitude was computed based on the matched modes. Finally, a root mean square (rms) estimate of the frequency and amplitude precision were calculated for pulsation modes detected with both the BIC and SNR criterion. This procedure was repeated for each star, while the (noise) peak of smallest amplitude was additionally recorded. We note that rms frequency precision in this project is a simple approximation, since the `real' frequency precision depends on many factors \citep[e.g. see][]{bowman2021towards}. In the next section we present the results of the combined frequency extraction for the \gdor{} and SPB star samples.

\section{Results: g-mode pulsators}\label{sec:results}
  
\subsection{Noise budget in the frequency domain}

\begin{figure}[t!]
\center
\includegraphics[width=\columnwidth]{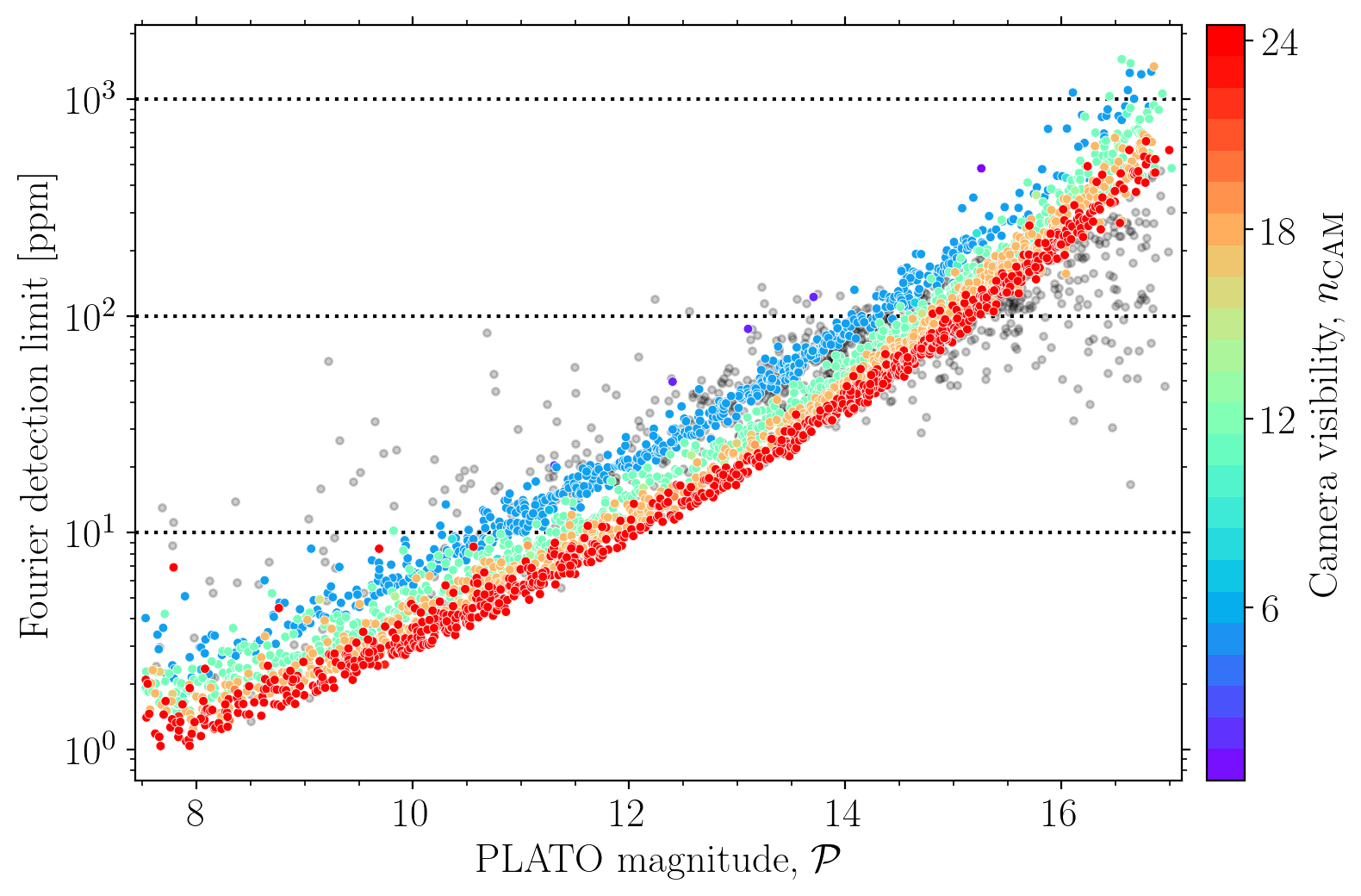}
\caption[]{Amplitude detection limit from a prewhitening strategy using the SNR stopping criterion and a high amplitude version of the \affogato{} SPB sample (c.f. Sect.~\ref{sec:results_limit_ampl}). This plot illustrates the general noise budget in the frequency domain enforced by random and systematic noise sources. Like in the time domain (see Fig.~\ref{fig:nsr}), the detection limit at mission level is a clear function of the camera observability (colour scale). The dark grey data points are stars with a SPR value larger than 6\% (a critical threshold explained in Sect.~\ref{sec:results_contamination}). The dotted horizontal lines are reference limits. We highlight that multiple stars have a camera visibility different from $\tx{n}{CAM}\in\{6, 12, 18, 24\}$ due to the usage of pointing error sources in our simulations (c.f. Sect.~\ref{sec:sim_spacecraft}). The same is true for Fig.~\ref{fig:result_detection_limit} and \ref{fig:result_limit_SPB_high}. While we illustrate the star count histogram (versus $\tx{n}{CAM}$) of this figure in Fig.~\ref{fig:star_count_histogram}, the effect in this plot is most noticeable from the four (upper-most) purple/blue data points with $\tx{n}{CAM}\in\{1, 2\}$.} 
\label{fig:result_limit_fourier}
\end{figure}

We start by considering the limiting amplitude that can be retrieved from prewhitening in the presence of random and systematic noise.  Figure~\ref{fig:result_limit_fourier} shows the result as computed from each stellar light curve of the \affogato{} (high amplitude, c.f. Sect.~\ref{sec:results_limit_ampl}) SPB sample as a function of magnitude and camera visibility. As expected from the NSR estimates in the time domain, the amplitude detection limit is a clear function of the camera visibility, with the smallest amplitudes being retrieved from light curves combined from 24 N-CAMs. 

Compared to the NSR estimate in the time domain, the amplitude detection limit is an order of magnitude lower when applying modern prewhitening strategies like that of \texttt{STAR\,SHADOW}. We highlight that the simulation batch \affogato{} is mainly dominated by Gaussian-like noise, i.e. the underlying noise model is well described by a jitter, photon, and sky/detector dominated noise regime \citep[c.f. Fig.~\ref{fig:nsr}; see also][]{borner2024plato}. With PLATO covering variable signals with amplitudes ranging from ppm to ppt level, Fig.~\ref{fig:result_limit_fourier} shows that for the expected operation of 24 N-CAMs, PLATO will be able to detect peaks in the periodogram below \SI{10}{\ppm}, \SI{100}{\ppm}, and \SI{1000}{\ppm} for $\Pb \lesssim 12$, $\Pb \lesssim 15$, and $\Pb \lesssim 17$, respectively.

Although we have illustrated that the underlying amplitude distribution for most pulsating stars is well described by a log-normal relation, Fig.~\ref{fig:result_limit_fourier} highlights that the dominant pulsation mode should be retrievable within a fairly large magnitude range, independent of the underlying asteroseismic sample. In fact, the dominant mode amplitude for the \gdor{} and SPB sample is recovered in most cases up to their simulation-limited magnitude of $\Pb=14$ and $\Pb=16$. We recover 98.9\% and 95.8\% for the \gdor{} sample and 98.2\% and 95.4\% for the SPB sample as per simulation batch \affogato{} and \cortado{}, respectively.  
 
\subsection{Limiting mode amplitude yields}\label{sec:results_limit_ampl}

\begin{figure*}[t!]
\center
\includegraphics[width=\columnwidth]{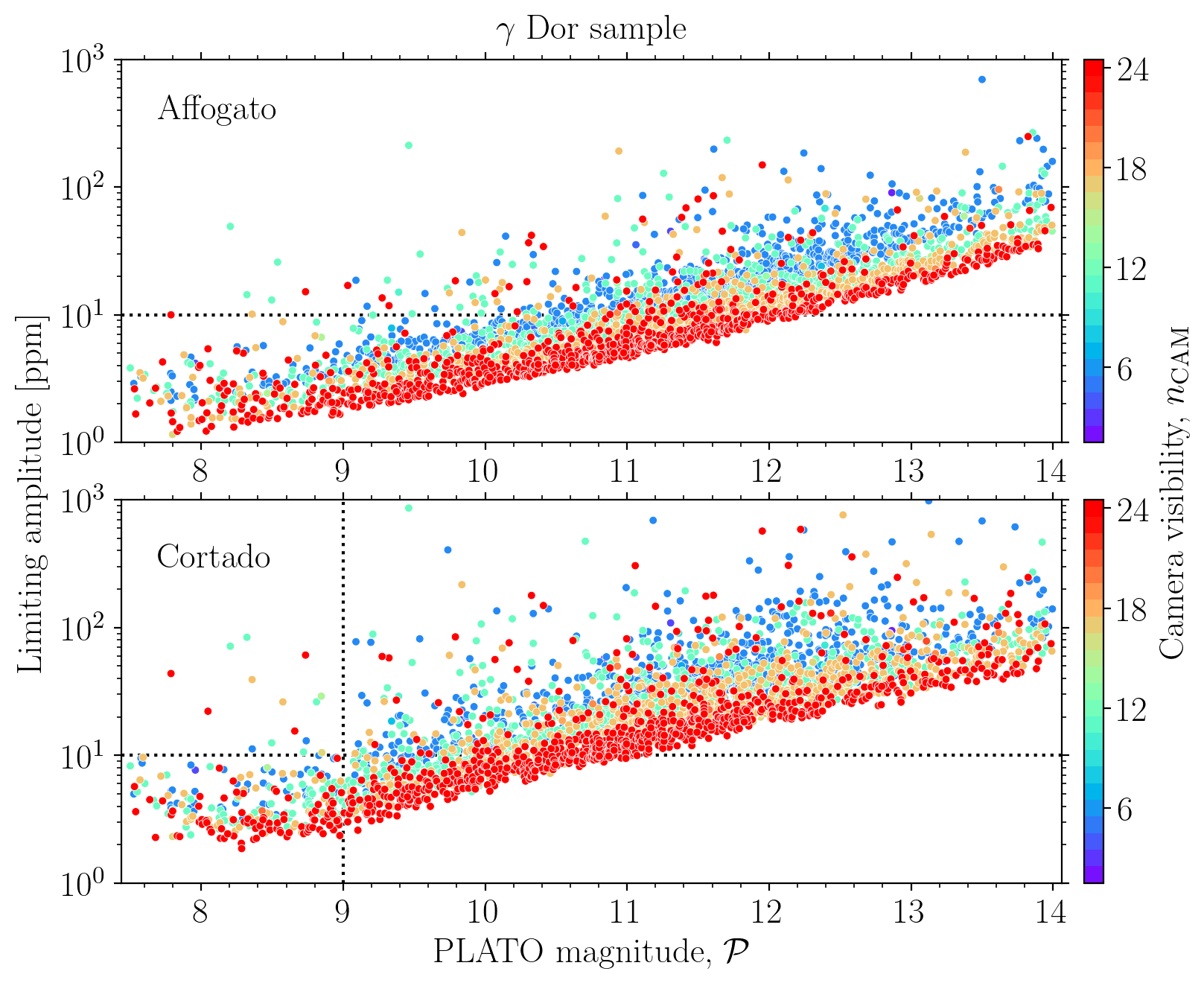}
\includegraphics[width=\columnwidth]{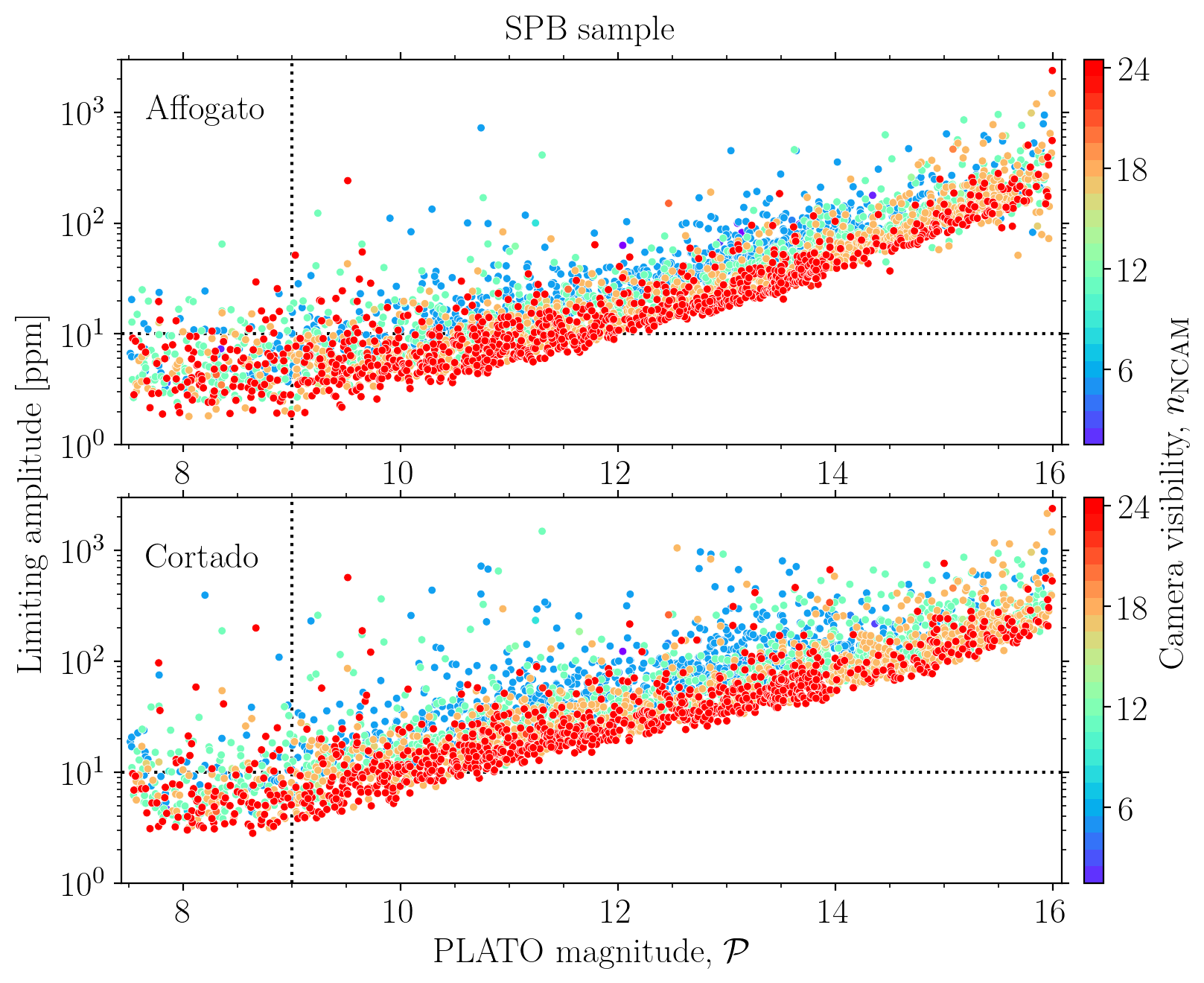}
\caption[]{Limiting mode amplitude detection for the \gdor{} sample (left panels) and the SPB sample (right panels). Top and bottom panel show the lowest mode amplitude detected for each star of the simulation batch \affogato{} and \cortado{}, respectively. The data points are colour-coded after camera visibility. The dotted horizontal lines are \SI{10}{\ppm} reference lines to compare the two batches. The dotted vertical line in the bottom panel marks a transition between a noise floor plateau (left of line) and an increasing detection limit.} 
\label{fig:result_detection_limit}
\end{figure*}

The smallest mode amplitudes detectable as a function of magnitude for each star of the \gdor{} and SPB sample are shown in Fig.~\ref{fig:result_detection_limit} (left and right panels, respectively), from the simulation batch \affogato{} (top panels) and \cortado{} (bottom panels). First of all, increased scatter is generally observed for \cortado{} as compared to \affogato{}. Secondly, as observed in the time domain, the noise budget of \affogato{} seems to follow the expected underlying noise model (c.f. Fig.~\ref{fig:result_limit_fourier}). On the other hand, the increased level of instrumental systematics of \cortado{} clearly impacts the general shape of this distribution, forming a single log-linear relation between $9<\Pb<13$. Using the dotted horizontal lines (at \SI{10}{\ppm}) as references, it is evident that the impact of instrumental systematics enforces that a similar detection limit of \affogato{} can only be reached if observing a magnitude brighter for \cortado{}. 

\begin{figure}[t!]
\center
\includegraphics[width=\columnwidth]{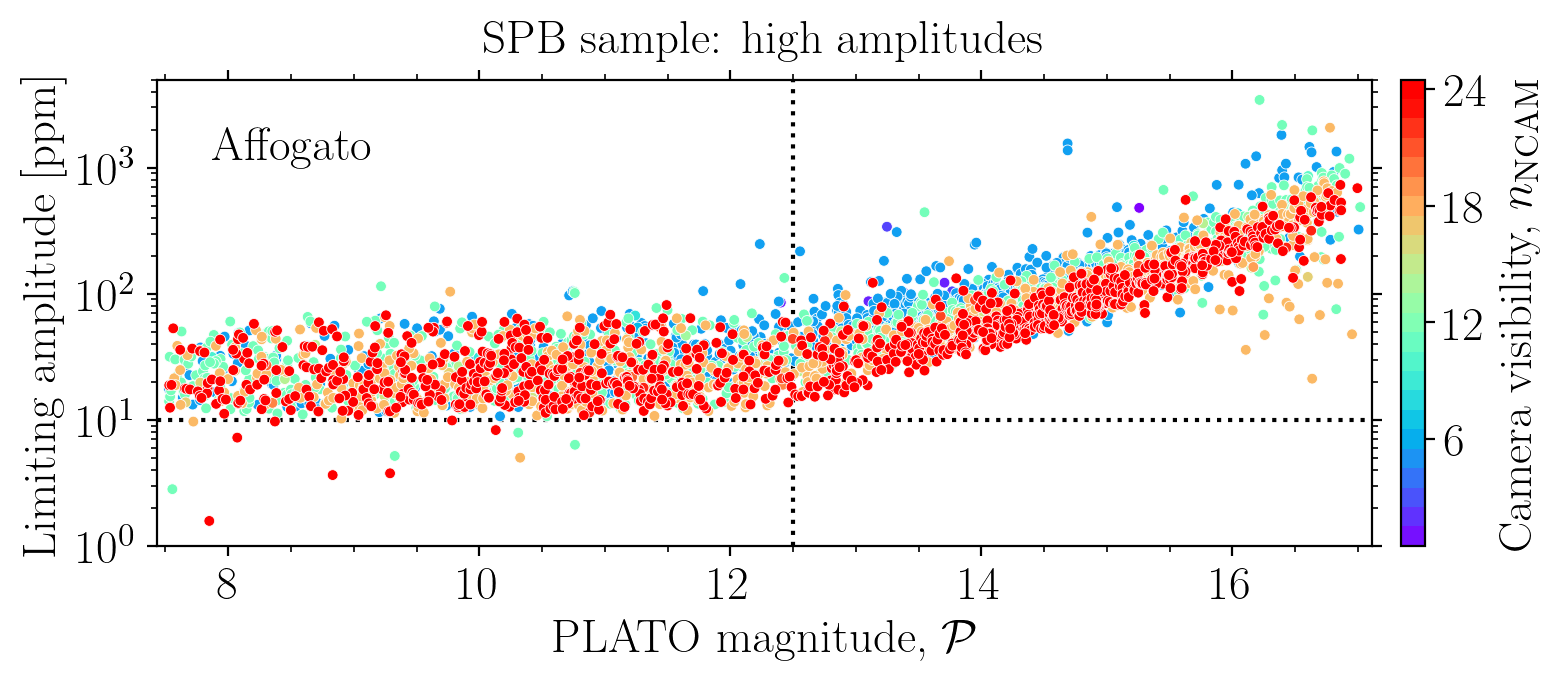}
\caption[]{Limiting mode amplitude detection for the SPB sample, simulation batch \affogato{}, and using an artificially increased mode amplitude distribution. The data points are colour-coded after camera visibility. The dotted vertical line marks a clear transition (at $\Pb \sim 12.5$) between a noise floor plateau in the bright regime and an increasing detection limit in the faint regime.} 
\label{fig:result_limit_SPB_high}
\end{figure}

The main differences of the \gdor{} and SPB sample is the magnitude distribution (with an upper magnitude cut of $\Pb = 14$ and $\Pb = 16$, respectively) and the mode amplitude distribution (with the SPB stars having generally larger amplitudes). The top right panel of Fig.~\ref{fig:result_detection_limit} shows the nominal distribution (of the simulation batch \affogato{}) constructed from the small \textit{Kepler} SPB sample of \cite{pedersen2021internal}. In contrast,  a noise plateau exists for the SPB sample for $\Pb < 9$ (top right panel). Similarly, a noise plateau is observed for both the \gdor{} and SPB sample of \cortado{} (lower left and right panel, respectively). Along each noise plateau, the limiting amplitude detection seems independent of the camera visibility.

The amplitude distribution of the \gdor{} sample is expected to be well representative and complete, as it includes all \gdor{} stars with identified modes in the \textit{Kepler} field \citep{gang2020gravity}. On the other hand, the amplitude distribution of the SPB sample is generated from a rather small core sample of stars with excellent seismic information \citep{pedersen2021internal}. More recent studies have shown that \textit{Gaia} too is able to detect g- and p-mode pulsators \citep[see e.g.][]{de2023gaia, hey2024confronting}. Combining the photometry of TESS and \textit{Gaia}, \cite{hey2024confronting} discovered thousands of new pulsating stars, across the \gdor{}, \dsct{}, SPB, \bcep{}, and hybrid classification, down to a detection threshold of $\sim$\SI{4}{\ppt}. In contradiction to the amplitude distribution recovered from the \gdor{} and SPB samples used in this work, \cite{hey2024confronting} found that statistically the two amplitude distributions were not significantly different. While these results are based on the first and second dominant pulsation modes, it is difficult to tell if the true underlying distribution of these two classes of pulsators is in fact identical or distinct, due to the observational biases of \textit{Gaia}. Moreover, a more pronounced difference in amplitude distribution between the two asteroseismic classes might emerge in PLATO's passband. Thus, to test how the limiting mode amplitude changes with the injected amplitude distribution, we generated another \affogato{} batch of simulations with higher mode amplitudes, more in line with \cite{hey2024confronting}. 

\begin{figure*}[t!]
\center
\includegraphics[width=\columnwidth]{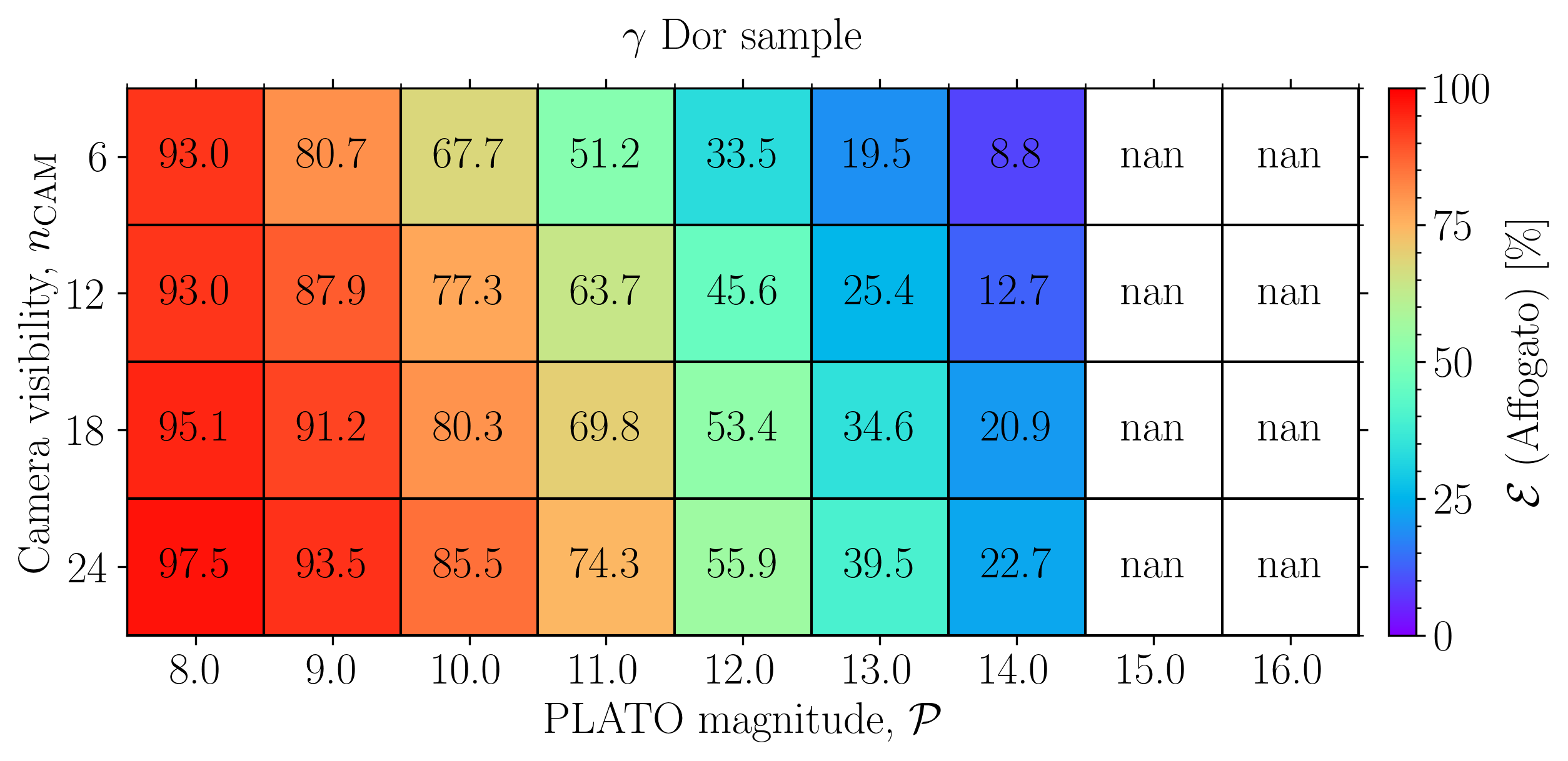}
\includegraphics[width=\columnwidth]{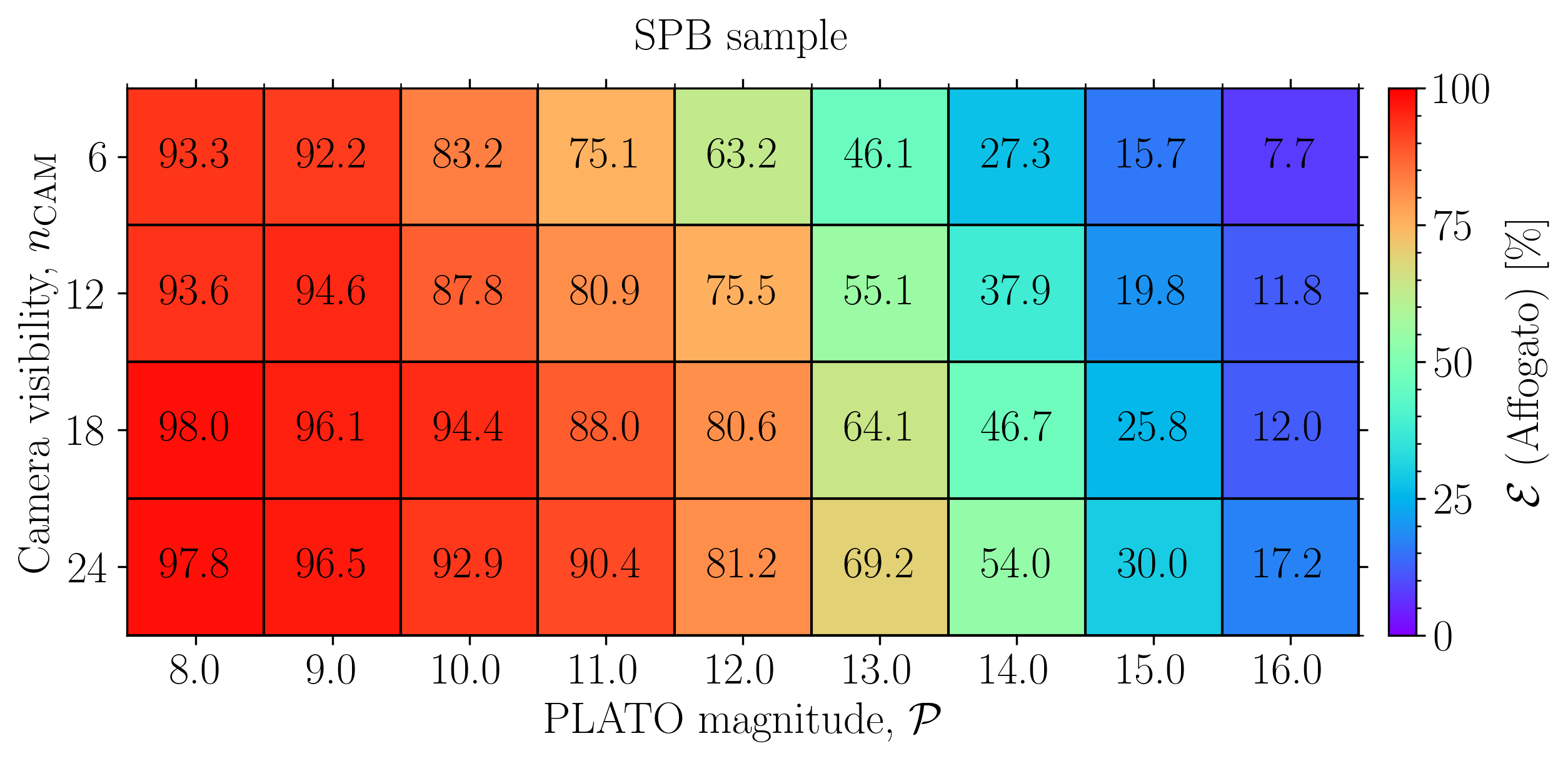}
\includegraphics[width=\columnwidth]{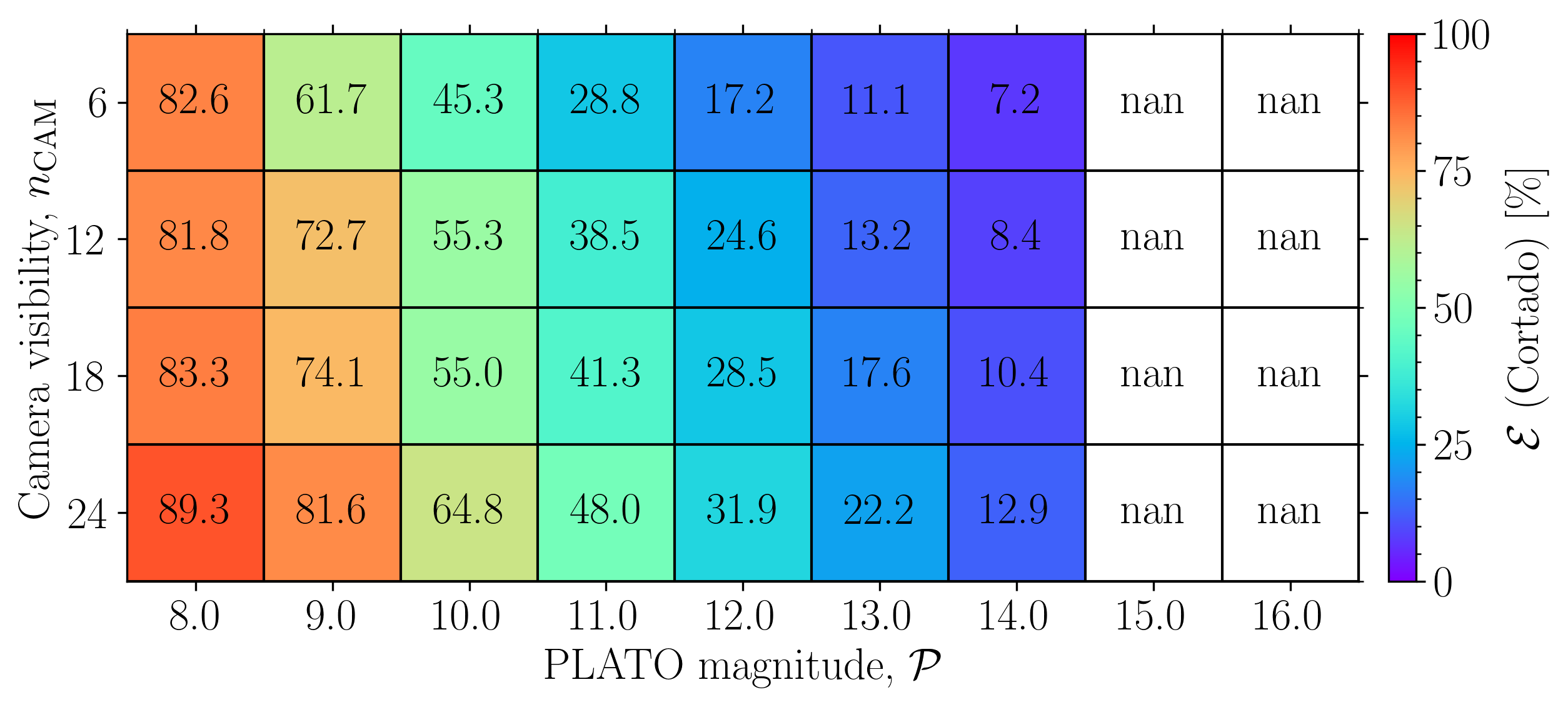}
\includegraphics[width=\columnwidth]{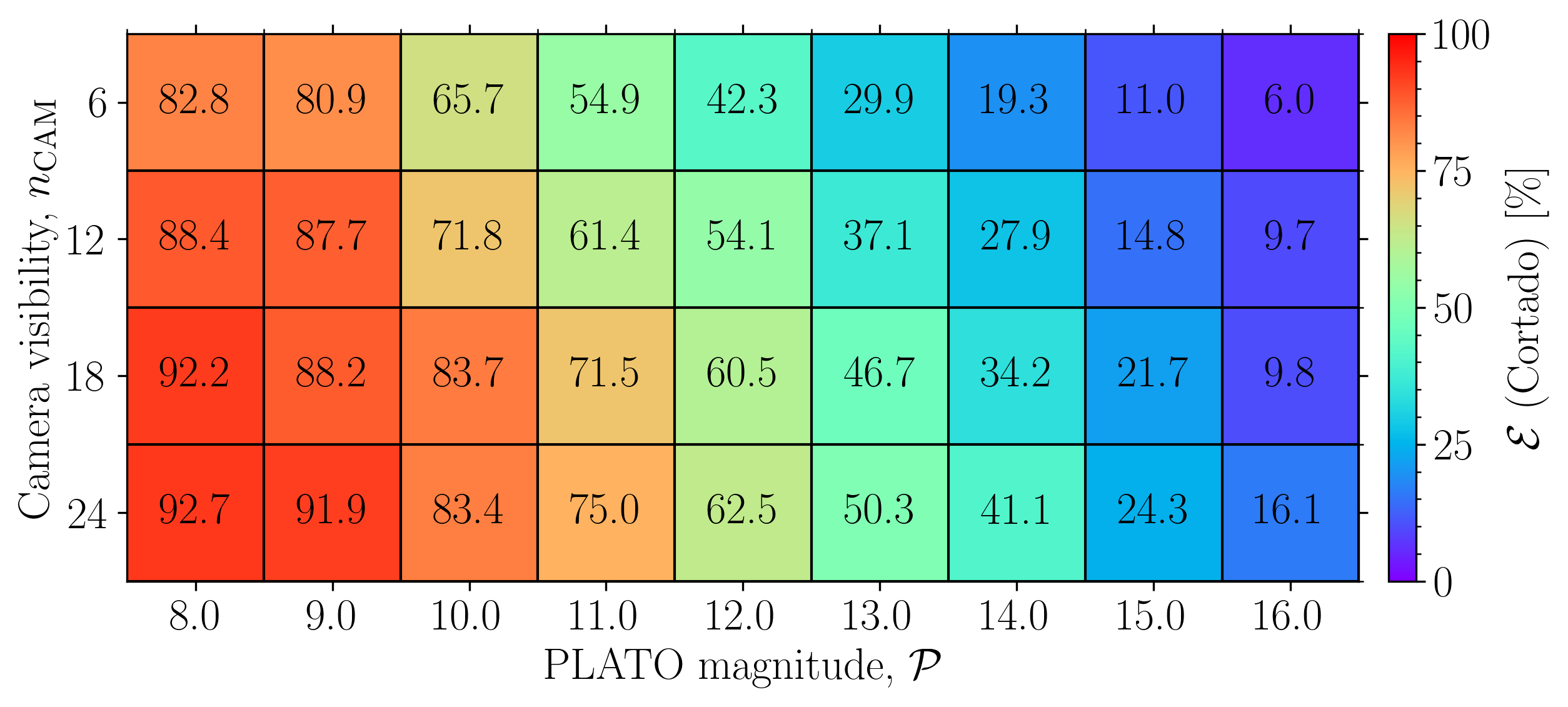}
\includegraphics[width=\columnwidth]{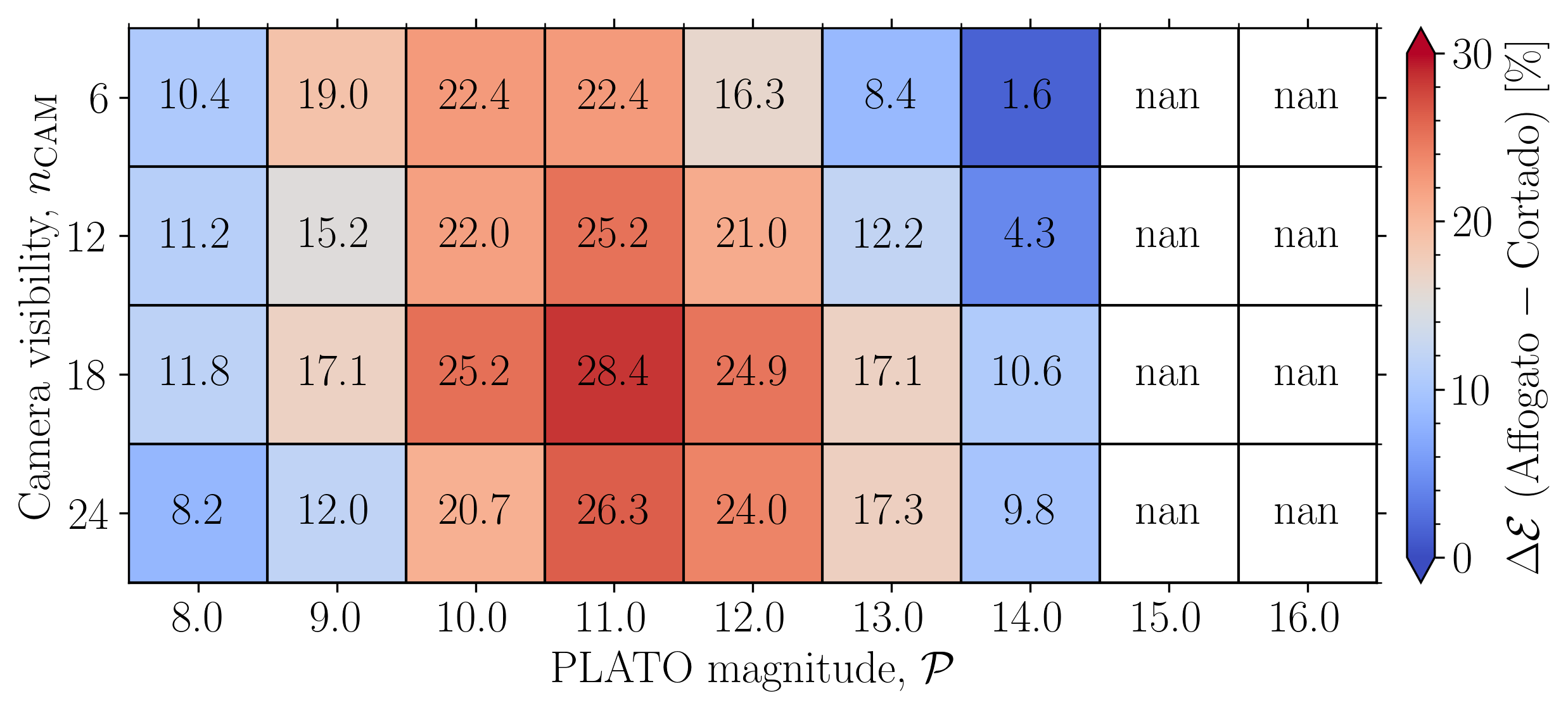}
\includegraphics[width=\columnwidth]{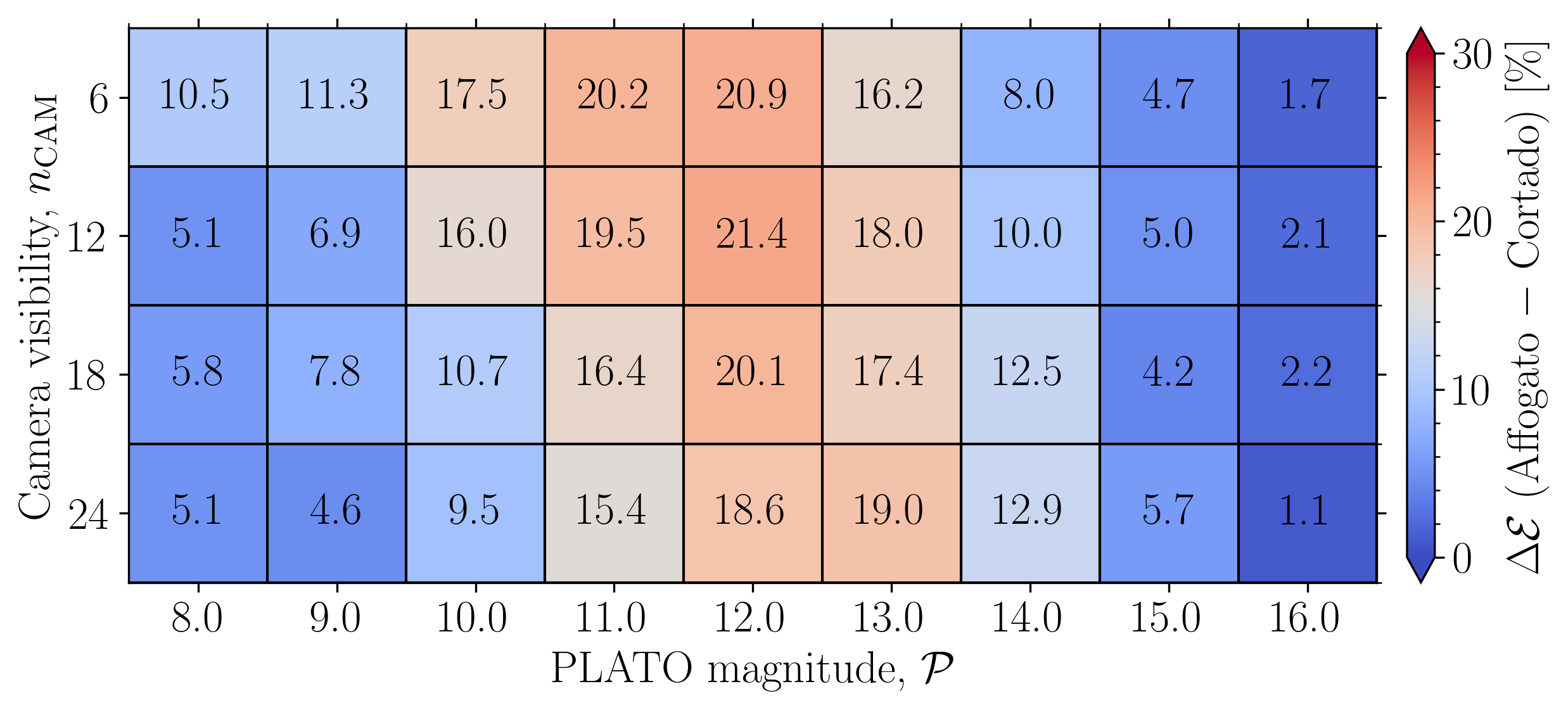}
\caption[]{Detection efficiency diagrams ($\mathcal{E}$ in \%) for the \gdor{} sample (left panels) and SPB sample (right panels). The top and middle panels show the detection efficiency as a function of the camera visibility for \affogato{} and \cortado{}, respectively. The bottom panels show the residuals ($\Delta\mathcal{E}$ in \%) between the results of \affogato{} and \cortado{}. White boxes marked as `nan' are outside the simulated parameter space. Note a direct comparison is possible between the two samples with respect to the colour scaling.} 
\label{fig:result_occurence_rates}
\end{figure*}

Figure~\ref{fig:result_limit_SPB_high} shows a few differences between the high amplitude simulations as compared to the nominal-amplitude simulations: i) the noise plateau now extends up to $\Pb \sim 13$; ii) the lower boundary of the noise plateau has increased to $\sim\SI{10}{\ppm}$ and; iii) the overall scatter has decreased (except in the very faint regime). The former observation is related to a saturation of detectability; left of the dotted vertical lines of Fig.~\ref{fig:result_detection_limit} and \ref{fig:result_limit_SPB_high}, almost all mode amplitudes are detected. The second observation simply reflects that the smallest amplitudes injected are significantly higher than those of the original sample. The latter observation can readily be understood considering a more precise determination of the mode amplitudes. Meanwhile in the faint end, beyond $\Pb > 16$, the increased scatter is in fact spurious results due to highly contaminated stars (see Fig.~\ref{fig:result_limit_fourier}).

We note that the detection limit presented in Fig.~\ref{fig:result_limit_SPB_high} is well aligned with the general PLATO noise budget (see Appendix~\ref{app:NSR}). This may reflect that at camera level, in the photon noise limit ($\Pb \lesssim 13$) the detection limit is independent on magnitude and all pulsation modes can be detected (if the $\text{SPR} = 0$). However, in the sky background/detector noise limit ($\Pb \gtrsim 13$) the pulsation modes start to be increasingly harder to detect, with the detection limit increasing as the square root of the magnitude in agreement with Fig.~\ref{fig:result_limit_fourier}.

\subsection{Mode occurrence rates}\label{sec:result_occurence_rates}   

To quantify the mode occurrence rates, we calculated the overall yield of detecting oscillation modes of g-mode pulsators as a function of the camera visibility and magnitude (in $\Delta\Pb=1$ bins). Figure~\ref{fig:result_occurence_rates} shows the results of the \gdor{} and SPB sample (left and right panels, respectively) for \affogato{} (top panel), \cortado{} (middle panel), and the residuals of the two simulation batches (bottom panel). Again, as expected, the detection rate is steadily decreasing as a function of magnitude and with decreasing camera visibility. As mentioned earlier, looking at the \gdor{} sample, for the brightest targets ($\Pb=8$), almost all pulsation modes are detected, with a recovery rate of 93.0--97.5\% for \affogato{} and  81.8--89.3\% for \cortado{}. However, for the faintest targets ($\Pb=14$), the recovery rates are only 8.8--22.7\% for \affogato{} and 7.2--12.9\% for \cortado{}. Similar results are recovered from the SPB sample, but with the gradient in the occurrence rate extending to $\Pb=16$. We note that the mode occurrence rate for $\Pb<11$ is generally below 50\%. This result is particularly promising as it means that the construction of period-spacing patterns, facilitating mode identification, would be possible for a large sample of \gdor{} and SPB stars in the LOPS2 \citep[given the 1455 pure g-mode and 1449 hybrid pulsators detected so far within the LOPS2, c.f.][]{hey2024confronting}

\begin{figure*}[t!]
\center
\includegraphics[width=\columnwidth]{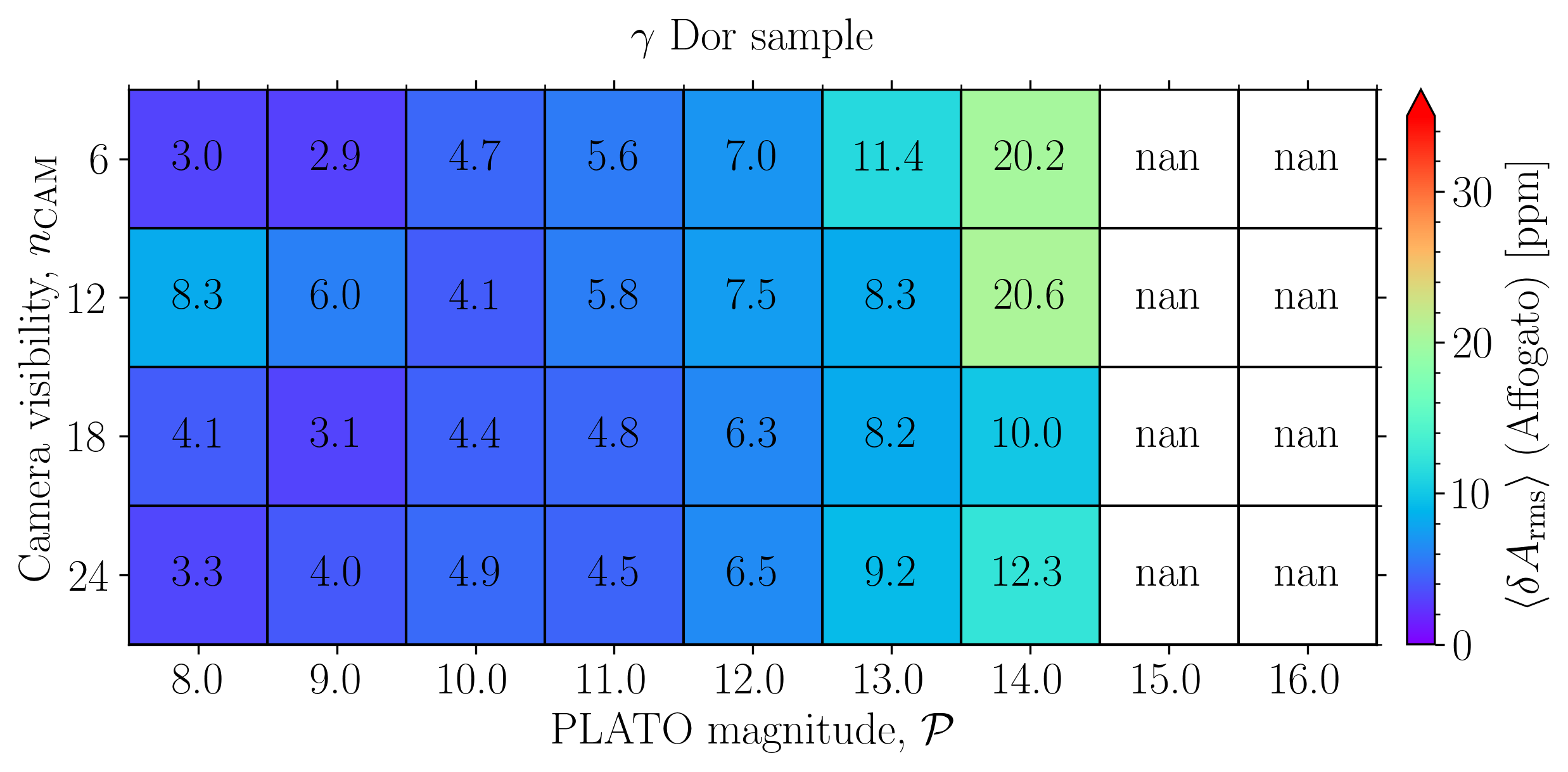}
\includegraphics[width=\columnwidth]{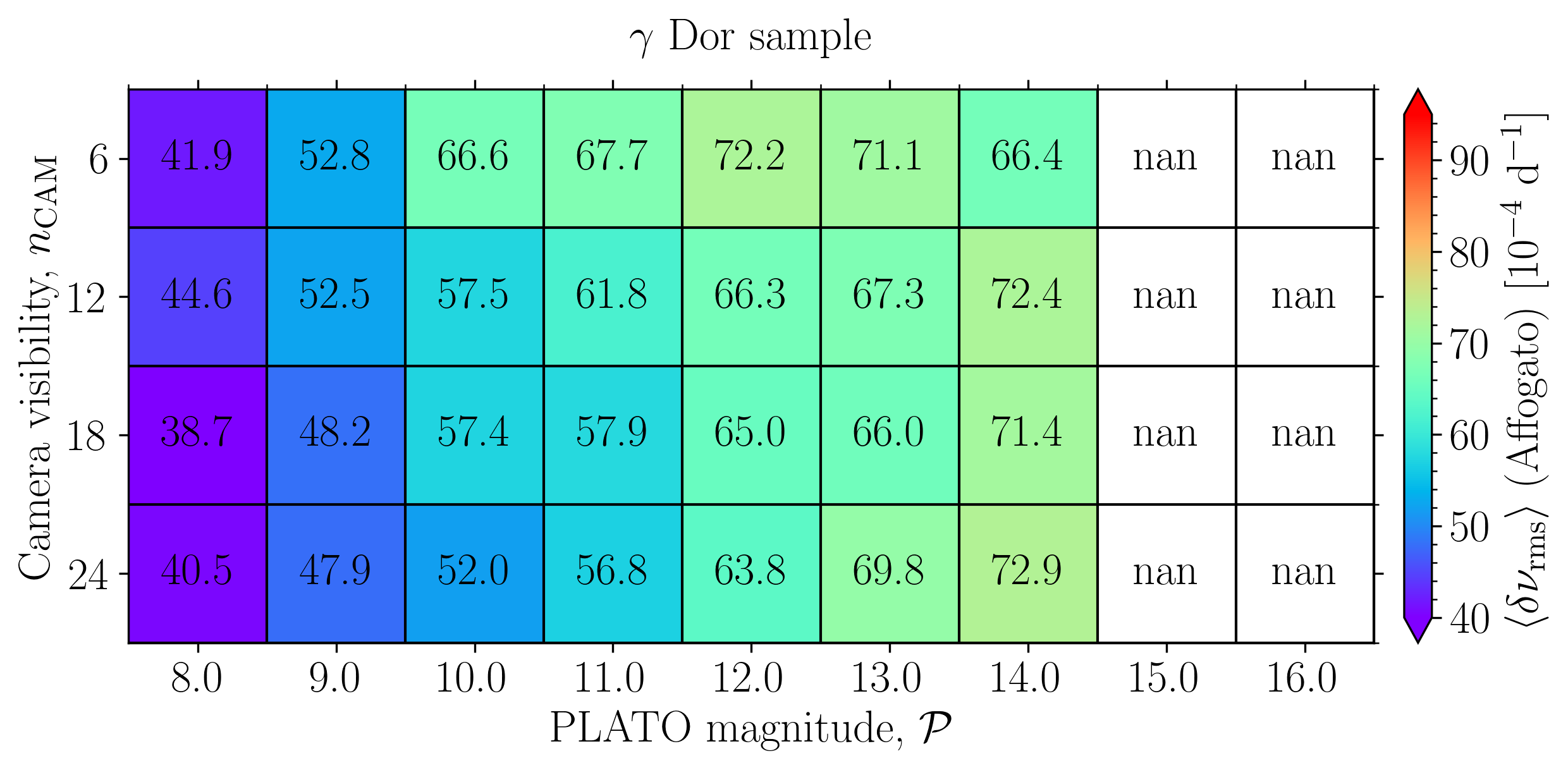}
\includegraphics[width=\columnwidth]{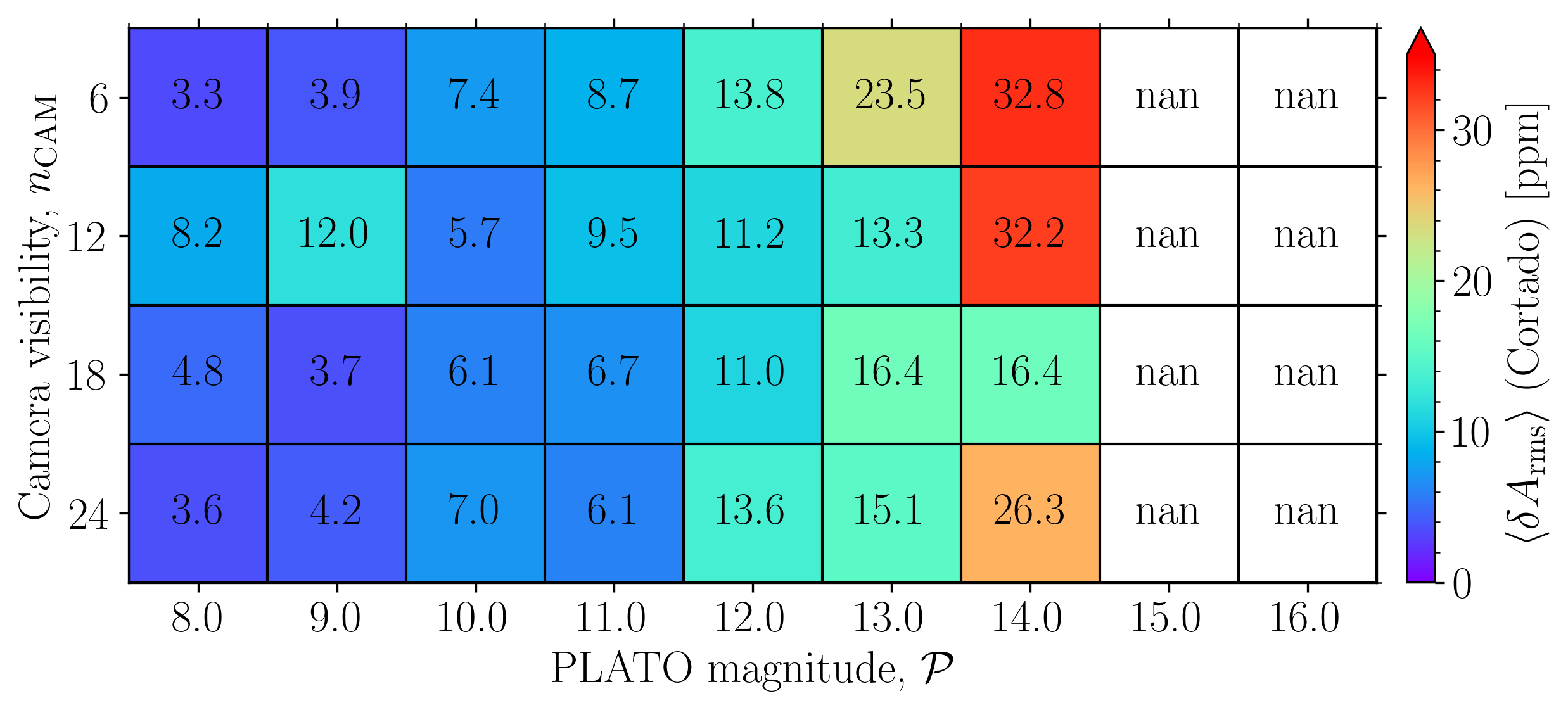}
\includegraphics[width=\columnwidth]{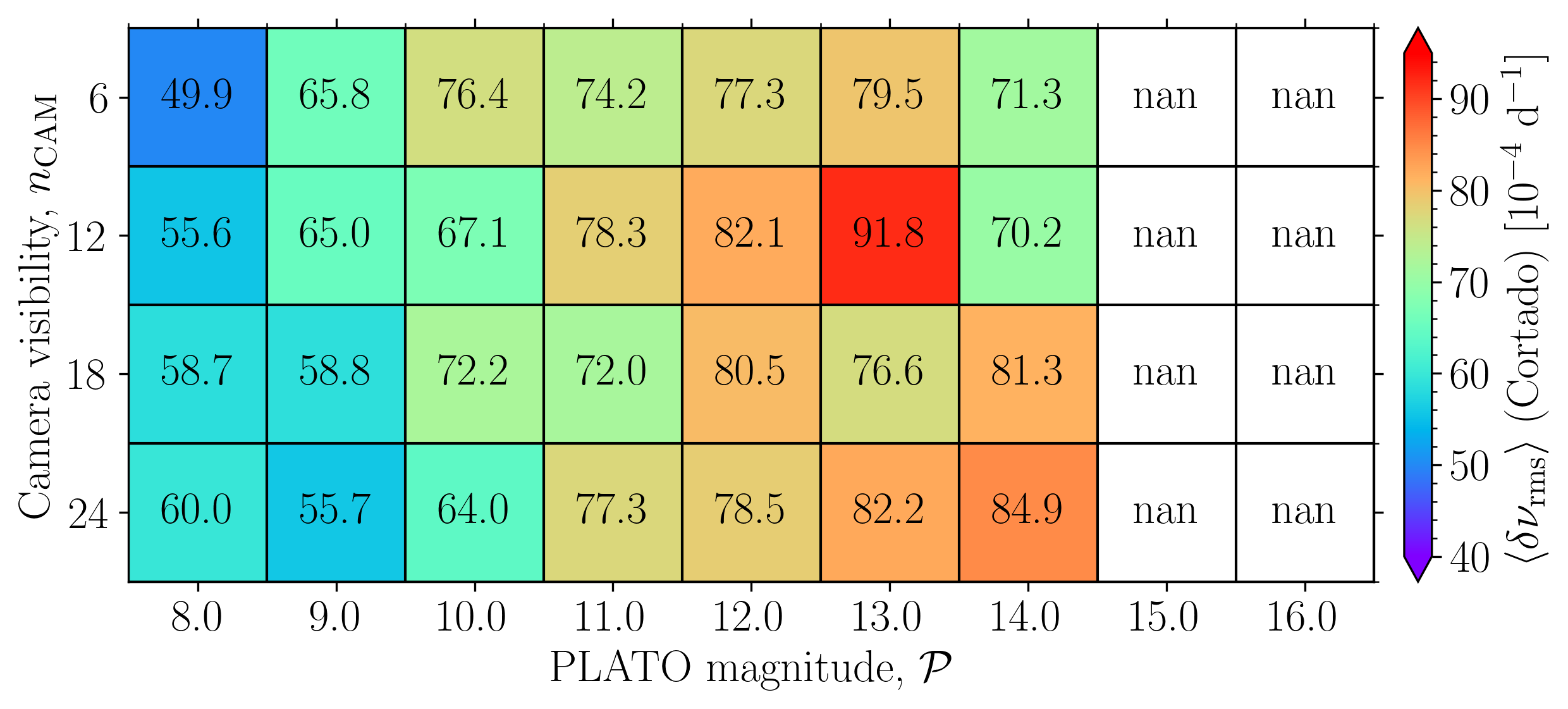}
\includegraphics[width=\columnwidth]{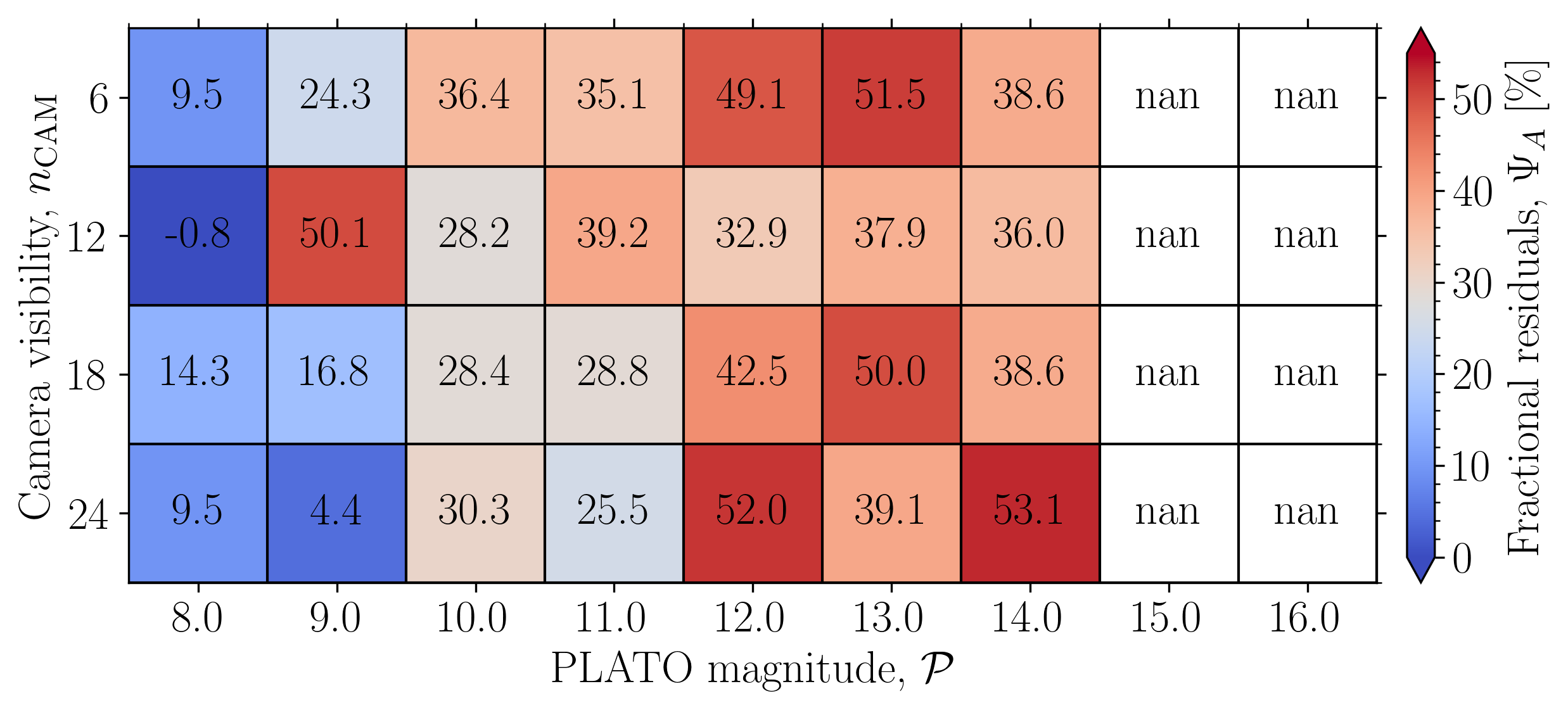}
\includegraphics[width=\columnwidth]{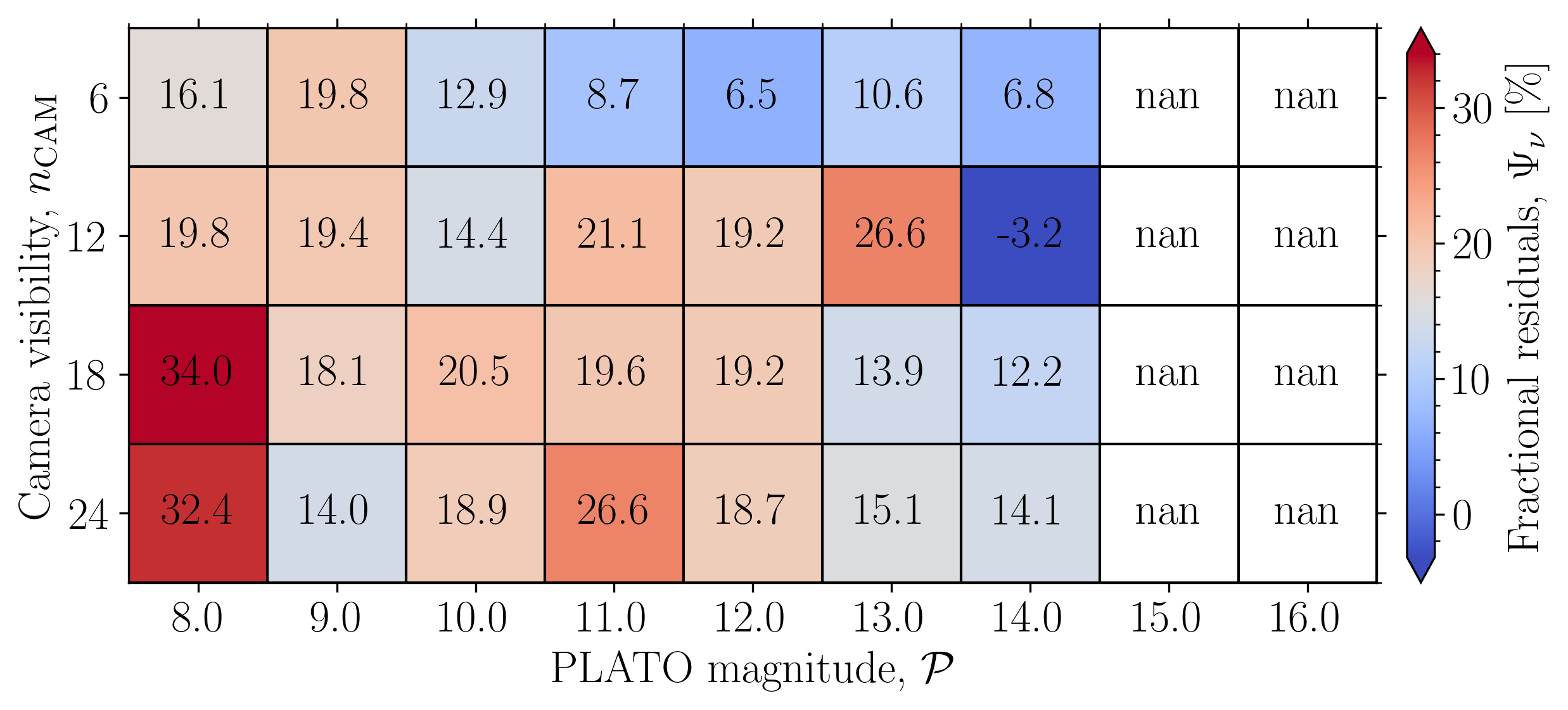}
\caption[]{Amplitude precision (left panels; in \si{\ppm}) and frequency precision (right panels; in \si{\per\day}) of detected pulsation modes as a function of magnitude and camera visibility for the \gdor{} sample. The top and middle panels show the results for \affogato{} and \cortado{}, respectively. The bottom panel shows the fractional residuals between the results of \affogato{} and \cortado{}. Note that compared to all other figures shown in this section, blue indicates the most desirable result. } 
\label{fig:result_amplitude_GDOR}
\end{figure*}

The impact of an increased instrumental noise budget of \cortado{} compared to \affogato{} for the \gdor{} and SPB sample is evident from the residual plot in the bottom left and bottom right panels of Fig.~\ref{fig:result_occurence_rates}, respectively. The biggest discrepancy between the two datasets is in the regime $9<\Pb<13$ where the photon noise typically is the dominant noise source. However, the increased dominance of the instrumental systematics (over photon noise) is particularly peaked at $\Pb\sim11$ for the \gdor{} sample and at $\Pb\sim12$ for the SPB sample. Interestingly, the residual trend increases with camera visibility for the \gdor{} sample, whereas the opposite is observed for the SPB sample. The former observation is anticipated as a higher level of spacecraft systematics is expected to leave a higher (uncorrected) systematic noise excess in the final light curve when combining uncorrelated observations of more and more cameras. The fact that the two residual distributions disagree indicates that the generally lower mode frequencies of the SPB sample are affected differently by the low frequency systematics than of the \gdor{} sample. It is, however, expected to find a smaller yield difference between \affogato{} and \cortado{} for the SPB sample compared to the \gdor{} sample, simply due to a smaller impact on the larger mode amplitudes.  

\subsection{Amplitude and frequency precision}\label{sec:result_precision}

We focus our discussion on the amplitude and frequency precision (c.f. Sect.~\ref{sec:sim_frequency}) of the \gdor{} sample, however, the trends are overall the same for the SPB sample. The left panels of Fig.~\ref{fig:result_amplitude_GDOR} shows the amplitude precision recovery as a function of camera visibility and magnitude. The top and middle panels display the results for \affogato{} and \cortado{}, respectively. The general trend in precision is (as expected) a decreasing function of magnitude and decreasing camera visibility. It is intriguing that an amplitude precision of $\lesssim\SI{100}{\ppm}$ is found below $\Pb<12$ for both simulation batches. If the two F-CAMs, having respectively a red and blue filter, can reach a similar precision a few orders of magnitudes brighter, this may be excellent news for improved mode identification \citep[e.g.][]{daszkiewicz2008identifying, berger2014identification, reese2017frequency}, which is notoriously hard for massive pulsators \citep[c.f.][]{bowman2020asteroseismology, aerts2024asteroseismic}.

Similar to the result of the amplitude precision, the frequency precision shown in the right panels of Fig.~\ref{fig:result_amplitude_GDOR} also shows a decreasing gradient as a function of magnitude and decreasing camera visibility for both \affogato{} (top panel) and \cortado{} (middle panel), respectively. The decreasing trend with decreasing camera visibility is, however, not obvious in the faint limit (e.g. for $\Pb>12$). Given that generally less than 15\% of the pulsation modes could be recovered in the faint regime, an overall frequency precision recovery of $\lesssim\SI{e-4}{\per\day}$ is a valuable result as forward modelling of high-order g modes requires a frequency precision better than \SI{e-3}{\per\day} \citep{aerts2019angular}.

Looking at the fractional difference in amplitude precision between \affogato{} and \cortado{} (shown in the bottom left panel of Fig.~\ref{fig:result_amplitude_GDOR}) the precision for which the mode amplitudes could be determined does not generally depend much on spacecraft systematics in the bright limit but does so in the faint limit. Comparing this to the fractional  difference in frequency precision between \affogato{} and \cortado{} (shown in the bottom right panel of Fig.~\ref{fig:result_amplitude_GDOR}), the largest difference is in the bright limit and the smallest difference is in the faint limit, with a slight dependence on the camera observability. The latter result is rather interesting, as one would generally expect systematics to play an increasing role the more noisy the light curve becomes.

\subsection{Impact of stellar contamination}\label{sec:results_contamination}
 
Out of 4000 stars, 2650 and 2957 stars (i.e. around 66.3\% and 73.9\%) from the \gdor{} and SPB sample, respectively, have a non-zero stellar pollution ratio (i.e. $\text{SPR}>0$). Here the SPR represents the average value over all aperture mask definitions from the co-added camera observations.

\begin{figure*}
\center
\includegraphics[width=\columnwidth]{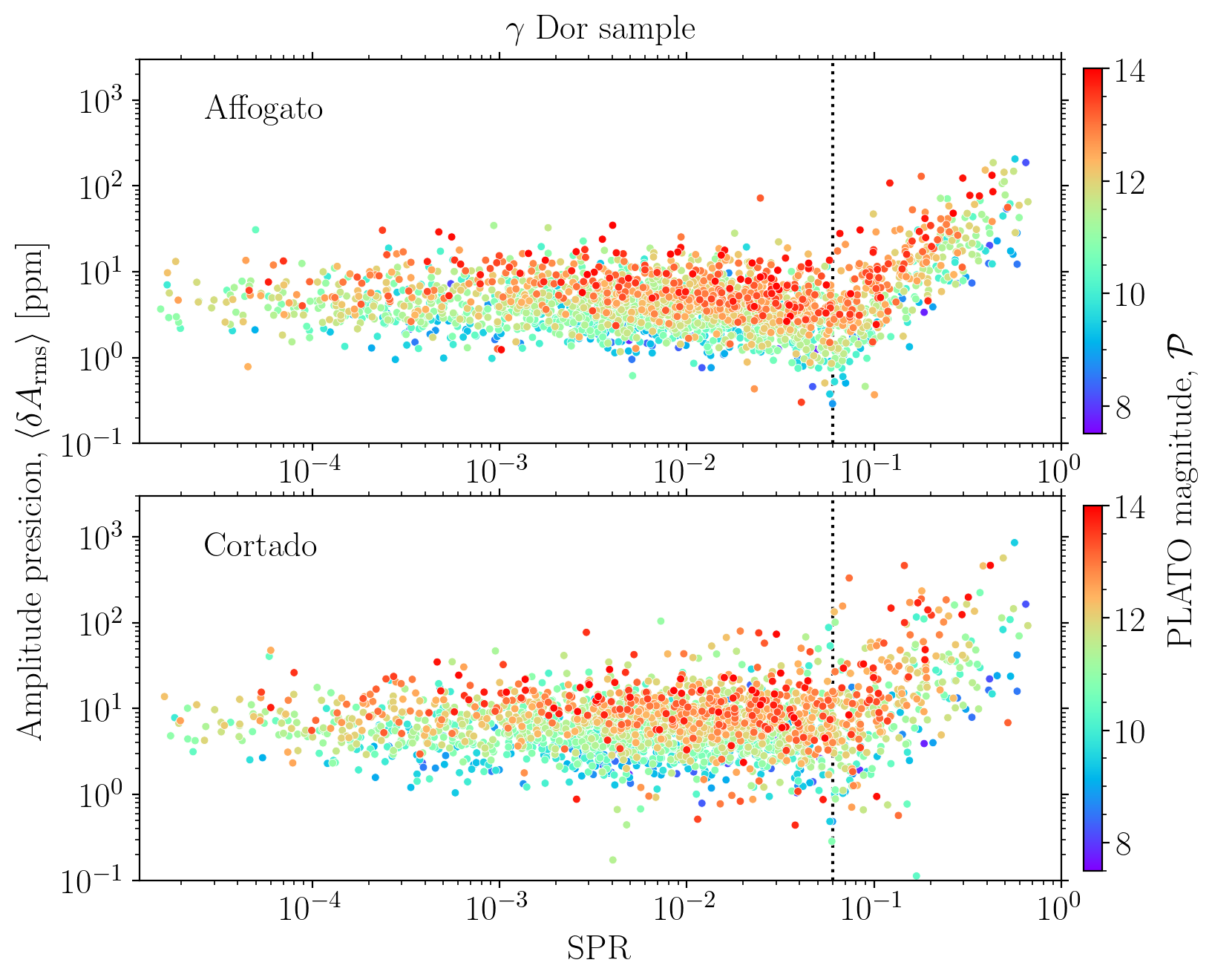}
\includegraphics[width=\columnwidth]{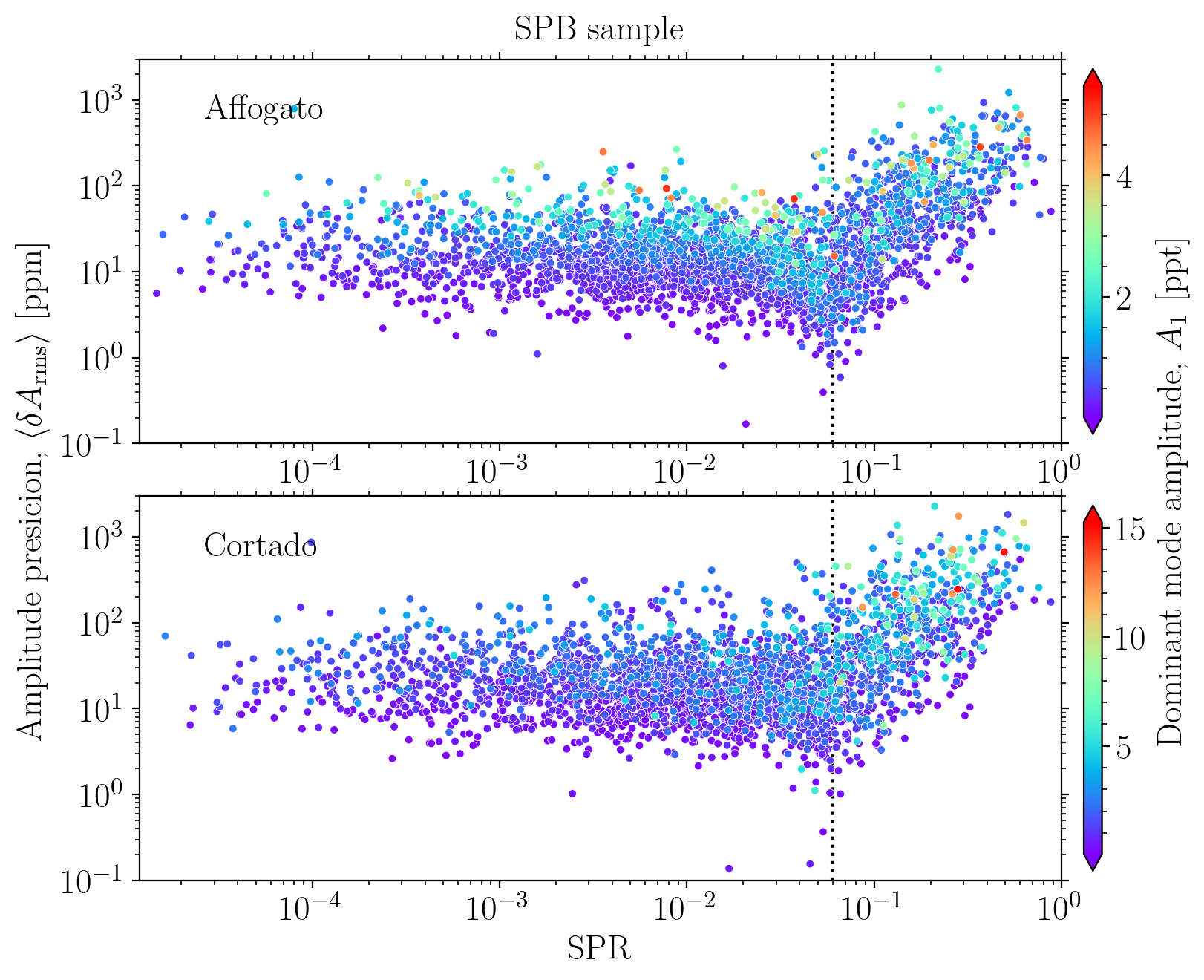}
\caption[]{Amplitude precision vs. SPR diagrams for the \gdor{} and SPB sample (left and right panels, respectively), and for the simulation batches \affogato{} and \cortado{} (upper and lower panels, respectively). The colour scaling of the \gdor{} sample shows the PLATO magnitude and for the SPB sample the dominant mode amplitude ($A_1$). The dotted vertical lines represent a SPR threshold between two amplitude precision regimes.} 
\label{fig:result_con_nonvariable}
\end{figure*}

The simplest case we can consider is flux dilution from non-variable contaminating stars. Figure~\ref{fig:result_con_nonvariable} explores the relationship between recovered amplitude precision as a function of the SPR and the magnitude for the \gdor{} sample (left plots) and as a function of the SPR and the dominant mode amplitude ($A_1$) for the SPB sample (right plots). The top and bottom panels are the results from \affogato{} and \cortado{}, respectively.

\begin{figure*}
\center
\includegraphics[width=\columnwidth]{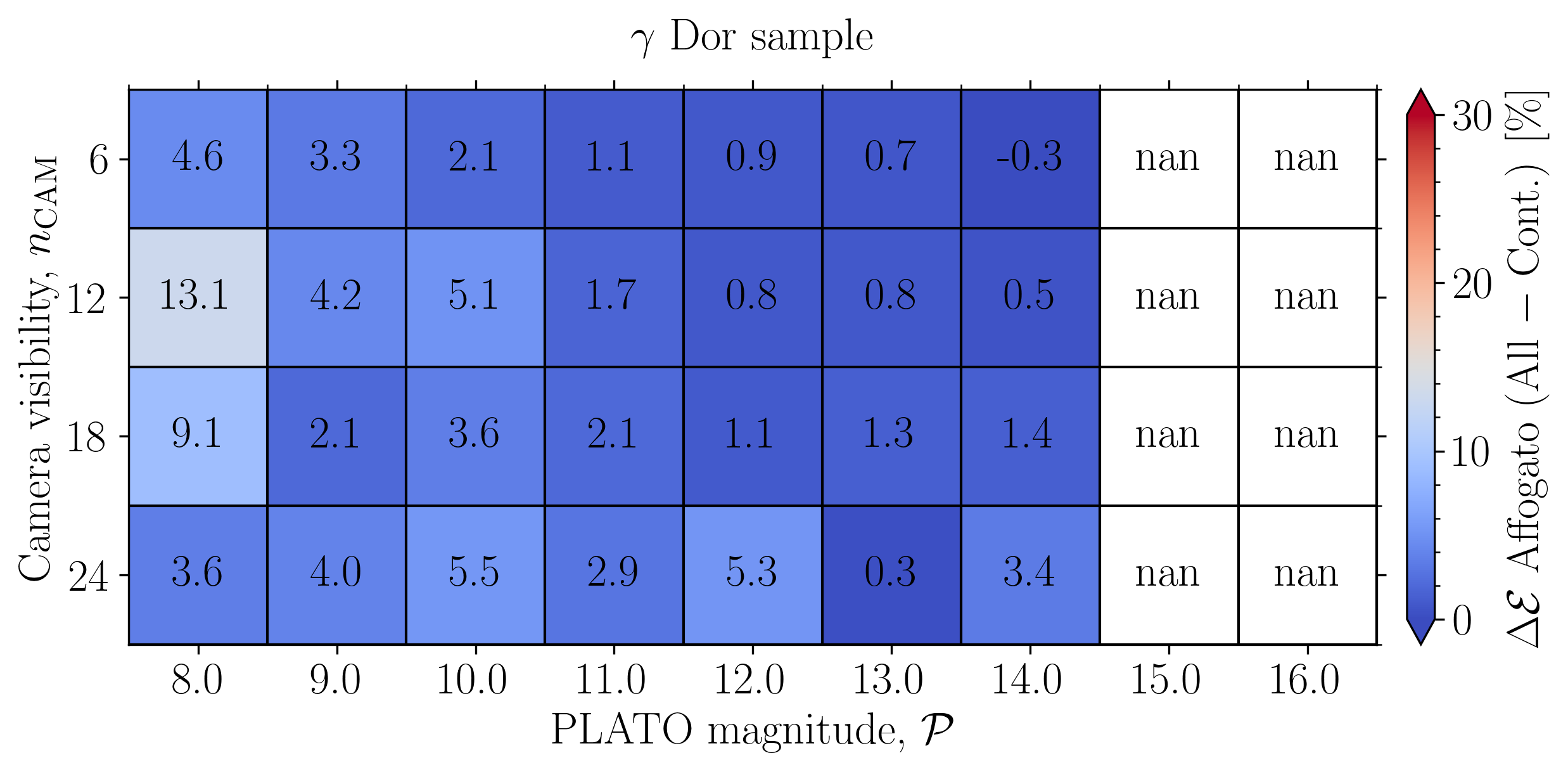}
\includegraphics[width=\columnwidth]{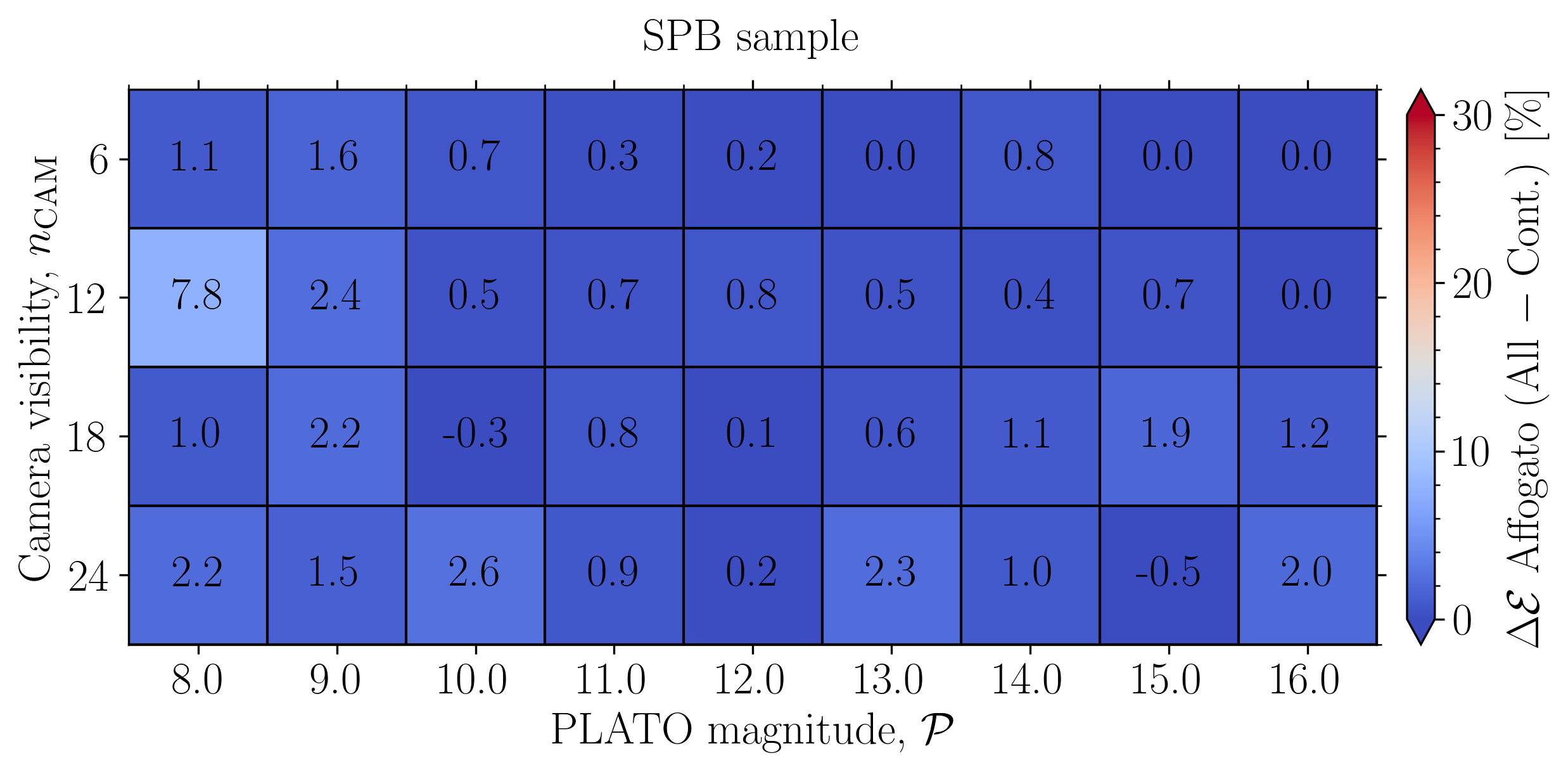}
\includegraphics[width=\columnwidth]{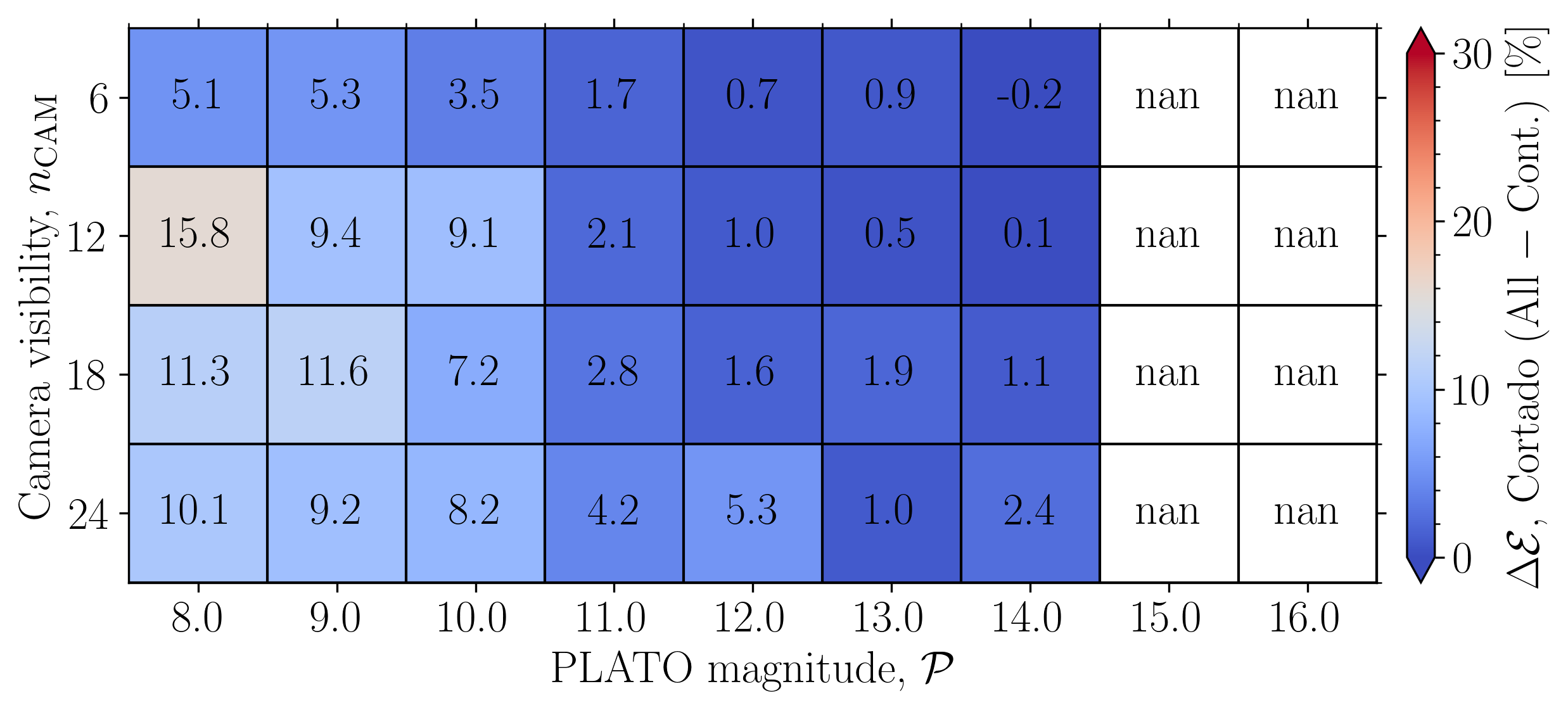}
\includegraphics[width=\columnwidth]{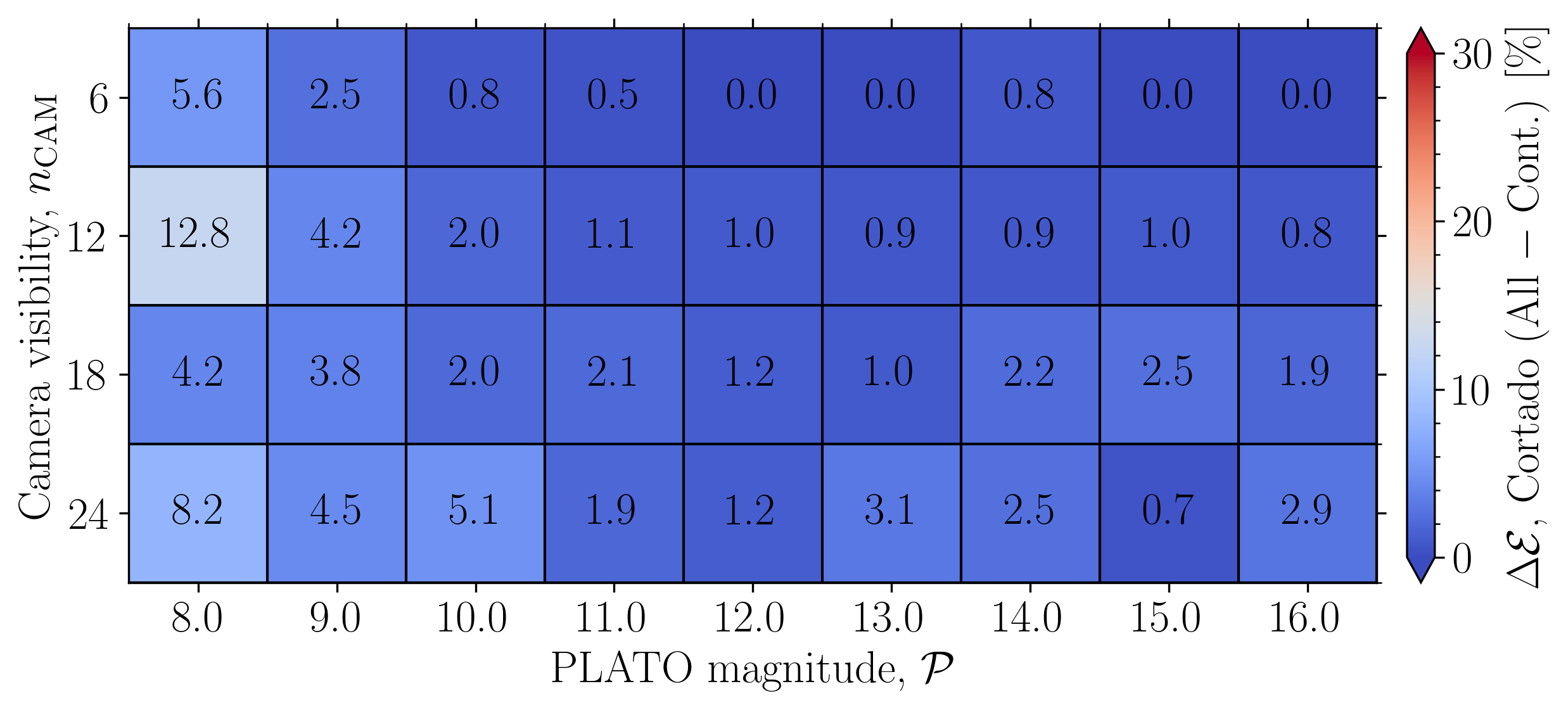}
\includegraphics[width=\columnwidth]{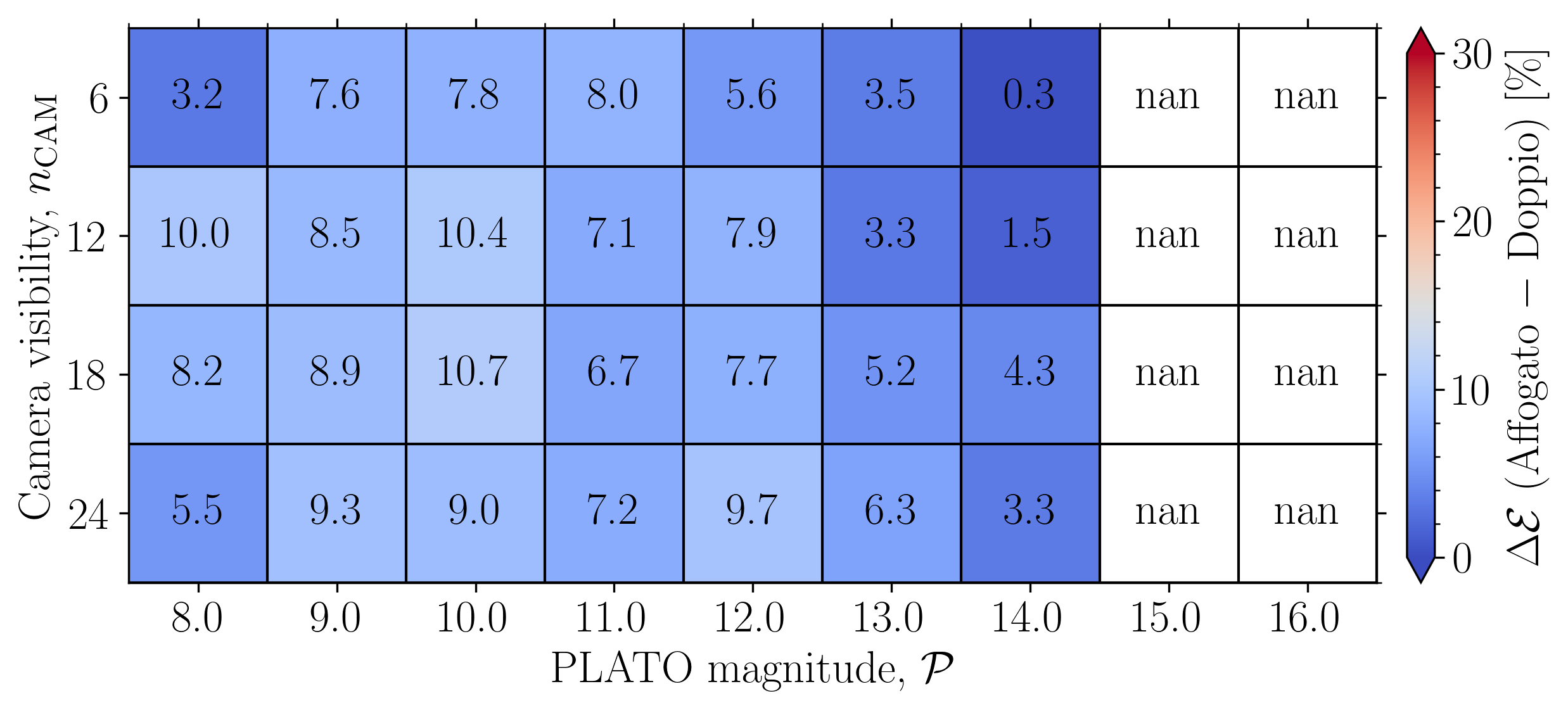}
\includegraphics[width=\columnwidth]{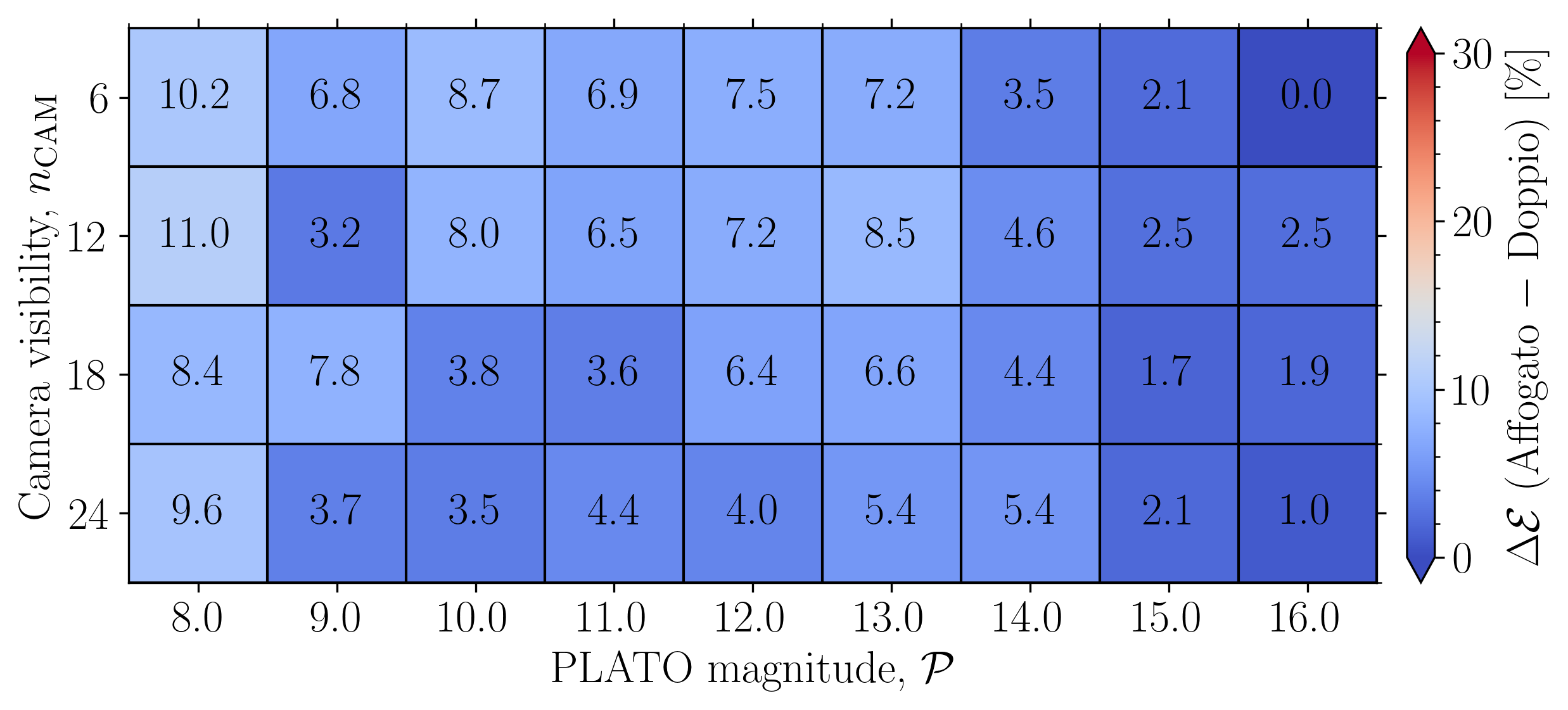}
\caption[]{Detection efficiency residual ($\Delta\mathcal{E}$ in \%) diagrams to test the impact of stellar contamination for the \gdor{} sample (left panels) and SPB sample (right panels). The top panels show the results for non-variable targets from \affogato{} when comparing all stars vs. contaminated stars only. The middle panels show the result for non-variable targets from \cortado{} when comparing all stars vs. contaminated stars only. The lower panels show the result considering only contaminated targets and comparing non-variable targets of \affogato{} vs. variable targets of \doppio{}. Note that the colour scaling of $[0,\,30]\,\%$ was chosen in order to make a direct comparison to the $\Delta\mathcal{E}$ diagram of \cortado{} displayed in the lower panels of Fig.~\ref{fig:result_occurence_rates}.} 
\label{fig:result_occurence_rates_con}
\end{figure*}

\begin{figure*}[t!]
\center
\includegraphics[width=\columnwidth]{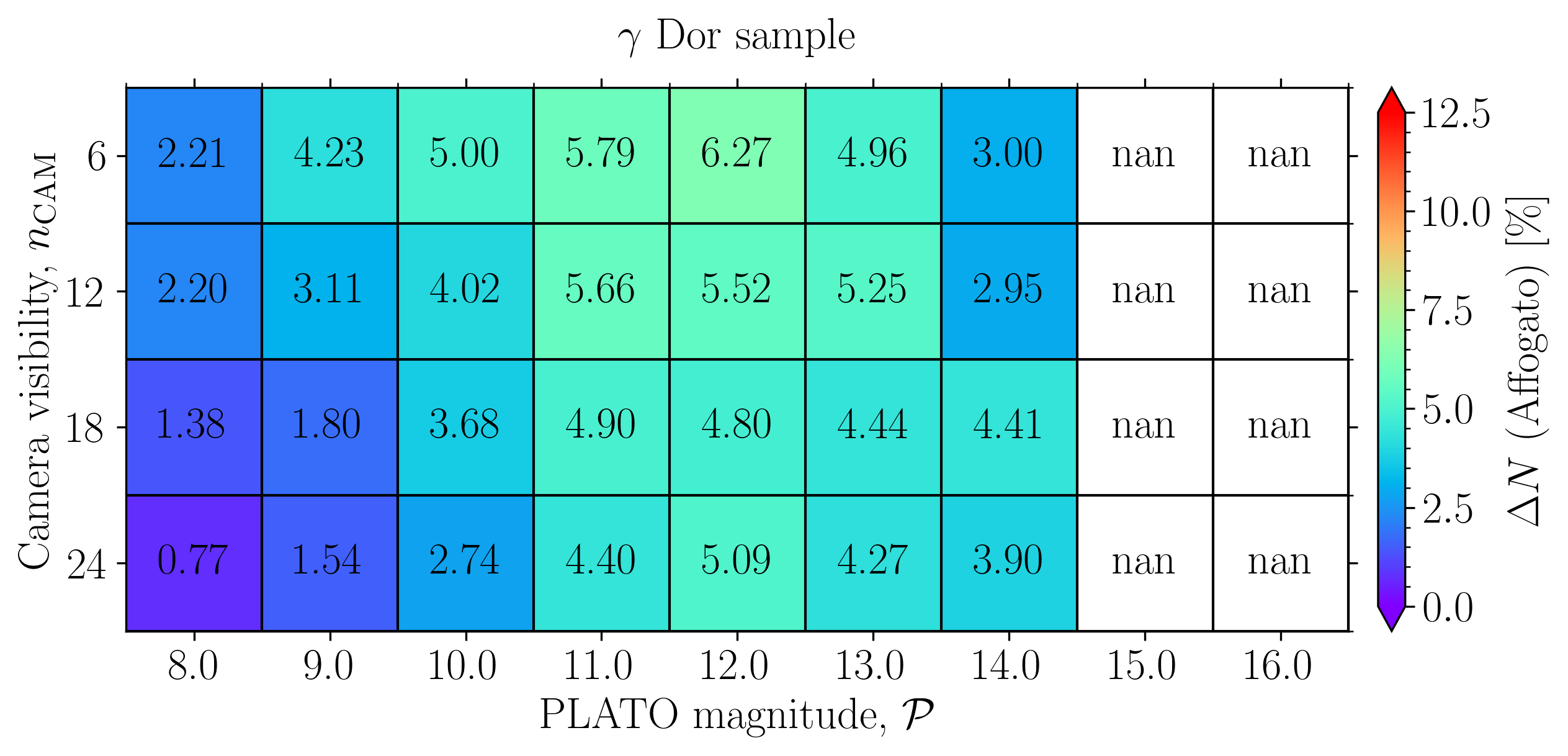}
\includegraphics[width=\columnwidth]{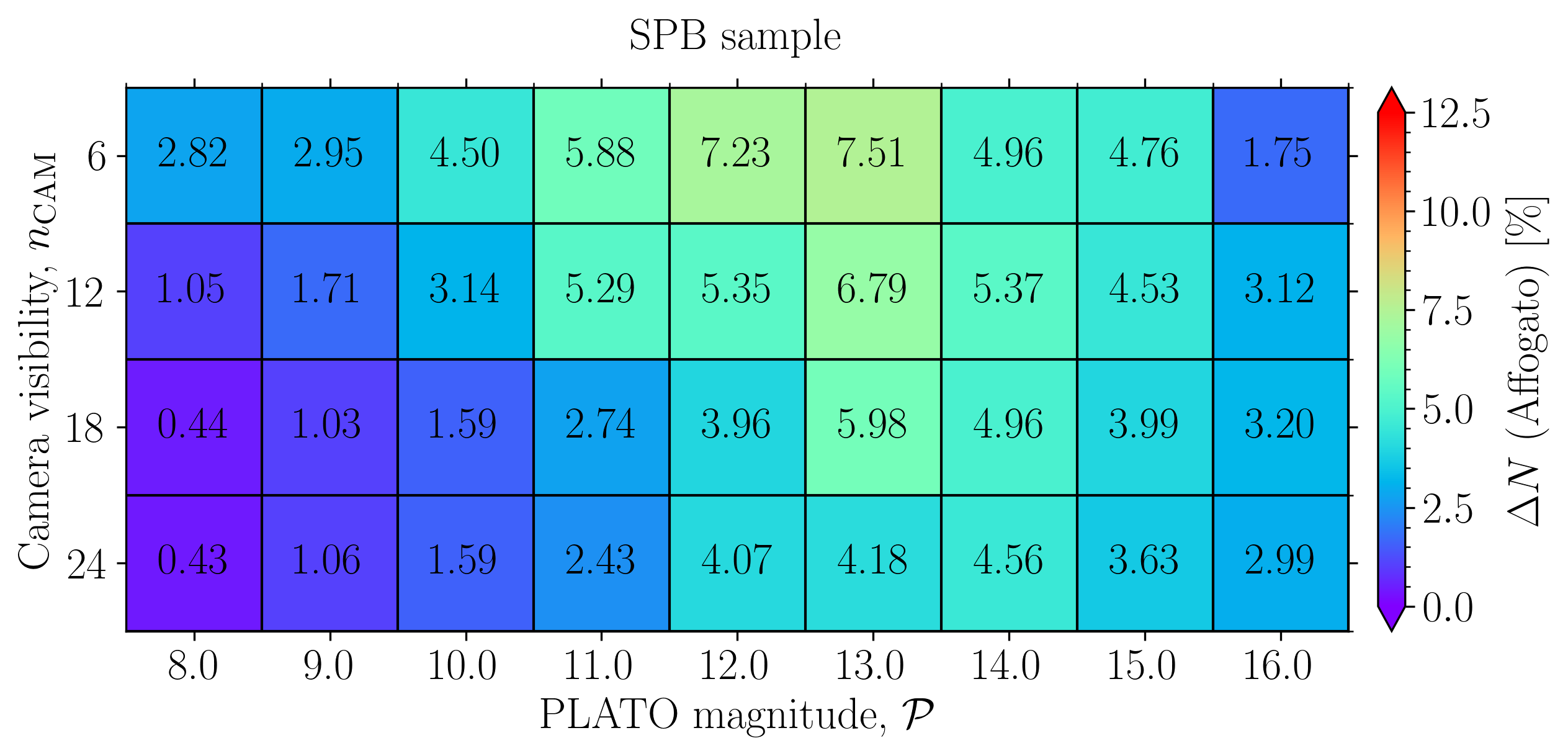}
\includegraphics[width=\columnwidth]{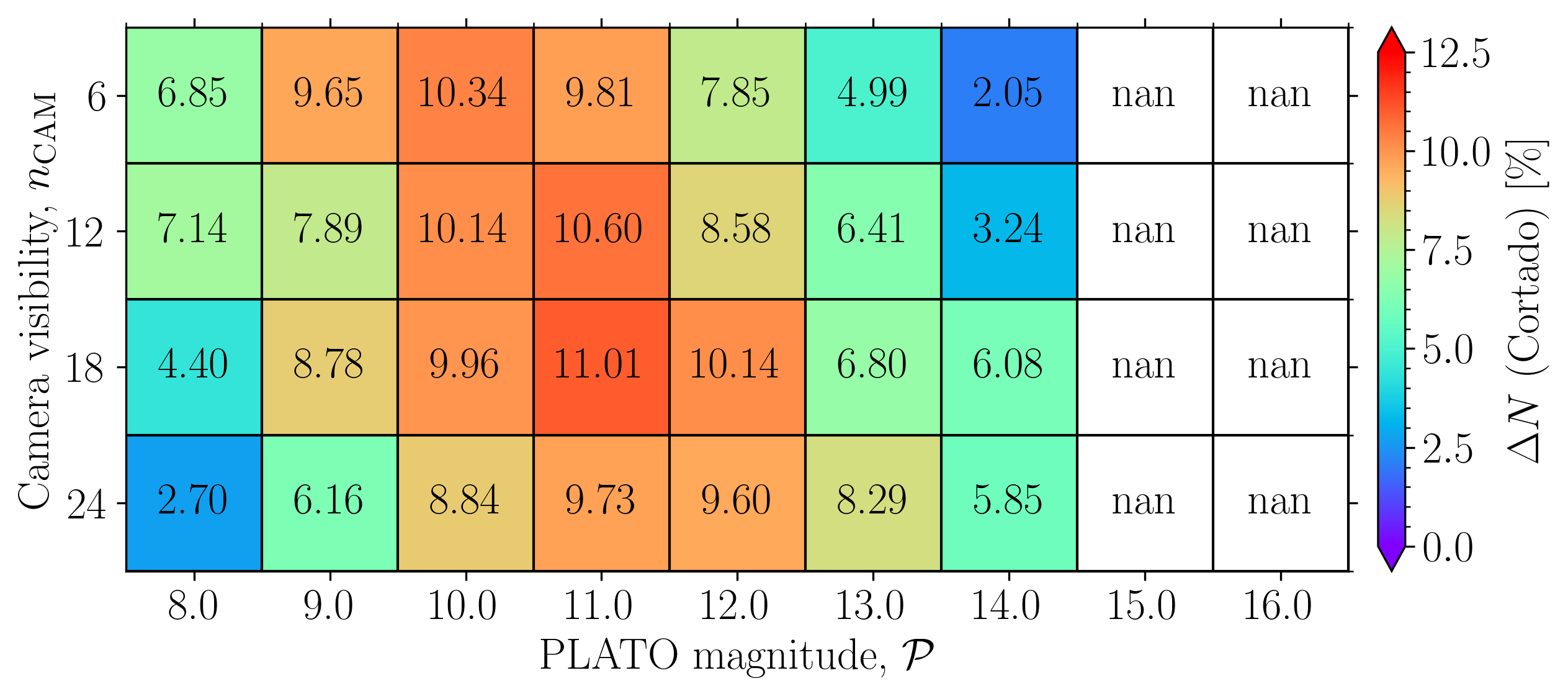}
\includegraphics[width=\columnwidth]{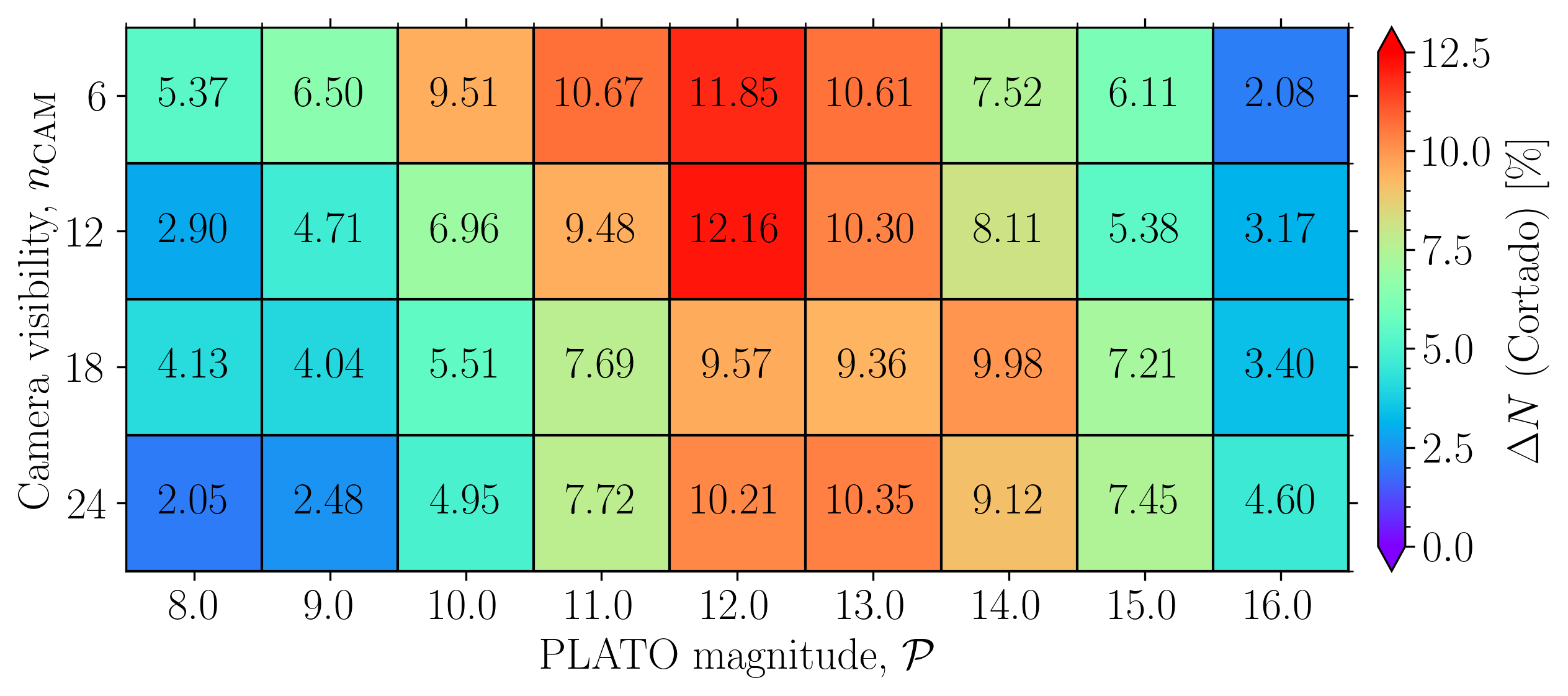}
\includegraphics[width=\columnwidth]{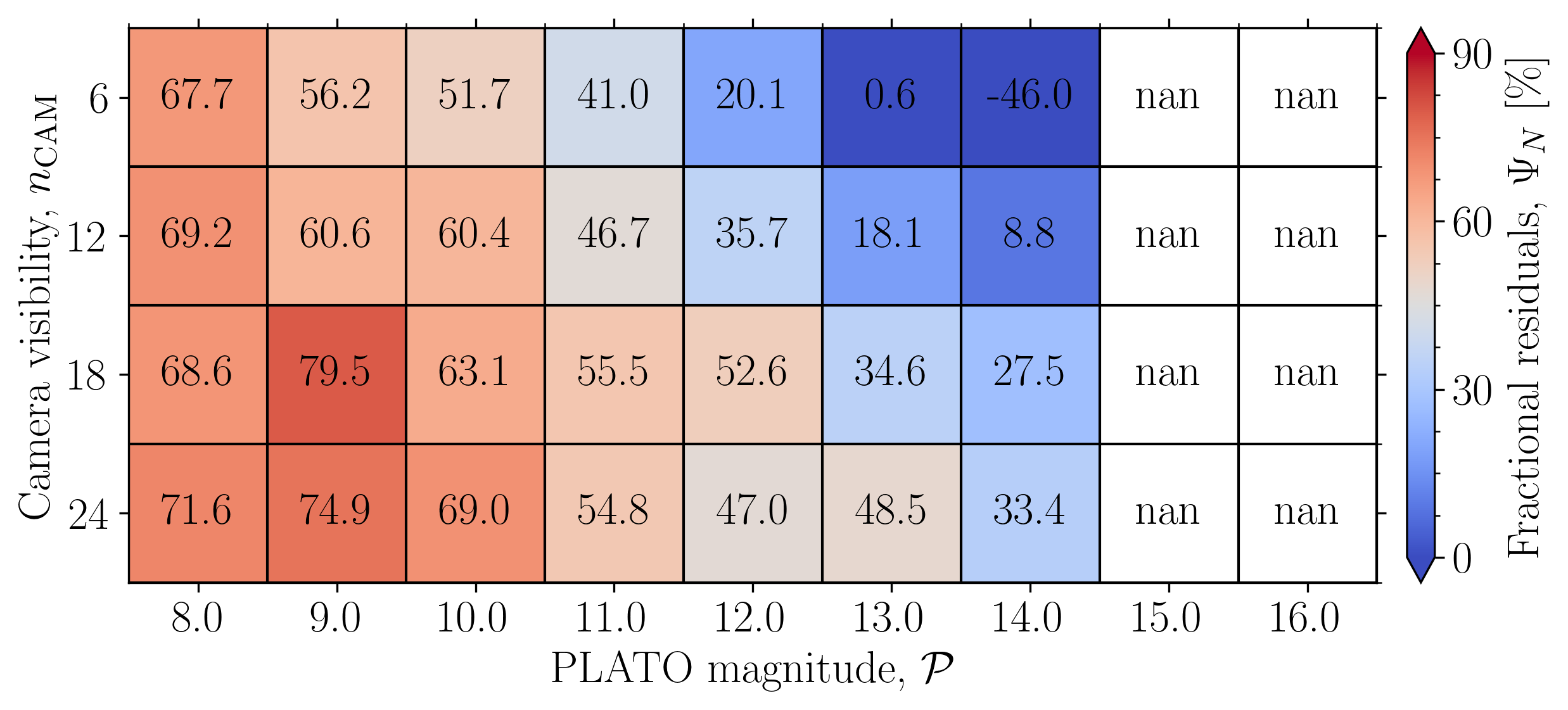}
\includegraphics[width=\columnwidth]{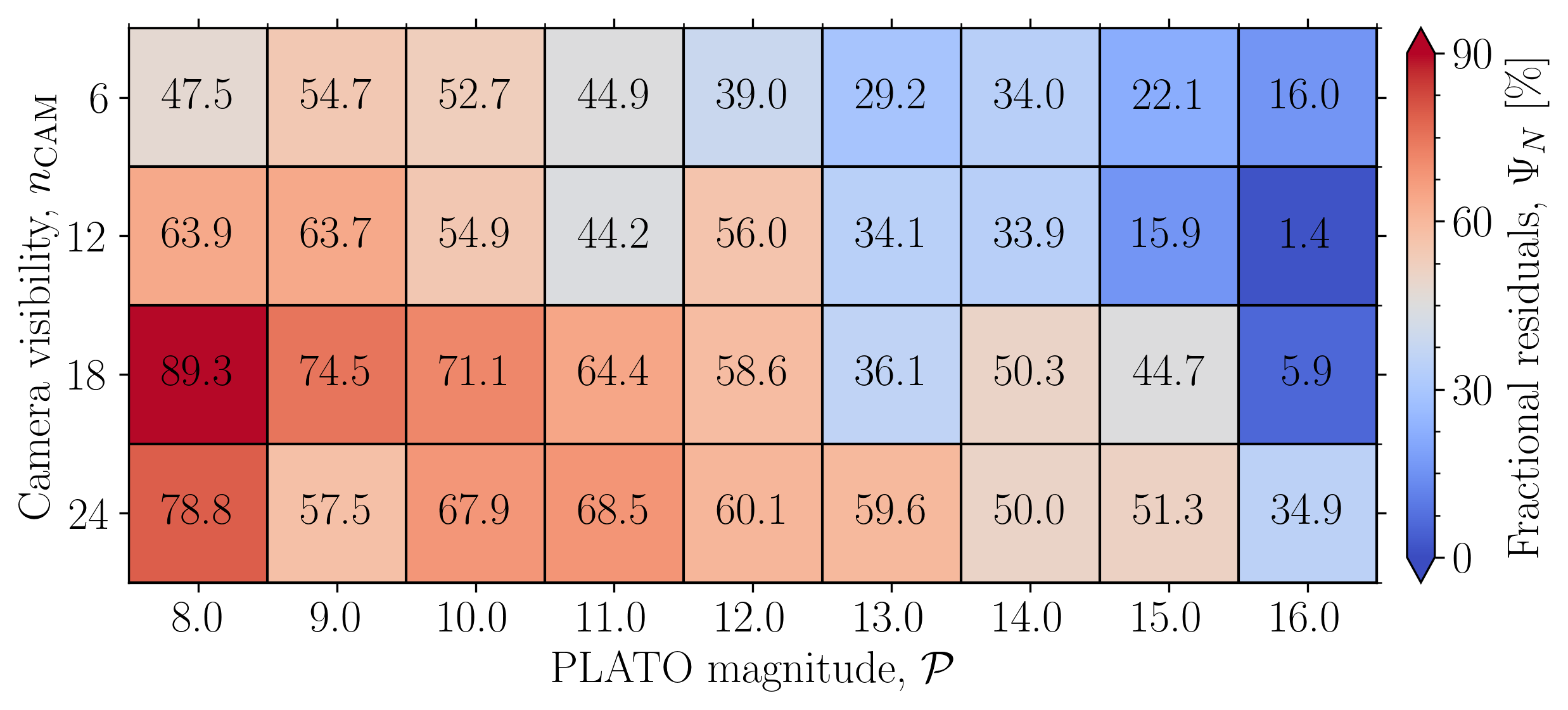}
\caption[]{Difference between pulsation mode detected using the BIC and the SNR prewhitening stopping criterion, $\Delta N = (\tx{N}{BIC}-\tx{N}{SNR})/\tx{N}{input}$, as function of magnitude and camera visibility for the \gdor{} sample (left panels) and SPB sample (right panels). The top panels show the results for \affogato{}, the middle panels the result for \cortado{}, and the bottom panels show the fractional residuals between \affogato{} and \cortado{}.} 
\label{fig:result_nmodes_all}
\end{figure*}

Both samples and datasets show a clear correlation between the amplitude precision and the SPR. Most notably is the kink at around $6\%$ third-light level (vertical dotted line), where a drastic increase (i.e. a worsening) of the amplitude precision is observed. While catching almost all spurious outliers of the noise trends shown in Fig.~\ref{fig:result_limit_fourier}, this appears to be a key threshold for on-board PLATO imagery. Additionally, for the dataset \affogato{} a slight non-linear decreasing trend is observed for increasing SPR values up to the kink, whereas for \cortado{} a plateau is observed for the \gdor{} sample but less so for the SPB sample. Both the decreasing trend and the local amplitude precision minimum at $\text{SPR}\sim0.06$ of \affogato{} are hard to physically motivate. Hence, it could in fact reflect an optimisation threshold of the optimal aperture mask algorithm itself.  

The colour coding in Fig.~\ref{fig:result_con_nonvariable} illustrates that the spread in amplitude precision is related to the brightness (see left plot). For the \affogato{} simulations (shown for the SPB sample in the right hand plot), this magnitude dependence results in an (well explained) amplitude precision ridge of high $A_1$ modes. On the other hand, for the \cortado{} simulations (here only illustrated for the SPB sample), the relation seems only to persevere for the most contaminated stars and wash out stars below the critical SPR threshold. This suggests that instrumental systematics dominate the noise budget up to $\text{SPR}\sim0.06$. Thereafter contamination of non-variable stars starts to dominate.

We can further investigate the impact of stellar contamination of non-variable stars in the presence of systematics. The upper and middle panels of Fig.~\ref{fig:result_occurence_rates_con} show the residuals of the mode occurrence rates for the \gdor{} sample (left plots) and the SPB sample (right plots). Specifically, the upper and lower panels are the residuals between the entire \affogato{} and \cortado{} samples compared to only stars within each sample with $\text{SPR} > 0$, respectively. All four plots show that contamination affects the occurrence rates mostly in the bright regime (i.e. $\Pb<11$). Moreover, it evident that the stellar contamination in the presence of increased spacecraft systematics has a larger impact on the occurrence rates (decreasing the detection efficiency as expected).  

The lower panels of Fig.~\ref{fig:result_occurence_rates_con} show the impact of contamination by variable stars (i.e. the batch \doppio{}) for the \gdor{} sample (left panel) and the SPB sample (right panel). As compared to results of non-variable contaminants shown in the four upper panels of Fig.~\ref{fig:result_occurence_rates_con}, the added variability overall washes out the strong gradient of high to low occurrences from the bright to faint regime. Instead, a rather smooth landscape of occurrence rates is observed. Except for a few magnitude/camera visibility configurations, overall the added variability decreases the occurrence rates, making it harder to detect pulsation modes as expected. Lastly, we highlight that instrumental systematics seem to overall decrease the occurrence rates by a factor of two or more, as compared to stellar contamination (variable or not). 

\subsection{Choice on prewhitening stopping criterion}

As a last point of attention, we investigated the difference in results when using either the \texttt{STAR\,SHADOW} BIC or the standardised SNR prewhitening stopping criterion with respect to the number of input modes. Figure~\ref{fig:result_nmodes_all} shows the average relative number difference between modes detected with the BIC and SNR threshold as a function of the camera visibility and magnitude.

Considering all samples, it is clear that for \affogato{} the choice of prewhitening stopping criterion is rather insignificant, with the largest relative difference of detected modes being $\sim$6.3\% and $\sim$7.5\% (corresponding to an actual number difference of $\sim$1.7 and $\sim$1.3) for the \gdor{} and SPB samples, respectively. The smallest difference is found in the bright and high $\tx{n}{CAM}$ regime, and in the faint and low $\tx{n}{CAM}$ regime. This is not surprising since \affogato{} is dominated by white-noise and the largest discrepancy is approximately in the photon noise dominated regime. 

Comparing the results of \affogato{} to that of \cortado{} (middle panels of Fig.~\ref{fig:result_nmodes_all}), the largest relative difference has now increased to $\sim$11.0 and $\sim$12.2\% (corresponding to an actual number difference of $\sim$2.8 and $\sim$2.2) for the \gdor{} and SPB samples, respectively. Similar to \affogato{}, the two samples of \cortado{} are generally affected similarly by the choice of the prewhitening strategy. However, as expected, the increased noise budget of \cortado{} introduce an offset of around one magnitude towards the bright regime as compared to \affogato{}.

While the results of the two samples are comparable among \affogato{} and \cortado{} themselves, an offset of around one magnitude for similar recovered values is observed. This is expected as the success of prewhitening depends on the underlying amplitude distribution. The same is true for the frequency distribution due to apodization. However, even for long-cadence PLATO data, the apodization for g-mode pulsators is rather insignificant (e.g. 0.1-1.9\% suppression for \SIrange{1}{5}{\per\day}). On a star-to-star basis, more low amplitude modes may be recovered when increasing the FAP for the choice of the SNR. However, given the rather low relative difference in detected modes between the two methods, this is discouraged for larger (statistical) stellar assemblies.

Lastly, we look at the fractional residuals of \affogato{} and \cortado{} for the two samples (lower panels of Fig.~\ref{fig:result_nmodes_all}). A clear trend can be seen towards larger values in the lower left corner (low $\Pb$ and high $\tx{n}{CAM}$) and smaller values in the upper right corner (high $\Pb$ and low $\tx{n}{CAM}$), with a rather smooth gradient in between. This result shows that the better the photometric precision, the more the instrumental systematics affects the choice of the prewhitening scheme. At first this may be surprising since spurious noise peaks should become increasingly likely to be mistaken for a mode peak in the periodogram. However, the a likely explanation may be that for low $\Pb$ and high $\tx{n}{CAM}$, the stellar signal usually dominates over any long-term systematics, resulting in a larger noise floor due to a higher chance of imperfect detrending. 

\section{Conclusions}\label{sec:conclusions}


In this work, we investigated PLATO's ability to detect pulsation modes of intermediate- to massive stars. Despite the fact that the PLATO passband has been optimised for finding transiting exoplanets around solar-type stars, our simulations predict that observing abundant frequency spectra of g-mode pulsators is achievable with the exquisite long-baseline photometry that PLATO is planning to deliver. For \gdor{} and SPB stars, we tested three noise regimes (i.e. \affogato{}, \cortado{}, and \doppio{} c.f. Sect.~\ref{sec:mocka}) and below we highlight a few key results:
\begin{itemize}
\itemsep0cm
\item The exact underlying amplitude distribution that will be observed plays a key role in how many modes can be detected and used for asteroseismic modelling. In this work we used a fully data-driven approach based on \textit{Kepler} data, as instability computations are inefficient. Below a limited magnitude of $\Pb<14$ for the \gdor{} sample and $\Pb<16$ for the SPB sample, the dominant mode amplitude is recovered in more than 95\% of the cases, even in the presence of strong residual spacecraft systematics.

\item The spacecraft systematics increase the noise budget in the photon limited regime of a PLATO observation ($9<\Pb<13$). The impact increases the amplitude detection limit with a difference of up to tens of ppm, being equivalent to a difference of around one magnitude in brightness between \affogato{} and \cortado{} in order to reach the same photometric precision.

\item Over the different scenarios of systematics, the general g-mode occurrence rates for both samples are above 80\% at the bright end and above 6\% at the faint end. Furthermore, we find that the g-mode occurrence rate for $\Pb<11$ is generally above 50\%. This highlight that the mode identification using period-spacing patterns would likely be possible for a large sample of \gdor{} and SPB stars in LOPS2.

\item Regarding the mean rms amplitude precision, we find that in most cases, an uncertainty of less than $\SI{20}{\ppm}$ is retrieved for the \gdor{} sample and less than $\SI{100}{\ppm}$ for the SPB sample (with exceptions found for faint targets and low camera visibilities). For the mean rms frequency precision, the uncertainty is always well below the required level of $\SI{e-3}{\per\day}$ needed for forward asteroseismic modelling \citep{aerts2019angular}. Spacecraft systematics affect the amplitude precision mostly for faint targets and low camera visibilities, whereas the opposite is observed for the frequency precision. 

\item Our simulations suggest that stellar contamination does not play a major role for on-board photometry if the third-light contamination is below 6\%. Above the 6\% level, the amplitude precision is heavily affected, whereas the frequency precision stays almost unaffected. We also showed that stellar contamination from variable stars generally has less of an impact on the yields compared to an increased budget of spacecraft systematics. 

\item We calculated a prewhitening stopping criteria for each PLATO cadence using a significance criterion surface. We showed that there is a difference of $\sim7\%$ in the number of detected modes when using a classic SNR stopping criterion compared to the BIC stopping criterion. Although the difference is small, the methodology of recovering mode frequencies in this work is biased towards finding pulsations and not noise peaks. Hence, the computed SNR criterion is a safer and more consistent strategy for prewhitening prior to any peak-pattern recognition as applied to high-order g modes \citep[e.g.][]{papics2014asteroseismology, papics2015asteroseismology, papics2017signatures, vanreeth2015detecting, gang2020gravity}.
\end{itemize}   


While \cite{jannsen2024platosim} discussed the strengths and shortcomings of \platosim{}, this work presents the most realistic end-to-end simulations for g-mode asteroseismology done to date for the PLATO space mission. MOCKA is the first legacy of simulations for PLATO-CS, meanwhile we plan to perform simulations in the nearby future for each work-package. With \platosim{} continuously being improved, future simulations may for example include a time-dependent model of scattered light from the Earth and the Moon. A more critical milestone for the next-generation simulations is to test PLATO's reduction pipeline (both on-board and on-ground) with realistic \platosim{} simulations. The expectation is that the PLATO pipeline will provide cleaner light curves as compared to the post-processing pipeline developed for this work.

This work constructs models of oscillations from observational \textit{Kepler} samples that themselves suffer from observational biases to some degree. Moreover, we avoided working with hybrid pulsators in order to unambiguously determine the yield for pure g-mode pulsators. Despite the theoretical instability regions for pulsators, it is clear that most main-sequence stars oscillate and can be hybrid g- and p-mode pulsators \citep[see][]{de2023gaia, hey2024confronting}. Thus, future studies may address the added complexity of observing hybrids.

In this project we did not apply a barycentric correction, which would alter the timings of the \SI{25}{\second} cadence of PLATO. Such irregular but periodically modulated data sampling effectively splits Nyquist aliases into multiplets. Since the shape of these aliases are different from real pulsation frequencies in amplitude spectra, \cite{murphy2013super} showed that for \textit{Kepler} time series longer than a single orbit around the barycenter, there is no ambiguity in distinguishing between the two. For stars with pulsation frequencies approaching the Nyquist frequency, this will become a problem for increasingly higher frequencies as their amplitudes would steadily reduce due to the irregular sampling. This in turns affects the amplitude SNR and thus the amplitude precision. As the SNR of a mode amplitude depends on the exposure time \citep{eyer1999variable}, we expect the effect to be insignificant for our analysis as the Nyquist frequency is well beyond the highest frequency injected for each sample and the effective exposure time compared to the time sampling for all three PLATO cadences is as high as $\sim 84\%$. While irrelevant for this project due to a choice of constant cadences, the irregular sampling of PLATO will allow super-Nyquist asteroseismology \citep{murphy2013super}.

Although the focus in this work encapsulates the potential for g-mode pulsators on or near the main sequence, MOCKA is an abundant open-source catalogue of pulsators (c.f. Table~\ref{tab:variables}) ready to be used by the community. With MOCKA marking the unset to estimate the yield for PLATO as a whole, simulation studies can be performed across multiple complimentary scientific disciplines to uncover the full potential of the mission.


\begin{acknowledgements}

This work presents results from the European Space Agency (ESA) space mission PLATO. The PLATO payload, the PLATO Ground Segment and PLATO data processing are joint developments of ESA and the PLATO mission consortium (PMC). Funding for the PMC is provided at national levels, in particular by countries participating in the PLATO Multilateral Agreement (Austria, Belgium, Czech Republic, Denmark, France, Germany, Italy, Netherlands, Portugal, Spain, Sweden, Switzerland, Norway, and United Kingdom) and institutions from Brazil. Members of the PLATO Consortium can be found at \url{https://platomission.com/}. The ESA PLATO mission website is \url{https://www.cosmos.esa.int/plato}. We thank the teams working for PLATO for all their work. 

The research behind these results has received funding from the BELgian federal Science Policy Office (BELSPO) through PRODEX grant PLATO: ZKE2050-01-D01.

CA and MV acknowledge financial support from the European Research Council (ERC) under the Horizon Europe programme (Synergy Grant agreement N$^{\circ}$ 101071505: 4D-STAR). While partially funded by the European Union, views and opinions expressed are however those of the author(s) only and do not necessarily reflect those of the European Union or the European Research Council. Neither the European Union nor the granting authority can be held responsible for them.

EP has been supported by the ‘SeismoLab’ KKP-137523 \'Elvonal grant of the Hungarian Research, Development and Innovation Office (NKFIH) and by the NKFIH excellence grant TKP2021-NKTA-64.

MU gratefully acknowledges funding from the Research Foundation Flanders (FWO) by means of a junior postdoctoral fellowship (grant agreement No. 1247624N).

DMB gratefully acknowledges funding from UK Research and Innovation (UKRI) in the form of a Frontier Research grant under the UK government's ERC Horizon Europe funding guarantee (SYMPHONY; grant number: EP/Y031059/1), and a Royal Society University Research Fellowship (URF; grant number: URF{\textbackslash}R1{\textbackslash}231631). 

DJF acknowledges financial support from the Flemish Government under the long-term structural Methusalem funding program by means of the project SOUL: Stellar evolution in full glory, grant METH/24/012 at KU Leuven.

LIJ acknowledges funding from the Research Foundation Flanders (FWO) under grant agreement 1124321N.

MGP acknowledges support from the Professor Harry Messel Research Fellowship in Physics Endowment, at the University of Sydney.

RH acknowledges support from the German Aerospace Agency (Deutsches Zentrum f\"ur Luft- und Raumfahrt) under PLATO Data Center grant 50OO1501.

This work has made use of data from the European Space Agency (ESA) mission \textit{Gaia} (\url{https://www.cosmos.esa.int/gaia}), processed by the \textit{Gaia} Data Processing and Analysis Consortium (DPAC, https://www.
cosmos.esa.int/web/gaia/dpac/consortium). Funding for the DPAC has been provided by national institutions, in particular, the institutions participating in the Gaia Multilateral Agreement. This research has made use of NASA’s Astrophysics Data System Bibliographic Services, and the SIMBAD and VizieR databases, operated at CDS, Strasbourg, France.

This project additionally made use of the following published Python packages: \texttt{NumPy} \citep{harris2020array}, \texttt{Numba} \citep{lam2015numba}, \texttt{Pandas} \citep{mckinney2011pandas, team2022pandas}, \texttt{SciPy} \citep{virtanen2020scipy}, \texttt{Matplotlib} \citep{hunter2007matplotlib}, and \texttt{Astropy} \citep{price2022astropy,price2018astropy,robitaille2013astropy}.

\end{acknowledgements}


\bibliographystyle{aa}
\bibliography{bibliography} 


\begin{appendix}

\onecolumn

\section{The LOPS2--\textit{Gaia} DR3 catalogue}\label{app:catalogue}

The stellar library of MOCKA was constructed from full-frame CCD images produced by \platosim{} while injecting a queried sky region slightly larger than the FOV of the LOPS2. In total, we produced full-frame simulation for each of the four CCDs in the focal plane for all the 24 N-CAMs (i.e. 96 images in total) to predict exactly where, and with how many cameras, a target is observed (taking into account for alignment errors altering the pointing of each camera slightly; see Appendix~\ref{app:NSR}). Figure~\ref{fig:full_frame_image} shows an example of the CCD focal plane for camera one in camera group four. The dominating features are the LMC (seen in the top left CCD), bright stars and their (CCD readout) smearing trails, and a few stellar associations and open clusters (with NGC 2516 being the easiest open cluster to spot close to the edge of the FOV in the lower left CCD). 

\begin{figure*}[h!]
\center
\includegraphics[width=1\columnwidth]{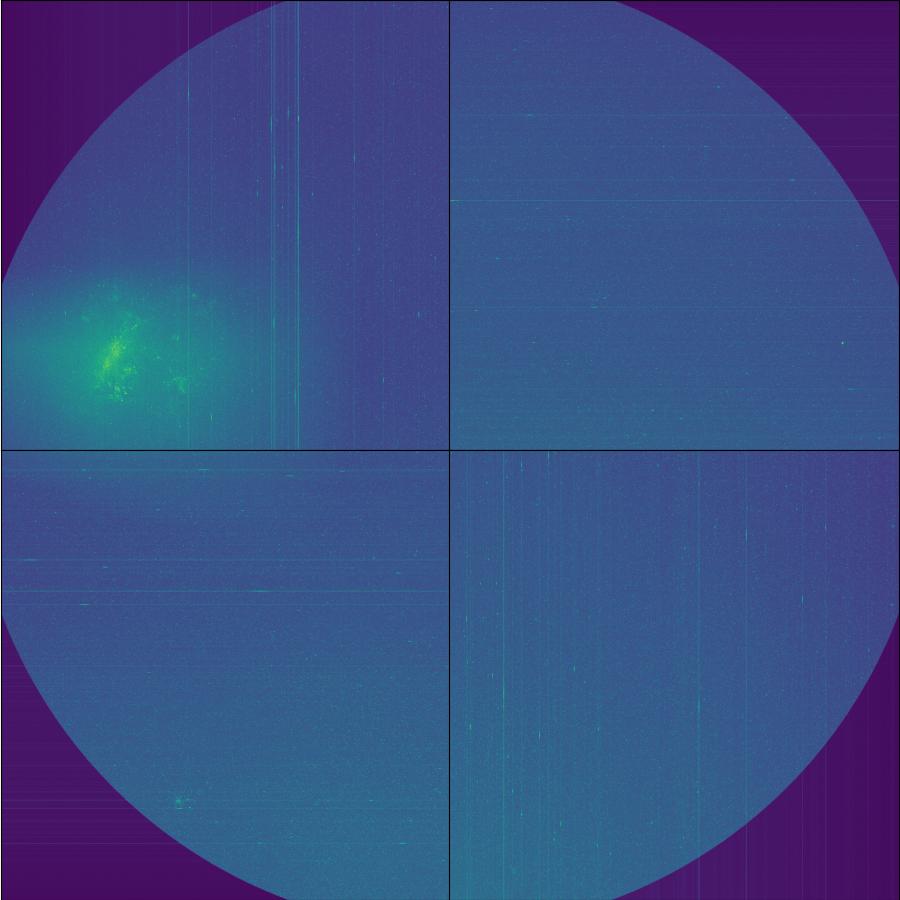}
\caption[]{Full-frame CCD images of the focal plane array of N-CAM 4.1 (i.e. camera \#1 in camera group \#4). The four images have been linearly stretched and scaled to visualise the brightest features in each image, being mainly the LMC, open clusters, stellar associations, and smearing trails from bright stars (magnitude limited to $G\gtrsim2$). The dark blue corner edges show the CCD regions unexposed to light. Note that the actual gaps between the CCDs has been minimised to enlarge the figure.} 
\label{fig:full_frame_image}
\end{figure*}

\begin{figure}[h!]
\center
\includegraphics[width=0.55\columnwidth]{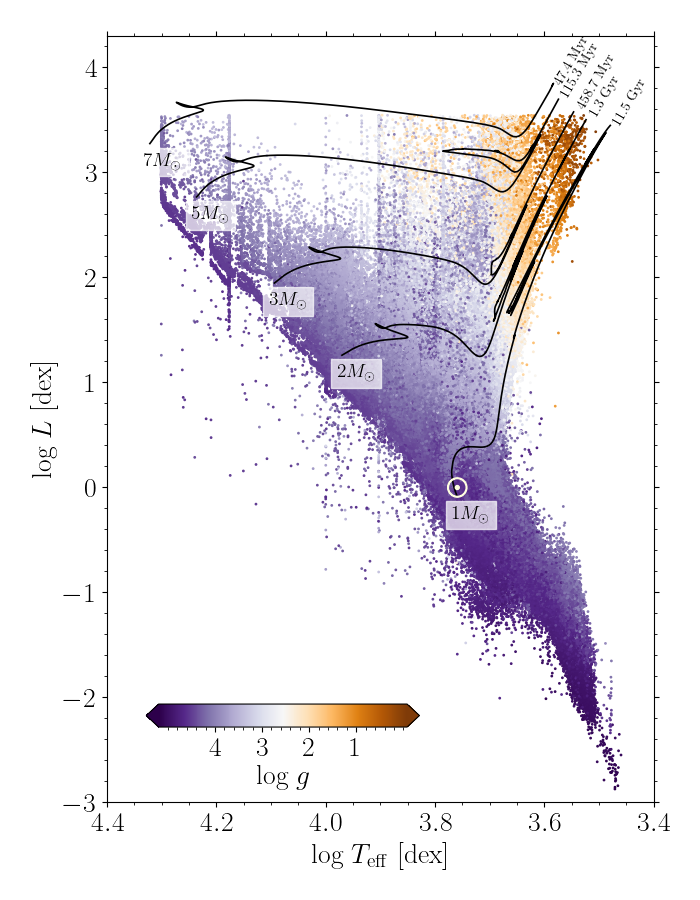}
\caption[]
{Hertzsprung--Russell diagram (HRD) of the LOPS2--\textit{Gaia} DR3 stars, which illustrates the systematic artefacts in \textit{Gaia}'s determination of astrophysical parameters including sharp luminosity and effective temperature cut-offs, constant temperature/surface gravity streaks, etc. As a reference, a few MIST model tracks with solar metallicity \citep{choi2016mesa} are shows for a few stellar masses between \SIrange{1}{7}{\Msun} and with their (arbitrary) age at the end of each track indicated.} 
\label{fig:gaia_HRD_LOPS2}
\end{figure}

Figure~\ref{fig:gaia_HRD_LOPS2} shows the LOPS2--\textit{Gaia} DR3 HRD. This diagram shows clear unphysical structures from observational and computational biases. In this work, we used the data products of the \textit{Gaia} \texttt{FLAME} pipeline, which do not include stars with $\log \tx{T}{eff} > 4.3$ and $\log L > 3.5$. In particular, stellar parameters for stars more massive than \SI{2}{\Msun} and post-TAMS stars are not reliable. This is clearly highlighted in Fig.~\ref{fig:gaia_HRD_LOPS2} from the obvious spurious $\tx{T}{eff}$ solutions forming vertical lines in $L$ at specific integers of $\tx{T}{eff}$. 

\section{Passband conversion: $(\Pb - G)_0$}\label{app:passband}

As part of the mission preparation, synthetic magnitudes in the photometric bands: $G$, $\tx{G}{BP}$, $\tx{G}{RP}$, and $\Pb$ were derived, with $\tx{G}{BP}$ and $\tx{G}{RP}$ being the blue and red \textit{Gaia} passband. This code was provided by M. Montalto in the PLATO technical note PLATO-UPD-SCI-TN-0019. The passband conversion from $\Pb$ to $G$ was established using the MPS-ATLAS%
\footnote{\tiny{\url{https://edmond.mpg.de/dataset.xhtml?persistentId=doi:10.17617/3.NJ56TR}}} %
\citep{witzke2021mps, kostogryz2022stellar}, MARCS%
\footnote{\tiny{\url{https://marcs.astro.uu.se/}}} %
\citep{gustafsson2008marcs}, POLLUX%
\footnote{\tiny{\url{https://pollux.oreme.org/}}} %
\citep{palacios2010pollux}, and the Coelho%
\footnote{\tiny{\url{http://svo2.cab.inta-csic.es/theory/newov2/index.php?models=coelho_highres}}} %
\citep{coelho2005library} stellar libraries, and has the polynomial representation: 
\begin{equation}\label{eq:P-G}
(\Pb - G)_0 = \sum_{i=1}^{i=6} b_i \, (\tx{G}{BP} - \tx{G}{RP})_0^i \, ,
\end{equation}
where the naught means intrinsic (i.e. extinction corrected) colours and $b_i$ are the best fit coefficients. The relation above is used for for dwarf and giant stars, each with their colour validity and fit coefficients being tabulated in Table~\ref{tab:passband}.
\begin{table}[h!]
\caption[]{Best fit coefficients of the passband conversion between the \textit{Gaia} and the PLATO N-CAM photometric band.}
\begin{center}
\begin{tabular}{lccccccc}
\hline\hline
Object & Valid colour range & $b_1$ & $b_2$ & $b_3$ & $b_4$ & $b_5$ & $b_6$ \\
\hline
Dwarfs & $-0.51 \leq (\tx{G}{BP} - \tx{G}{RP})_0 \leq 5.75$ &  -0.3613390 & 0.0632494 & 0.0301607 & -0.0163962 & 0.0027984 & -0.0001679 \\
Giants & $-0.51 \leq (\tx{G}{BP} - \tx{G}{RP})_0 \leq 6.50$ & -0.3586933 & 0.0598219 & 0.0244786 & -0.0119261 & 0.0017487 & -0.0000870 \\
\hline
\end{tabular}
\end{center}
\label{tab:passband}
\end{table}

\section{PLATO's noise budget}\label{app:NSR}

\begin{figure}[h!]
\center
\includegraphics[width=0.49\columnwidth]{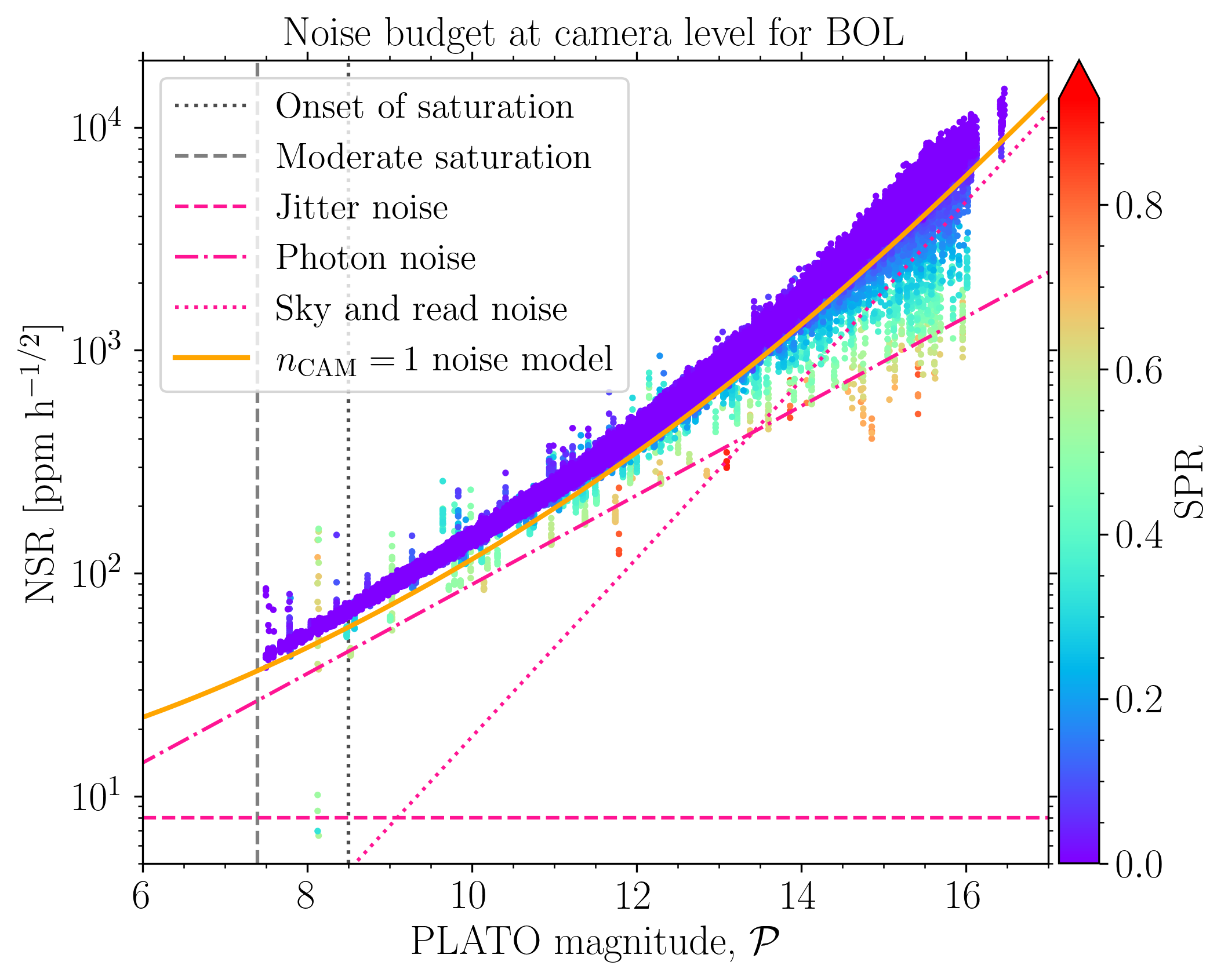}
\includegraphics[width=0.49\columnwidth]{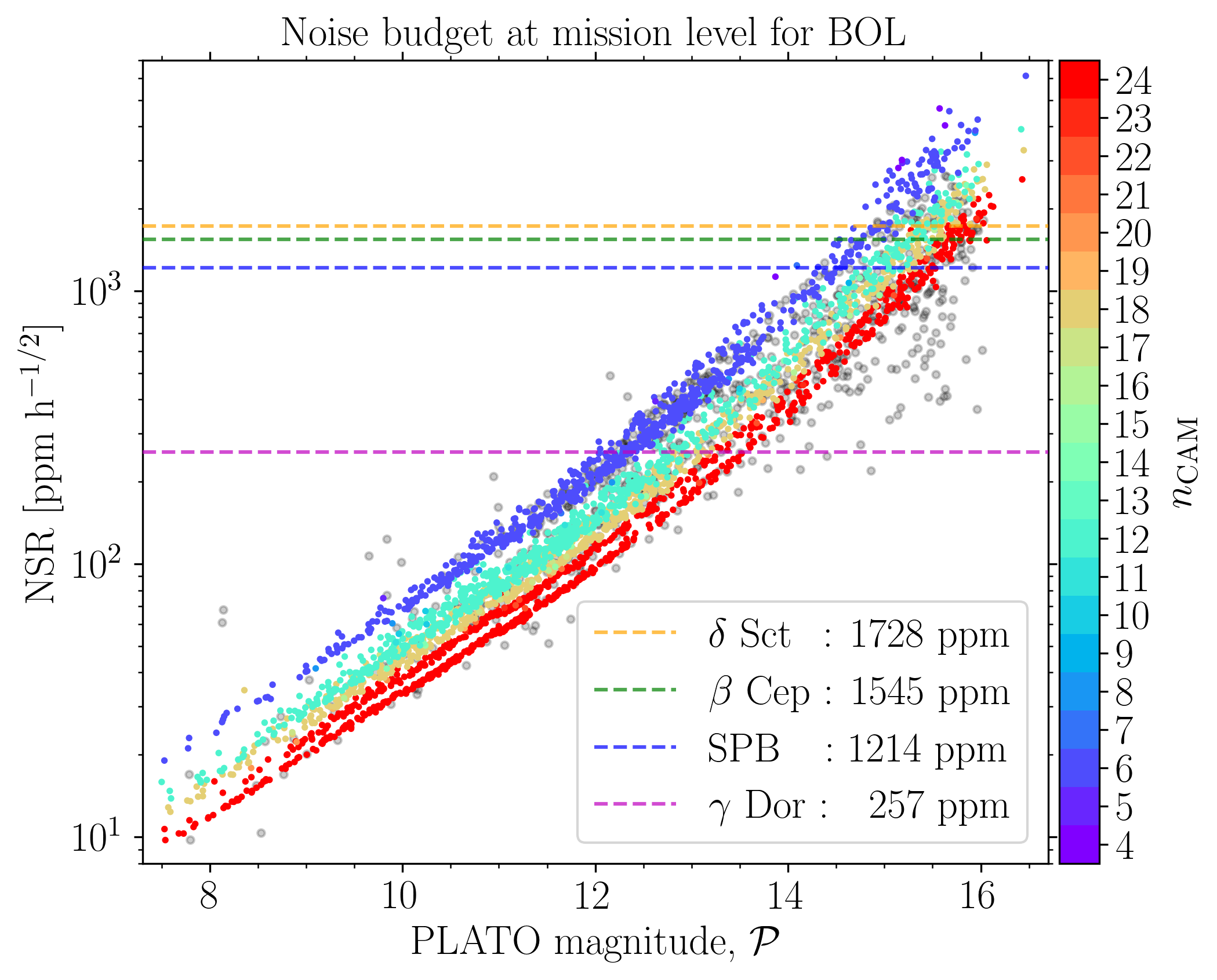}
\caption[]{The PLATO N-CAM noise budget at BOL for the camera level (left) and mission level (right). Left panel: The colour coding reflects an increased level of stellar contamination (from blue to red) given by the stellar pollution ratio (SPR) unity metric. This simulation illustrates that a \platosim{} simulation generally agrees with the expected three dominating noise limits: jitter noise below $\Pb<9$ (dashed pink line), photon noise between $9<\Pb<13$ (dashed-dotted pink line), and CCD read noise/sky background noise for $\Pb>13$ (dotted pink line). The vertical lines illustrate the unset of pixel saturation (dotted grey line) and non-conservative aperture photometry due to blooming (dashed grey line). Right panel: The colour scaling shows the camera visibility. The dashed horizontal lines show a first-order estimate of the expected detection limit for four different classes of pulsators. We define the detection limit as when more than 95\% of the injected amplitudes (draw from a log-normal distribution) can be recovered. This limit is used to define the limiting simulated magnitude.} 
\label{fig:nsr}
\end{figure}

To first approximation, it is clear that any signal detection rate can be computed from a simple estimate of the SNR (or the NSR) given a finite-size threshold. In our analysis, that is a threshold dictating when no pulsation modes can be detected to where a certain fraction of the pulsation modes are detectable. In fact, we can construct a base for such expectations regarding the detection efficiency of pulsation modes using the NSR vs. $\Pb$ diagram at camera and mission level as shown in Fig.~\ref{fig:nsr} (left and right plot, respectively). Due to the specific query of the input catalogue, the distribution of stars over the PLATO FOV means that several NSR($\Pb$) sequences exist per $\tx{n}{CAM}$ count, which typically are not seen \citep[for a detailed explanation for the specifics of the noise budget components, see e.g.][]{jannsen2024platosim}. Using the simulation batch \affogato{}, we show at mission level the approximate regions for which we should be able to detect the 95\% of all pulsation modes of each asteroseismic sample (i.e. below the dashed lines).

\begin{figure}[h!]
\center
\includegraphics[width=0.85\columnwidth]{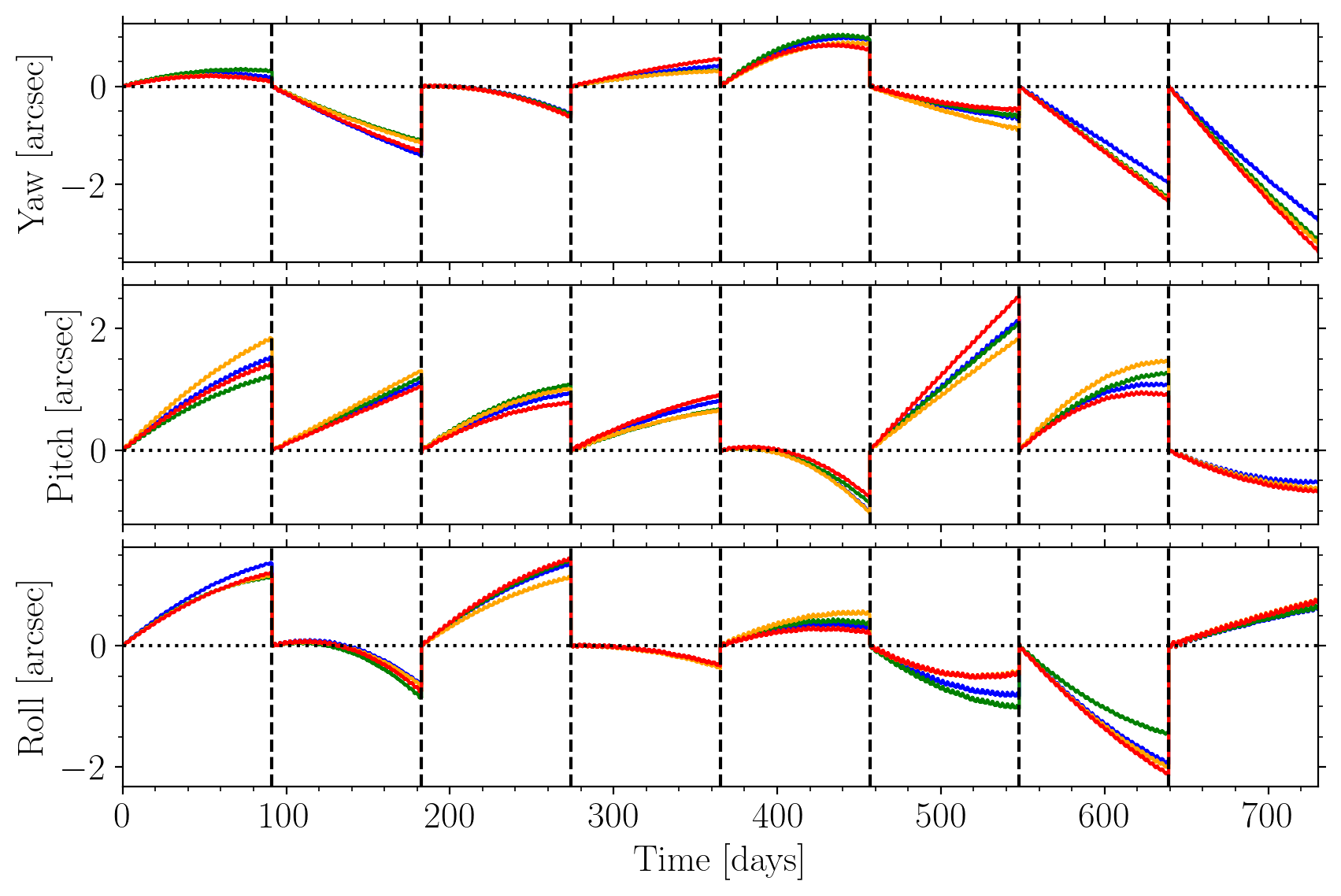}
\caption[]{Analytic thermo-elastic distortion (TED) model used for simulations. The panels (top to bottom) show the TED contribution in the angle yaw, pitch, and roll, respectively. The dashed vertical lines indicate quarterly rotations of the spacecraft. Each angle component is modelled for each camera group (c.f. the colouring), meaning that all six co-pointing cameras within the same group share the same model. The four models are generated using small perturbations to the original second order polynomial model randomly drawn for each mission quarter. A shorter period modulation to each signal is a dynamical model of reaction-wheel offloading events. This plot shows the amplitude distribution used for simulation batches \affogato{} and \doppio{}, while for \cortado{} the models are identical but the amplitudes have been inflated four times.} 
\label{fig:TED}
\end{figure}

\begin{figure}[h!]
\center
\includegraphics[width=0.5\columnwidth]{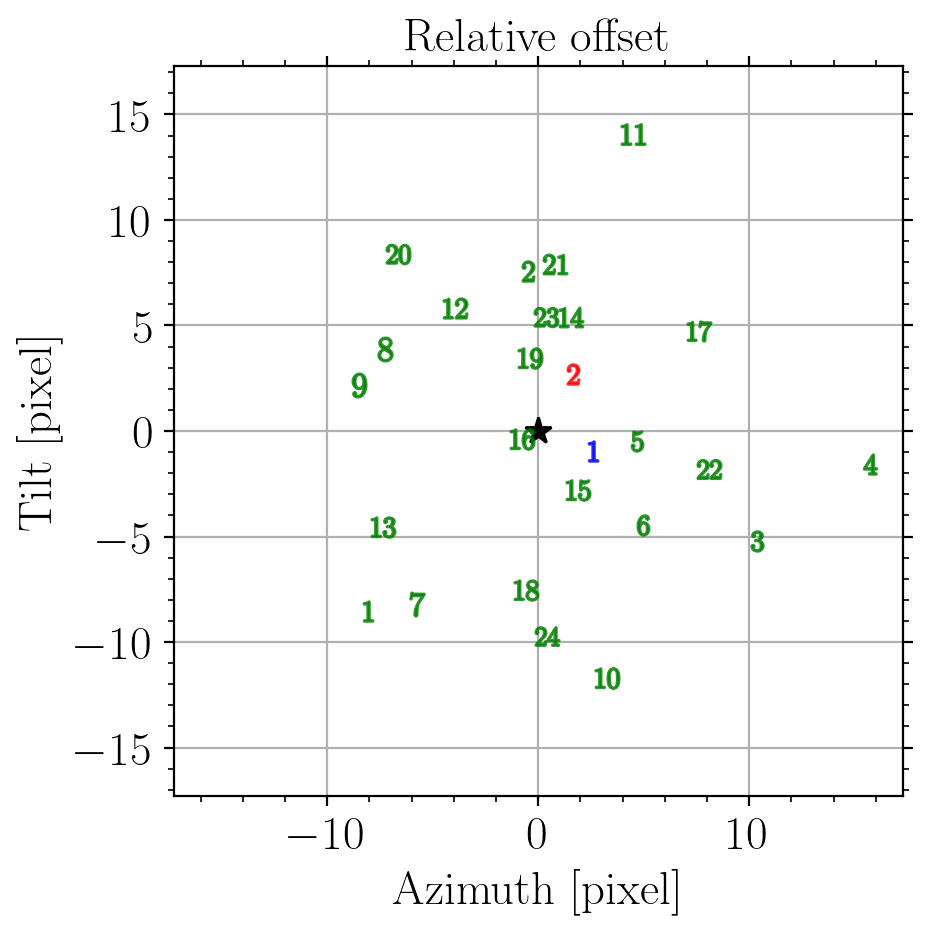}
\includegraphics[width=0.4\columnwidth]{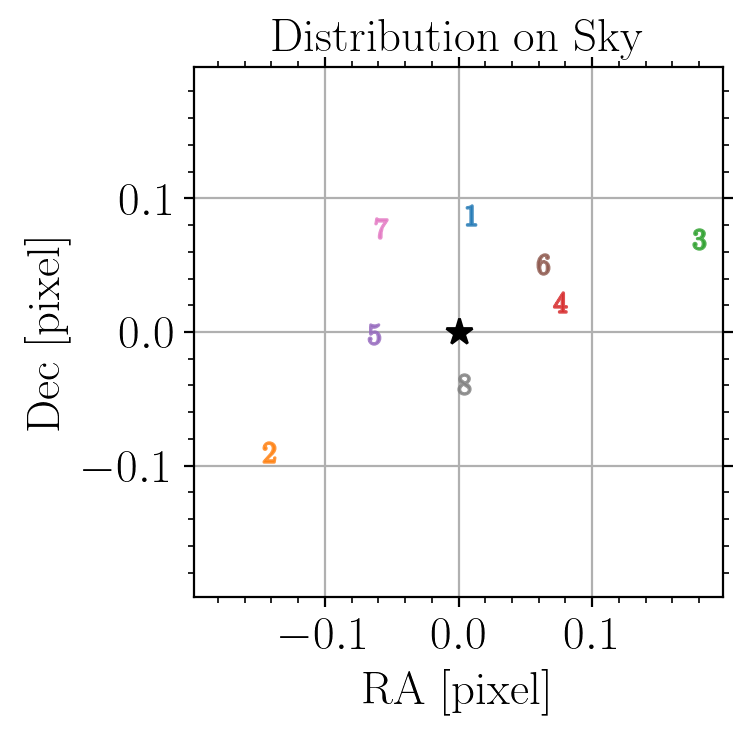}
\caption[]{Pointing error sources used for all simulations. The left panel shows the effective camera misalignment errors as seen in the focal plane. The number is the (\platosim{}) ID of each camera (green for the N-CAMs and blue/red for the blue/red filter of the F-CAMs). The right panel shows the effective payload pointing errors between consecutive mission quarters (indicated by number, with one being the first pointing towards the LOPS2) as seen in the focal plane.} 
\label{fig:APEandPRE}
\end{figure}

In order to make \platosim{} simulations as realistic as possible, one needs to apply different types of pointing error sources. Besides spacecraft jitter, we include an analytic model of thermo-elastic distortion (TED; mainly caused by a temperature gradient between the cameras and the optical bench) as shown in Fig.~\ref{fig:TED}. The model is a composition of a simple polynomial and a dynamic model of wheel-offloading events. Secondly, to make sure that all six co-pointing cameras of each camera group differ in instrumental systematics, we calculate and apply camera misalignments (caused by mounting errors during the integrating and small misplacements during launch). Lastly, (small) erroneously spacecraft pointing is taken into account between consecutive mission quarters. Figure~\ref{fig:APEandPRE} shows the effective offsets in the focal plane by the latter two models (left and right panel, respectively). 

\begin{figure}[h!]
\center
\includegraphics[width=0.7\columnwidth]{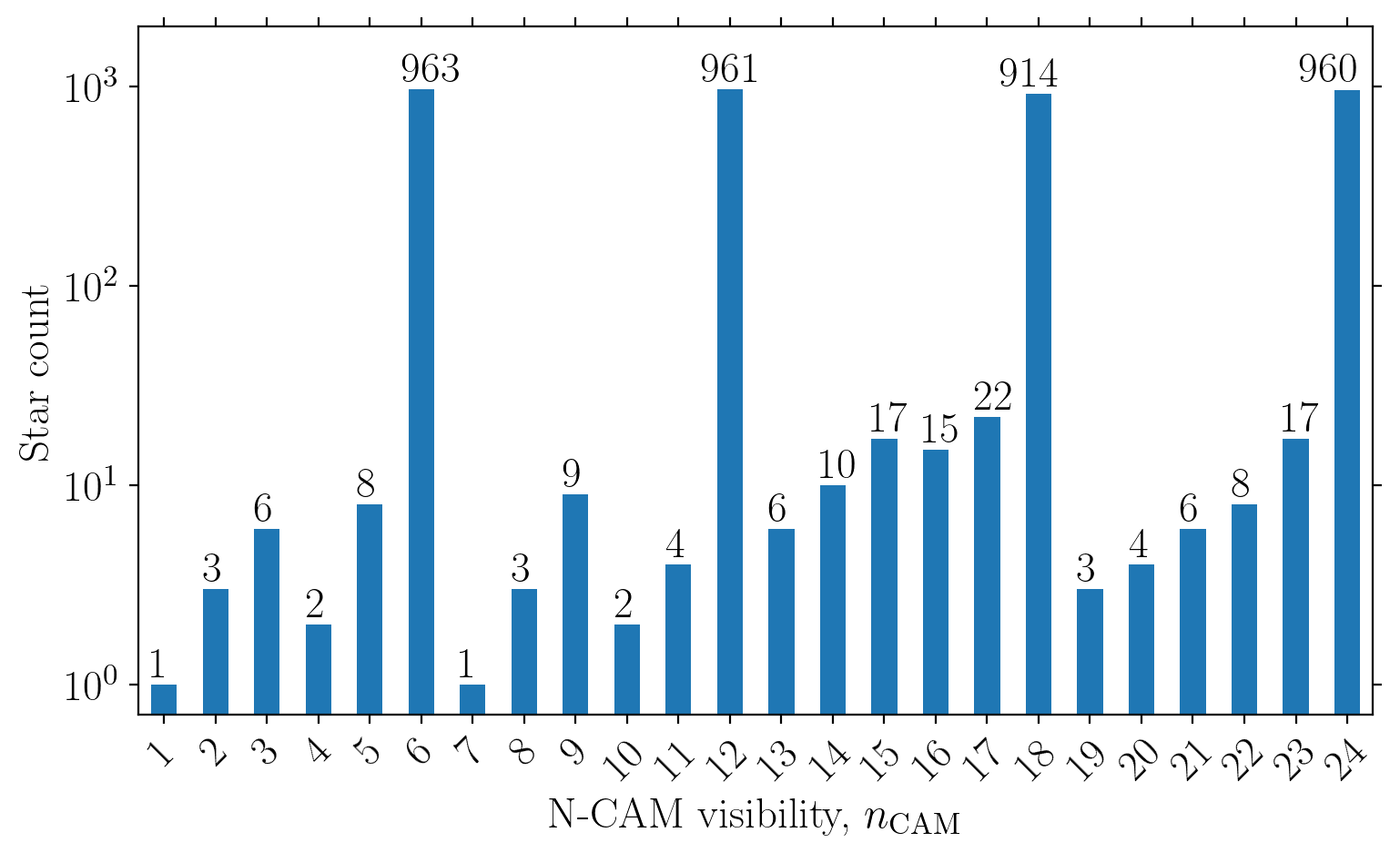}
\caption[]{Star count histogram as function of N-CAM visibility $\tx{n}{CAM}$ of the 4000 SPB sample stars.} 
\label{fig:star_count_histogram}
\end{figure}

It is important to notice that multiple stars (from each asteroseismic sample) are observed with a camera count different from $\tx{n}{CAM}=\{6, 12, 18, 24\}$. For example, considering the simulations of the 4000 (high-amplitude) SPB stars as shown in Fig.~\ref{fig:star_count_histogram} (i.e. the stars also shown in Fig.~\ref{fig:result_detection_limit}), it is clear that most stars are observed with 6, 12, 18, and 24 N-CAMs, however, the numbers above each histogram bin also show that many stars are not observed with the ``expected'' number of cameras. This is because we assemble the stellar catalogue from a single mission quarter simulation where all cameras are perfectly aligned. As discussed in Sect.~\ref{sec:sim_spacecraft}, we apply spacecraft pointing errors between mission quarters and camera misalignments in our pixel simulations. Thus, our simulations realistically reflect that not every star is observable for every single mission quarter/camera if it (i) falls between two CCDs, (ii) falls outside a CCD, or (iii) is blocked by the stray light mask. Again, this illustrates the force -- and importance -- of using pixel-based simulations for space missions like PLATO that utilise a (more complex) multi-camera concept.

\section{Stellar parameter space}\label{app:parameters}

\begin{figure}[h!]
\center
\includegraphics[width=0.50\columnwidth]{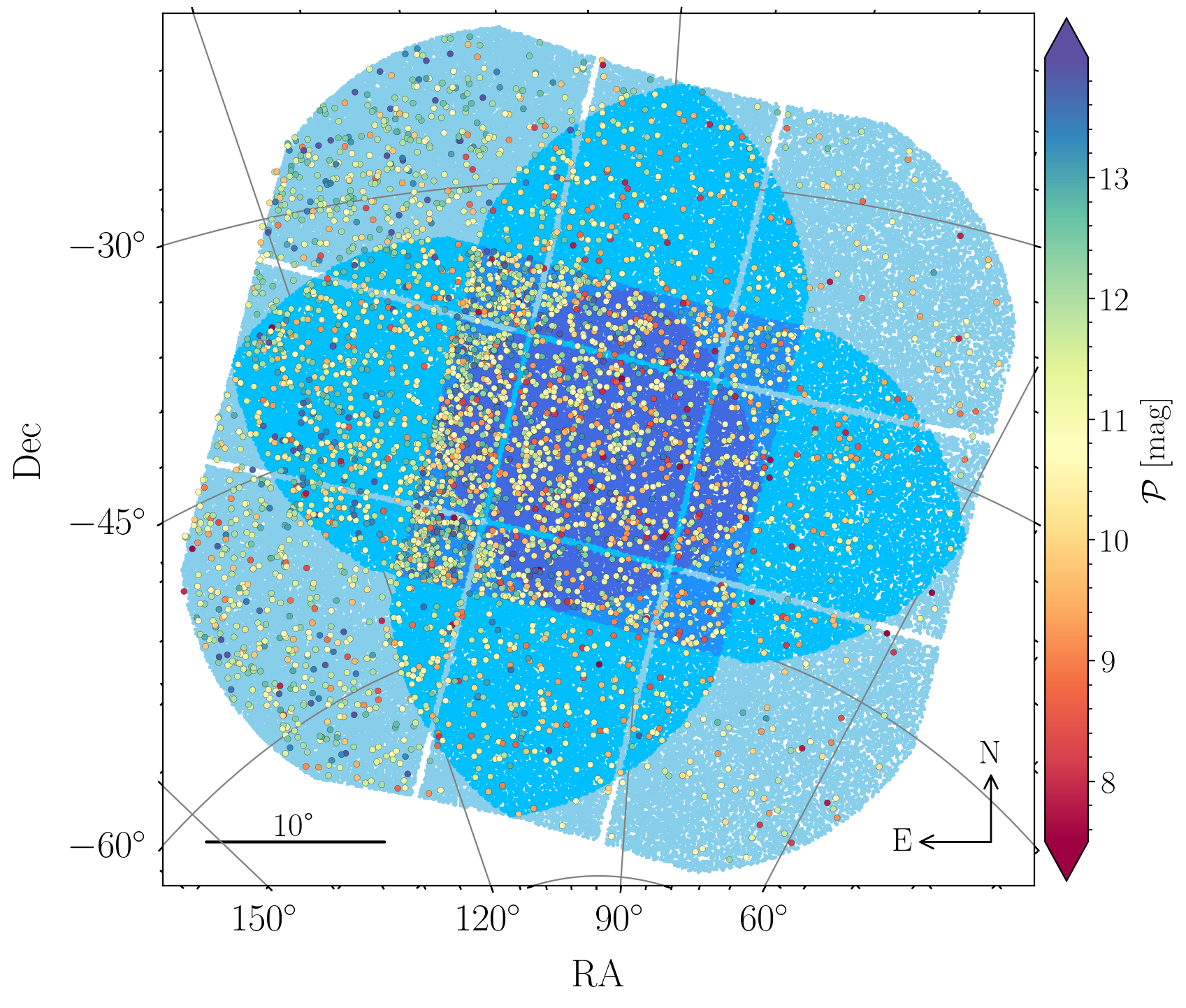}
\includegraphics[width=0.45\columnwidth]{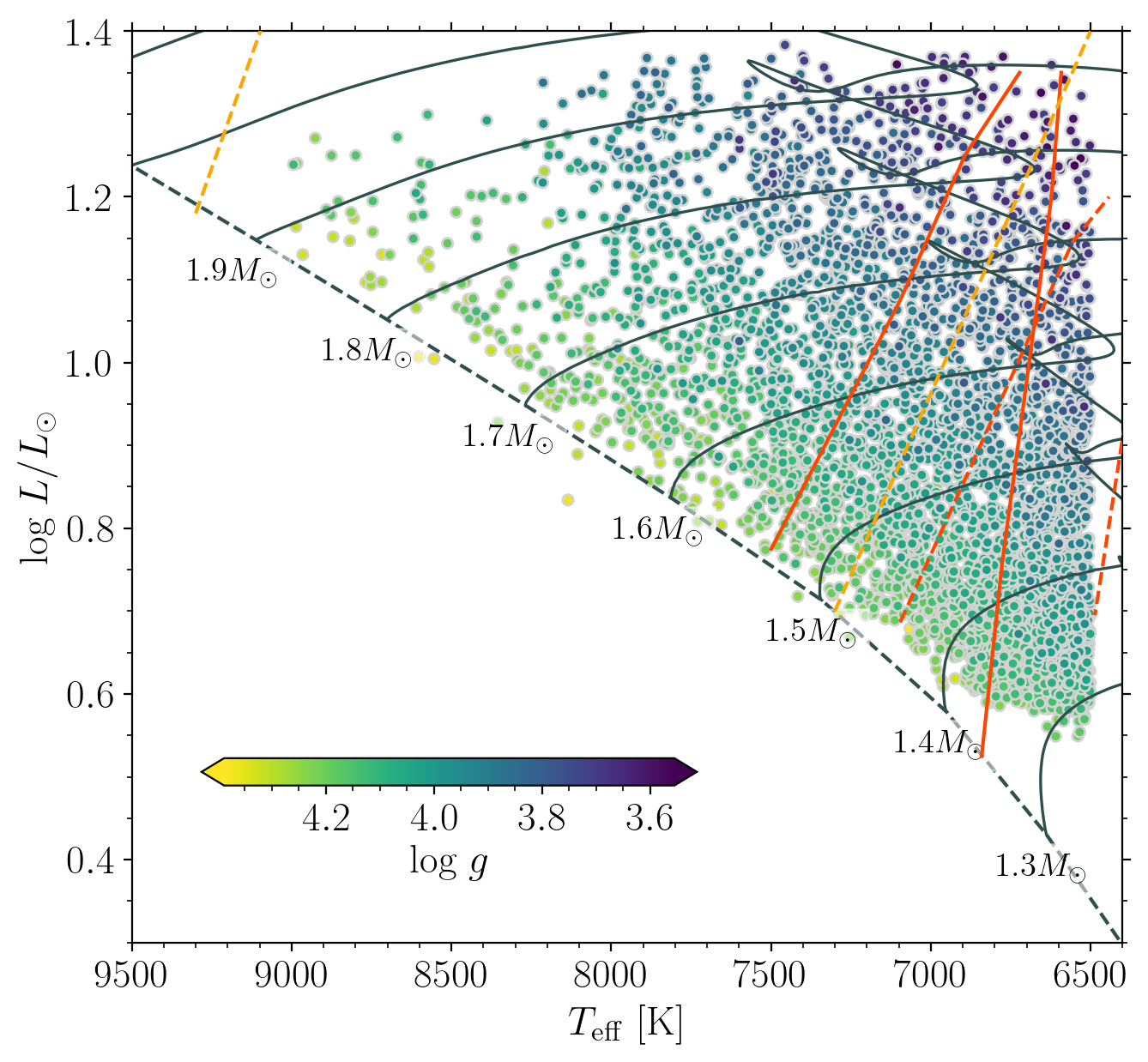}
\includegraphics[width=0.48\columnwidth]{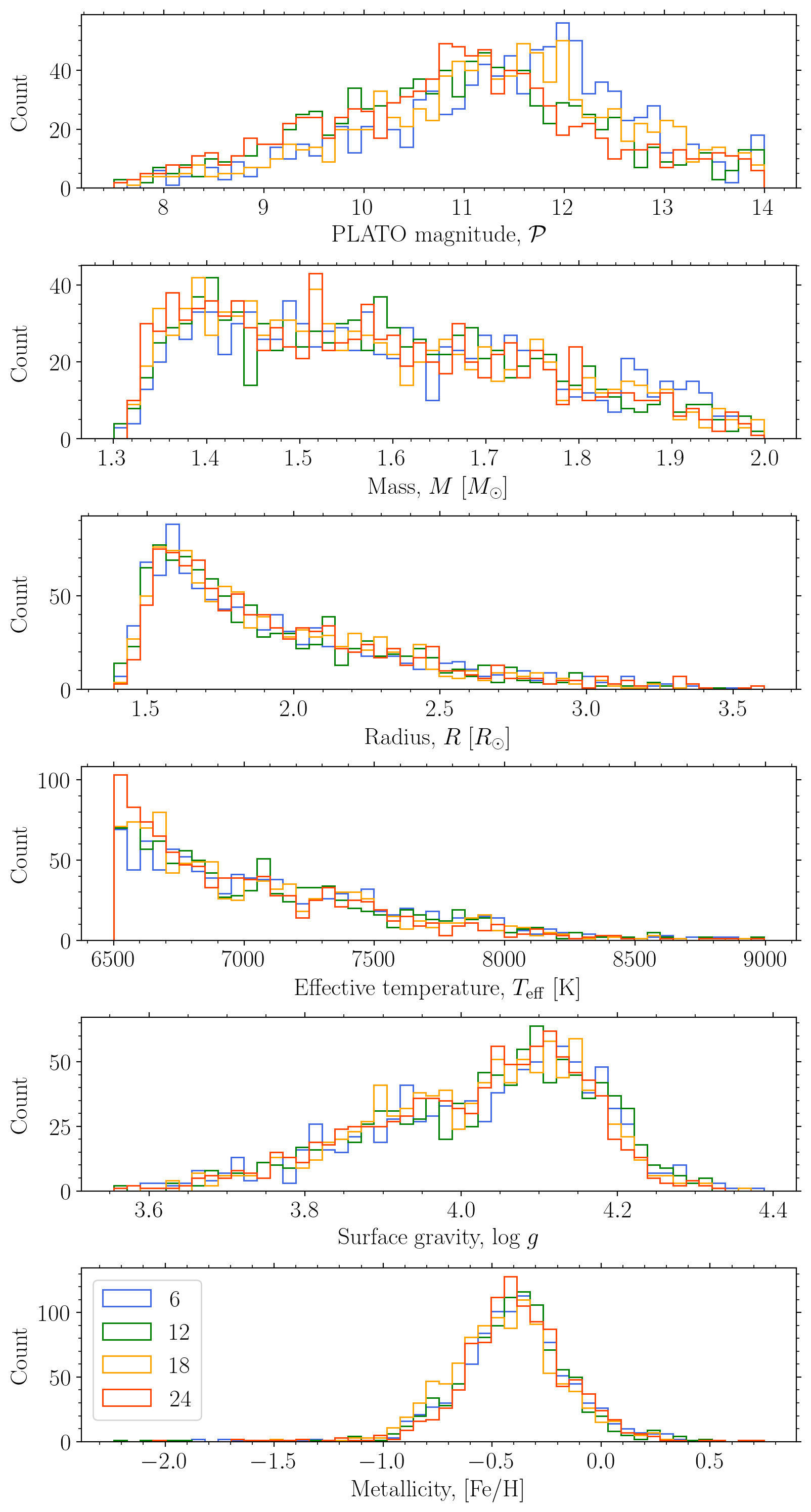}
\includegraphics[width=0.48\columnwidth]{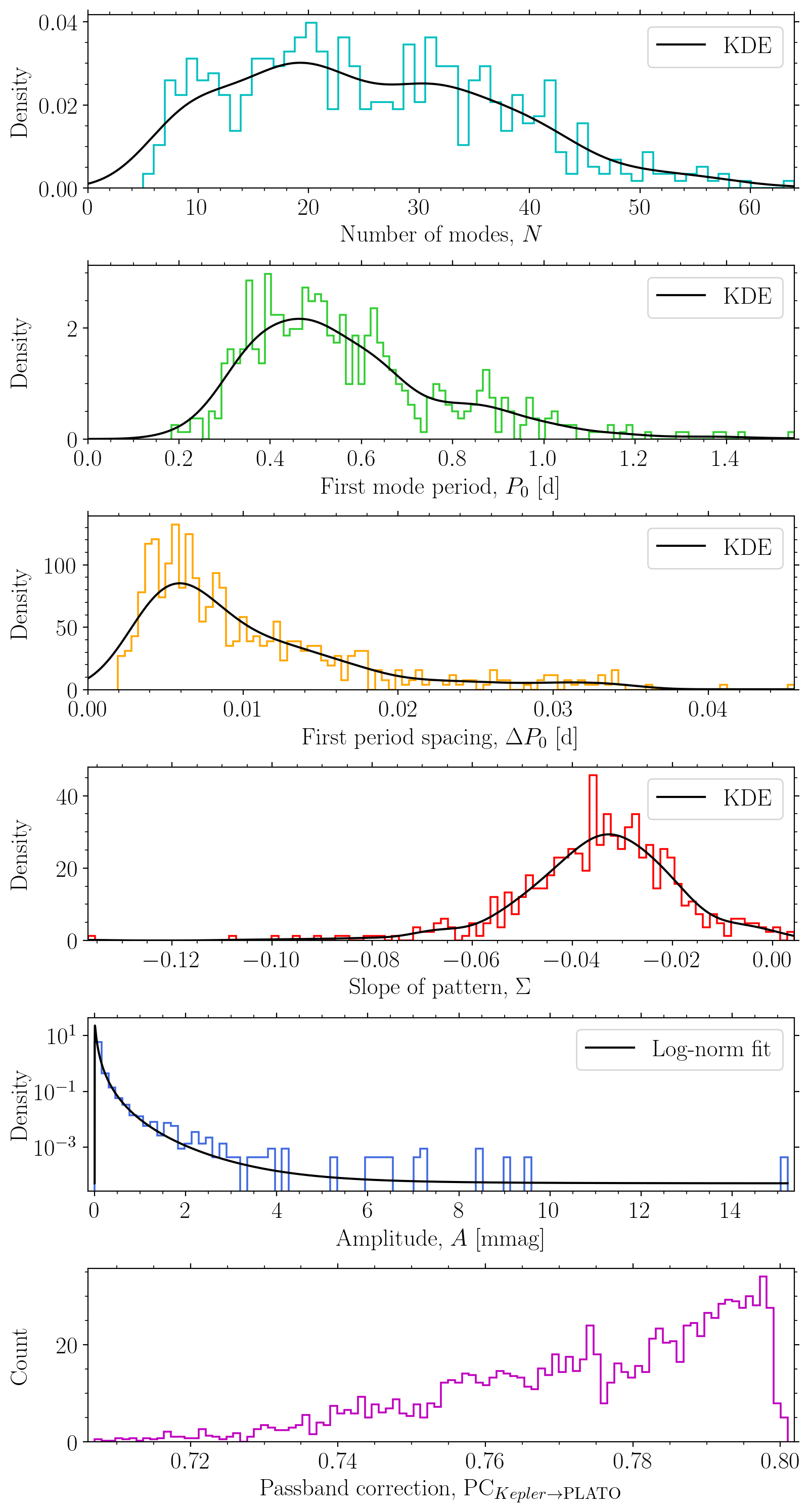}
\caption[]{Model parameters for the \gdor{} sample. See description of Appendix~\ref{app:parameters}.} 
\label{fig:sample_gdor}
\end{figure}

\begin{figure}[h!]
\center
\includegraphics[width=0.50\columnwidth]{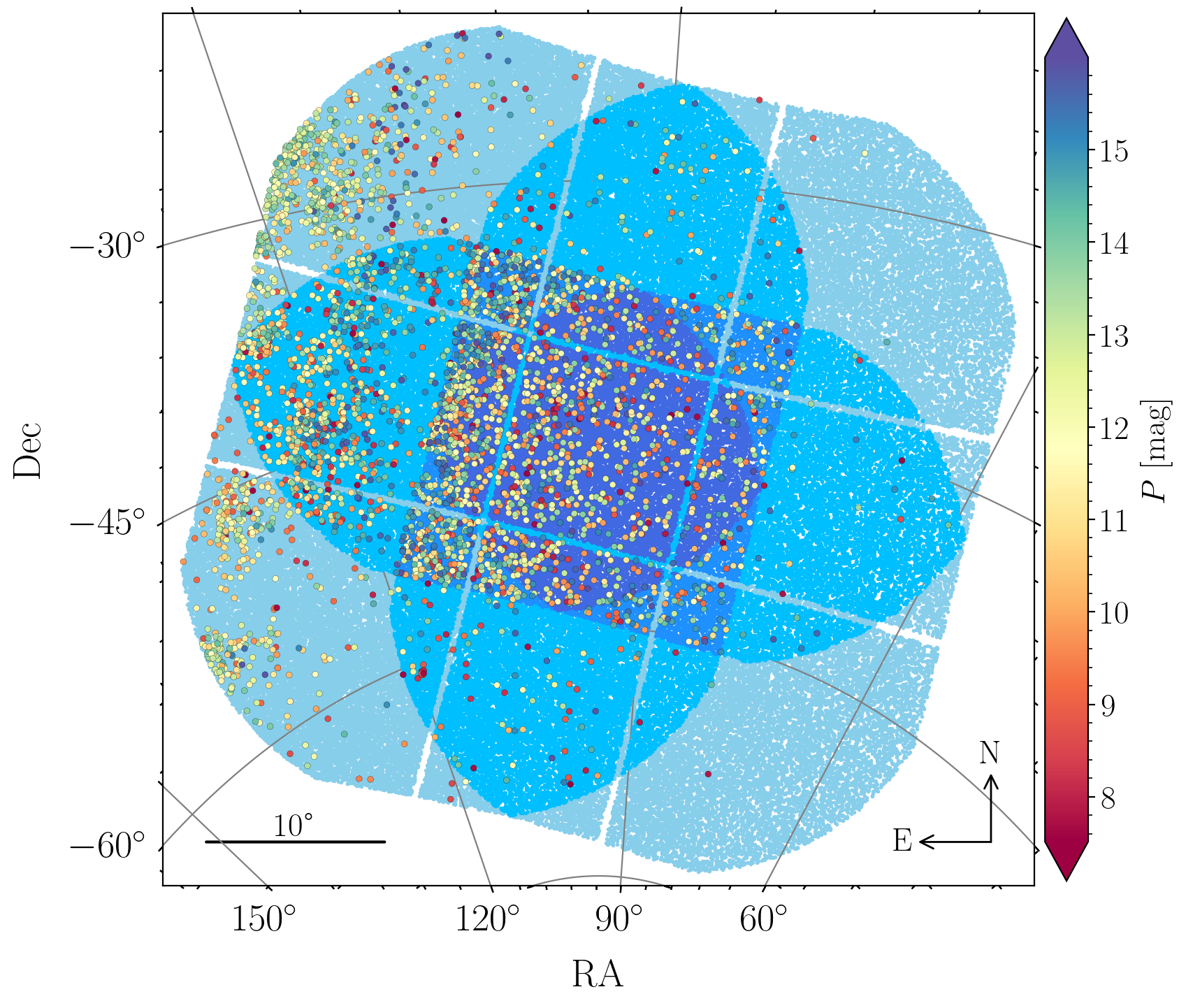}
\includegraphics[width=0.45\columnwidth]{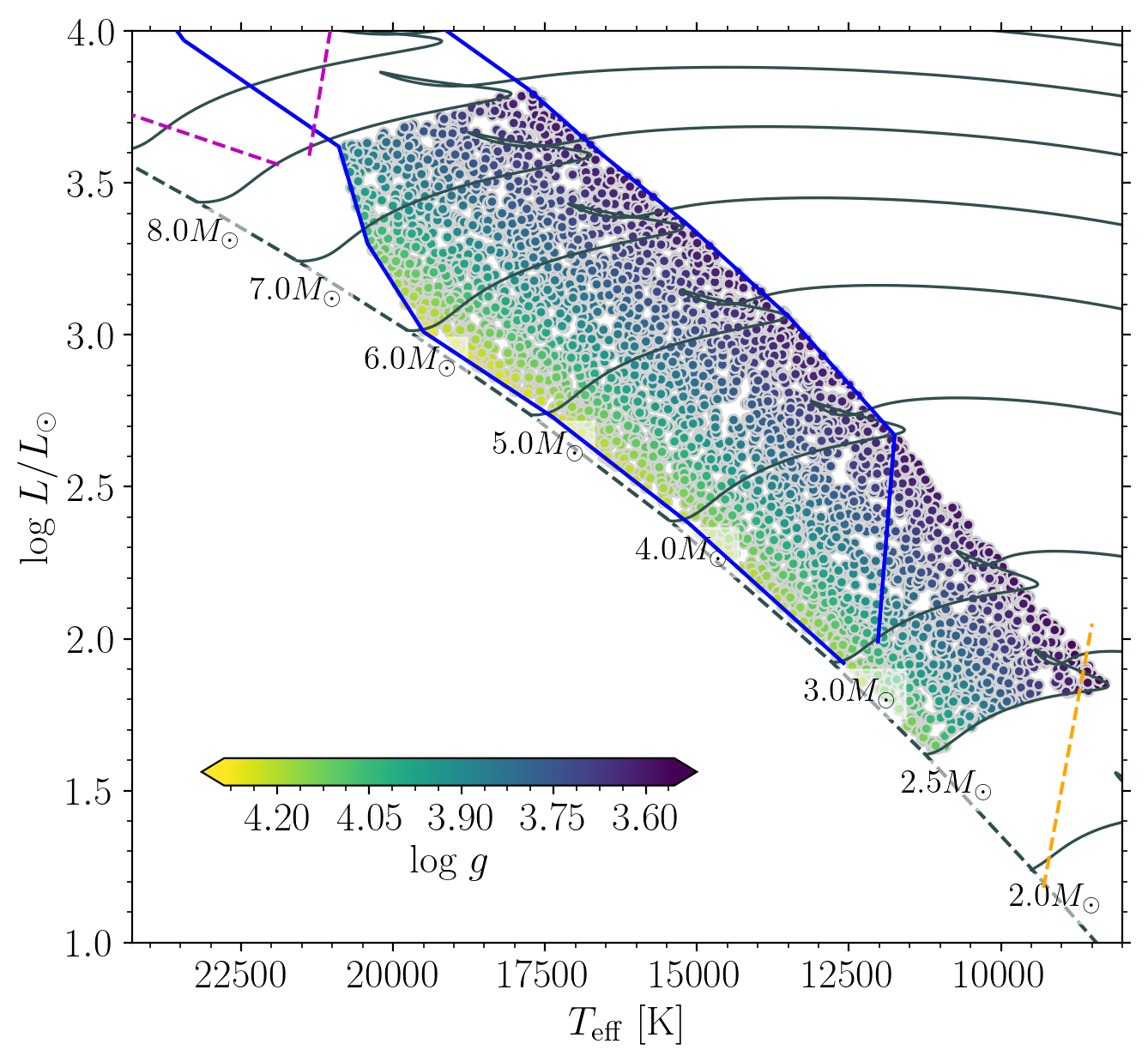}
\includegraphics[width=0.48\columnwidth]{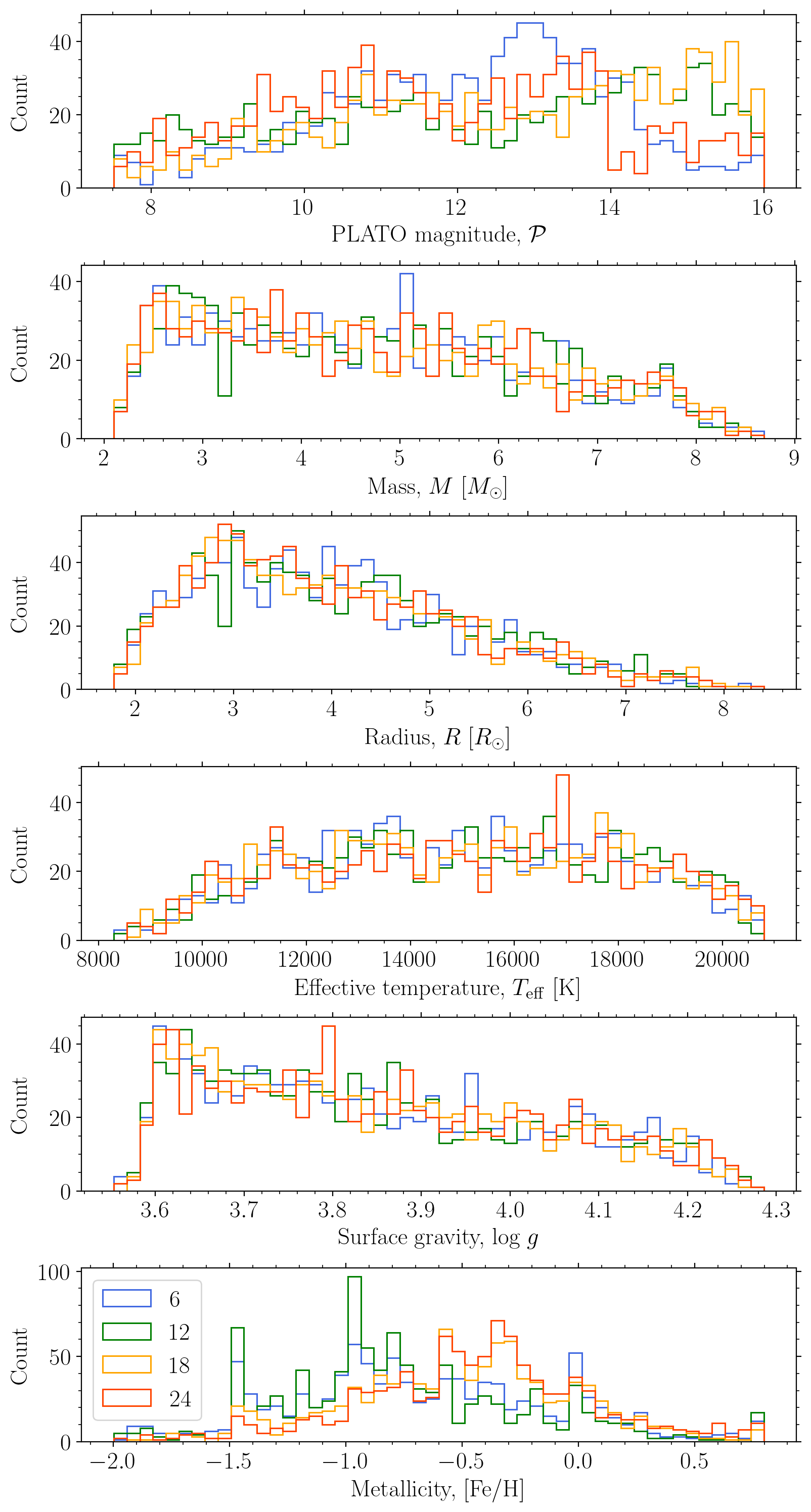}
\includegraphics[width=0.48\columnwidth]{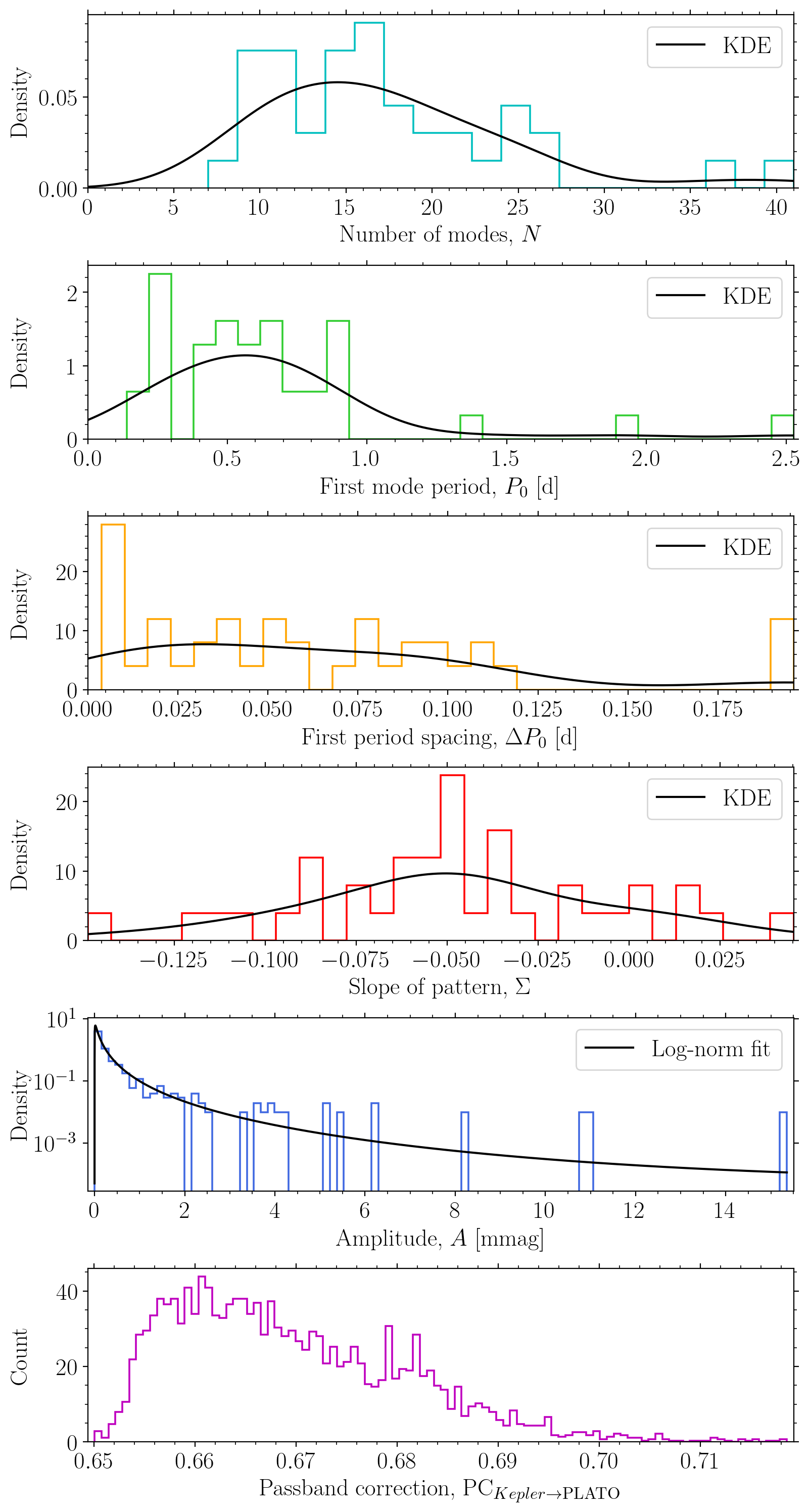}
\caption[]{Model parameters for the SPB sample. See description of Appendix~\ref{app:parameters}.} 
\label{fig:sample_spb}
\end{figure}

\begin{figure}[h!]
\center
\includegraphics[width=0.50\columnwidth]{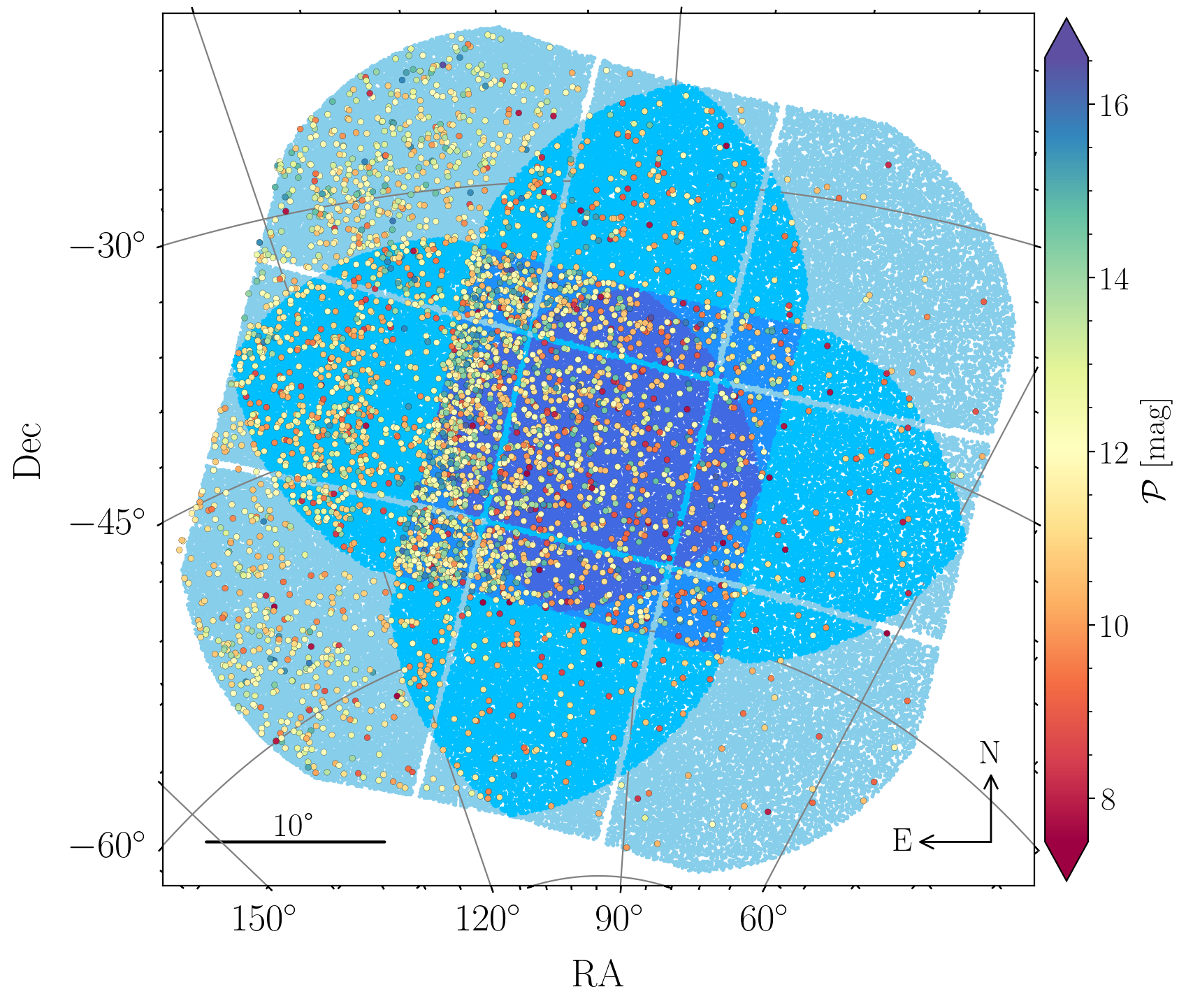}
\includegraphics[width=0.45\columnwidth]{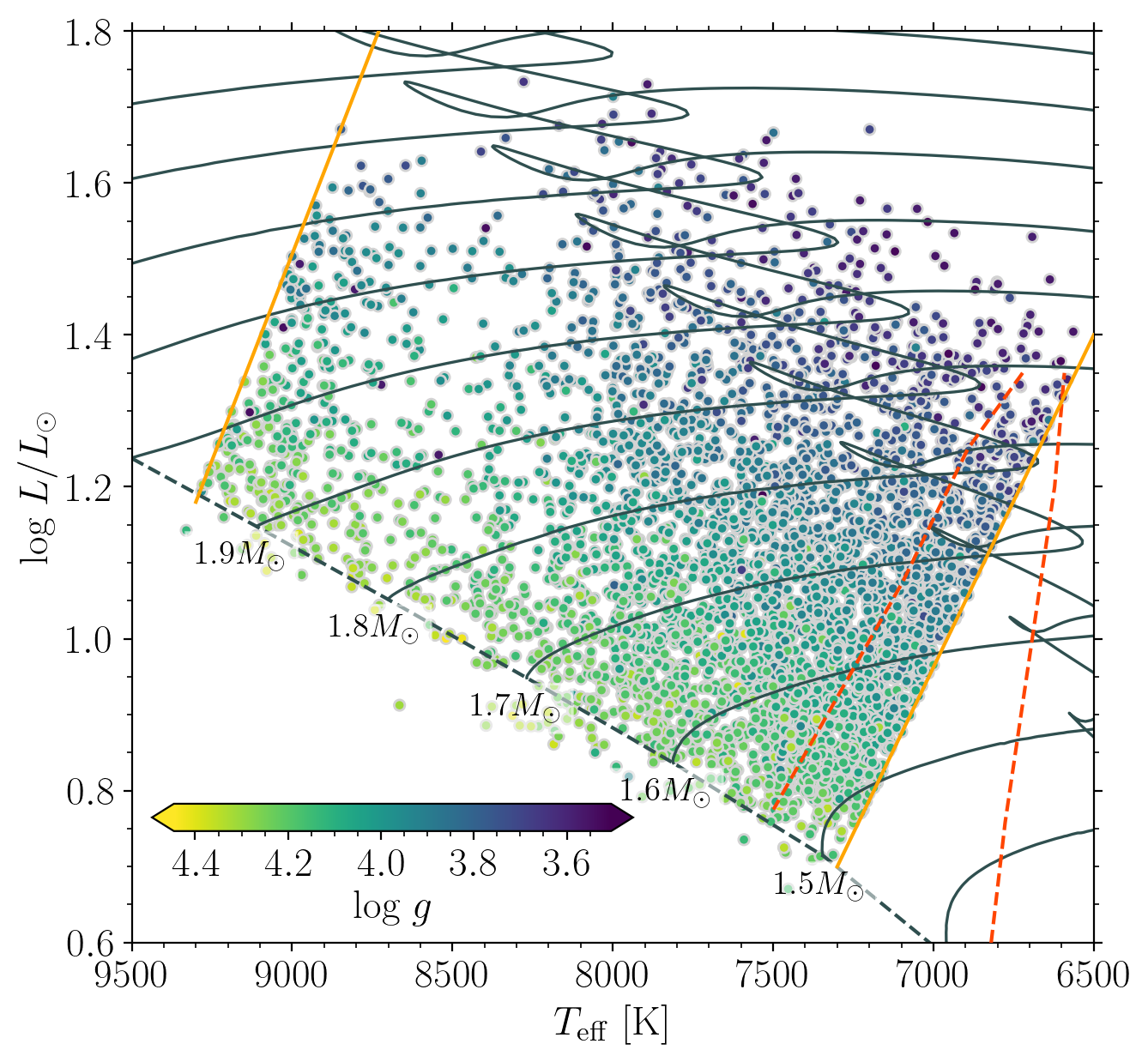}
\includegraphics[width=0.48\columnwidth]{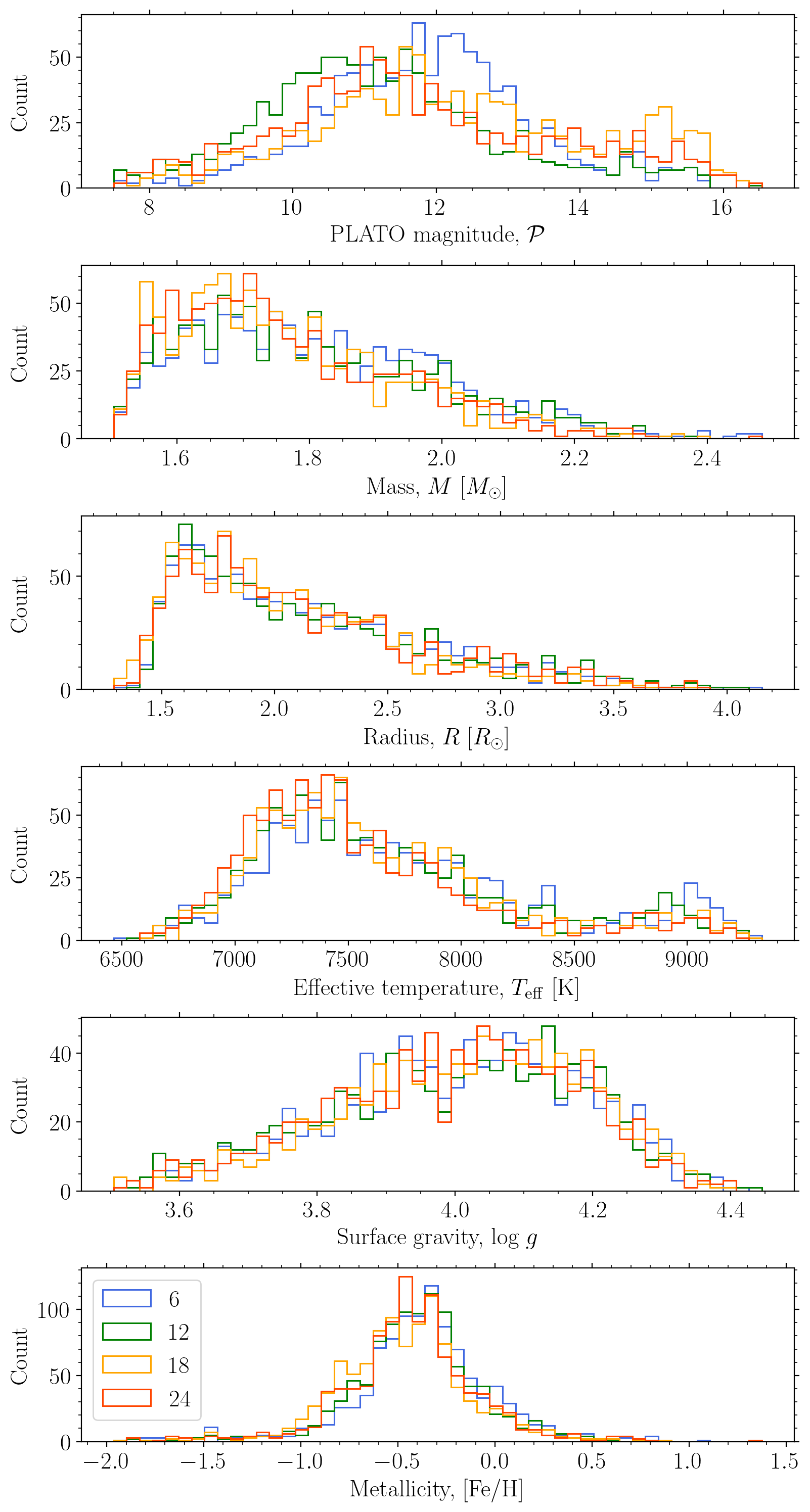}
\includegraphics[width=0.48\columnwidth]{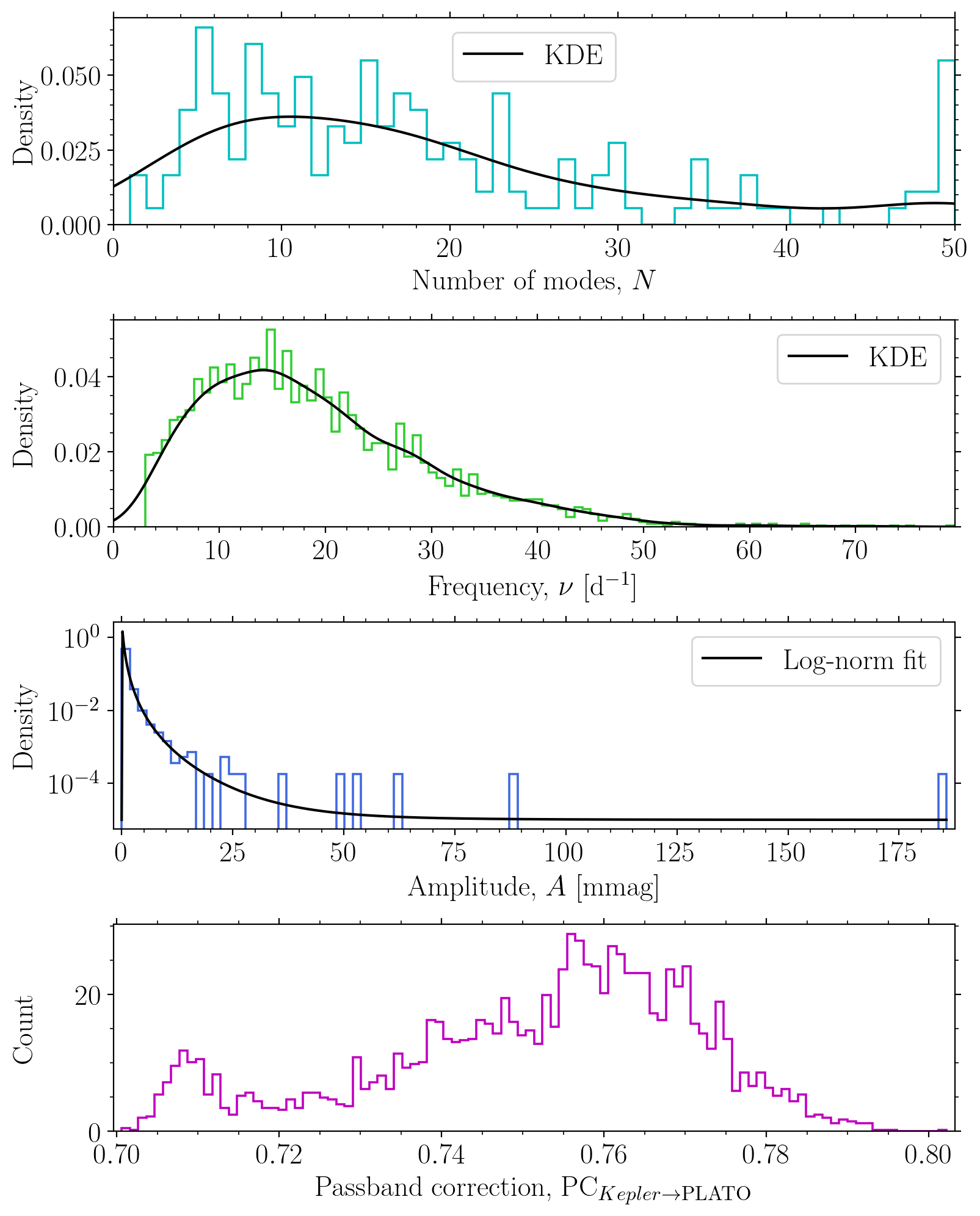}
\caption[]{Model parameters for the \dsct{} sample. See description of Appendix~\ref{app:parameters}.} 
\label{fig:sample_dsct}
\end{figure}

\begin{figure}[h!]
\center
\includegraphics[width=0.50\columnwidth]{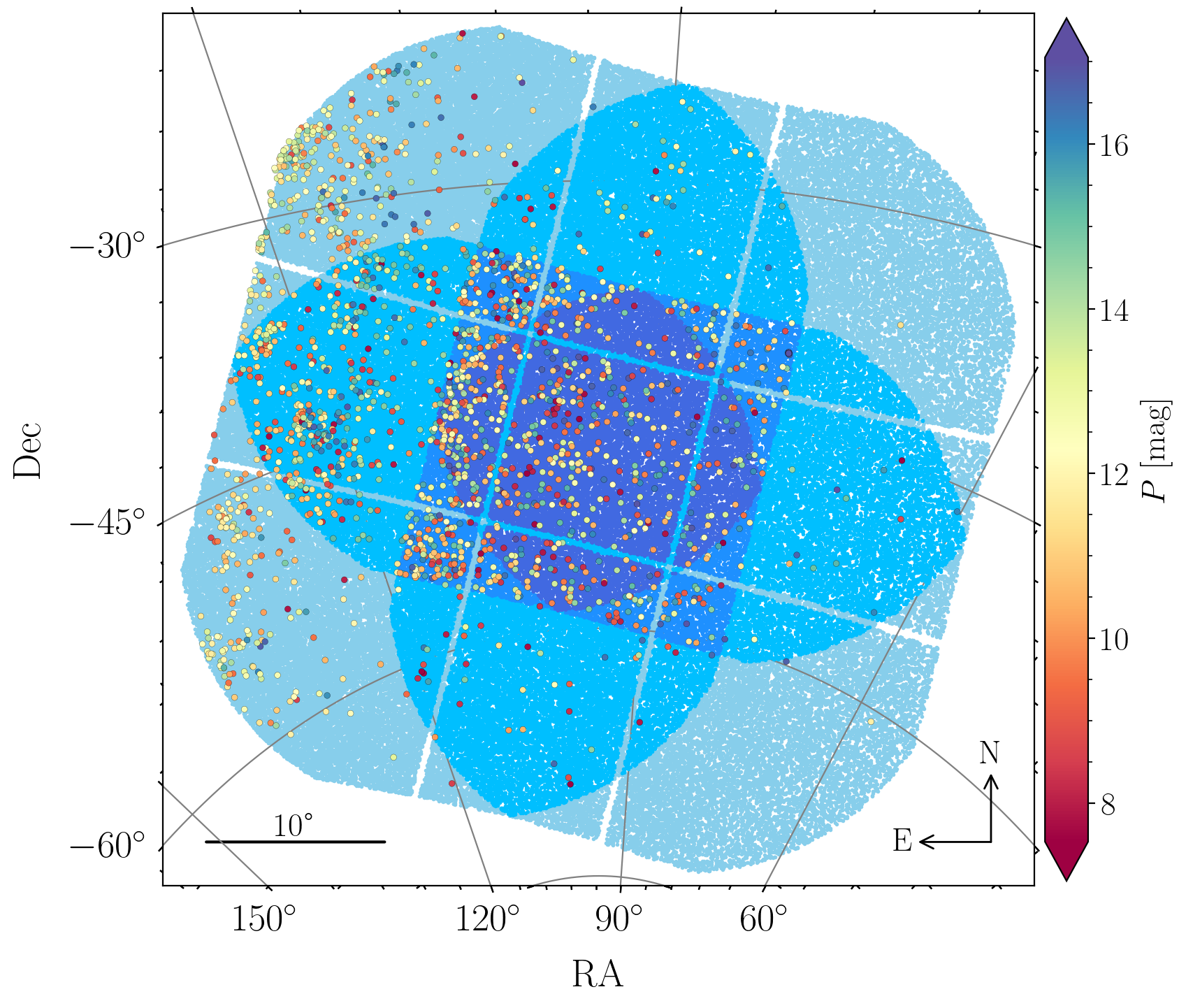}
\includegraphics[width=0.45\columnwidth]{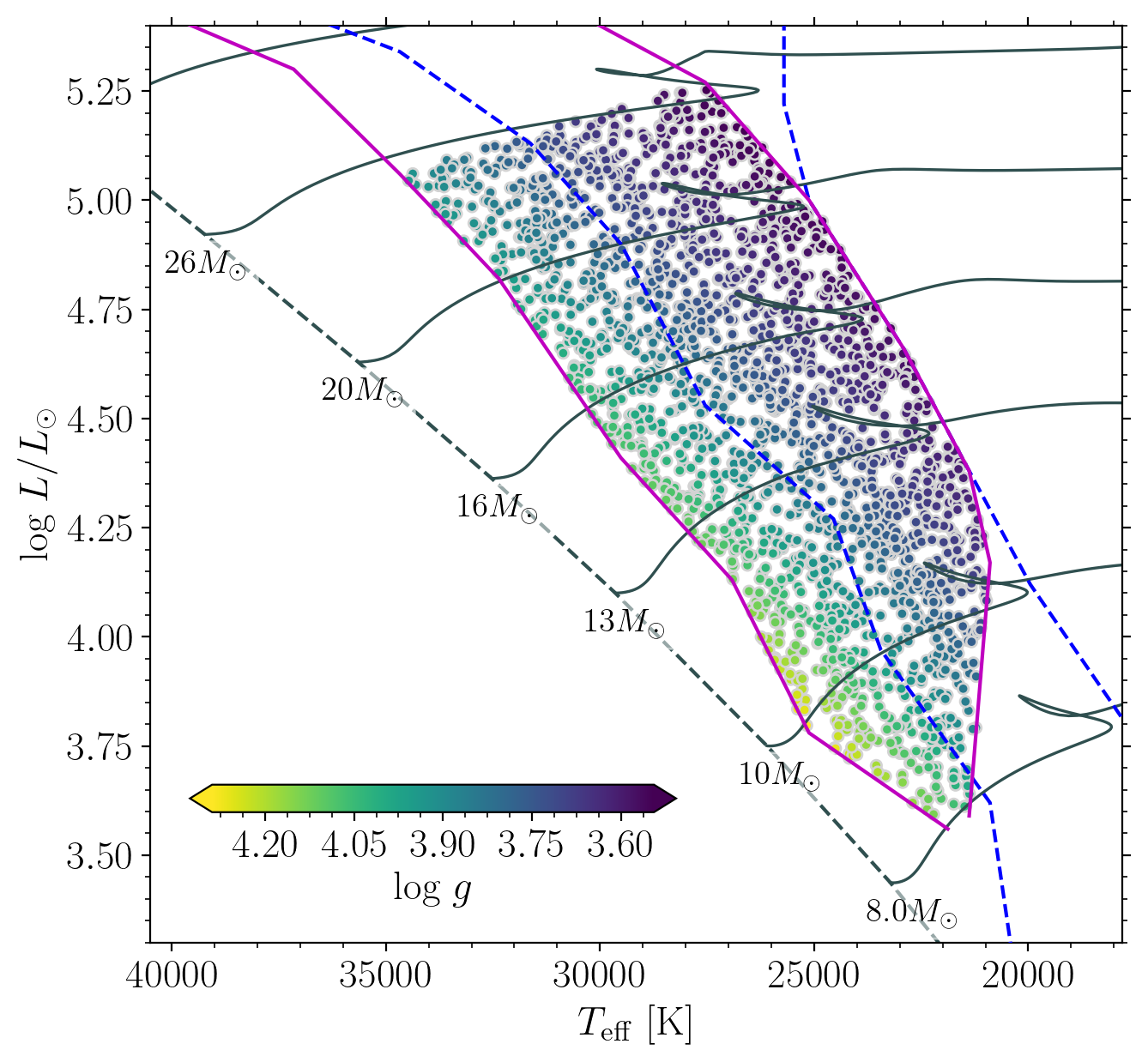}
\includegraphics[width=0.48\columnwidth]{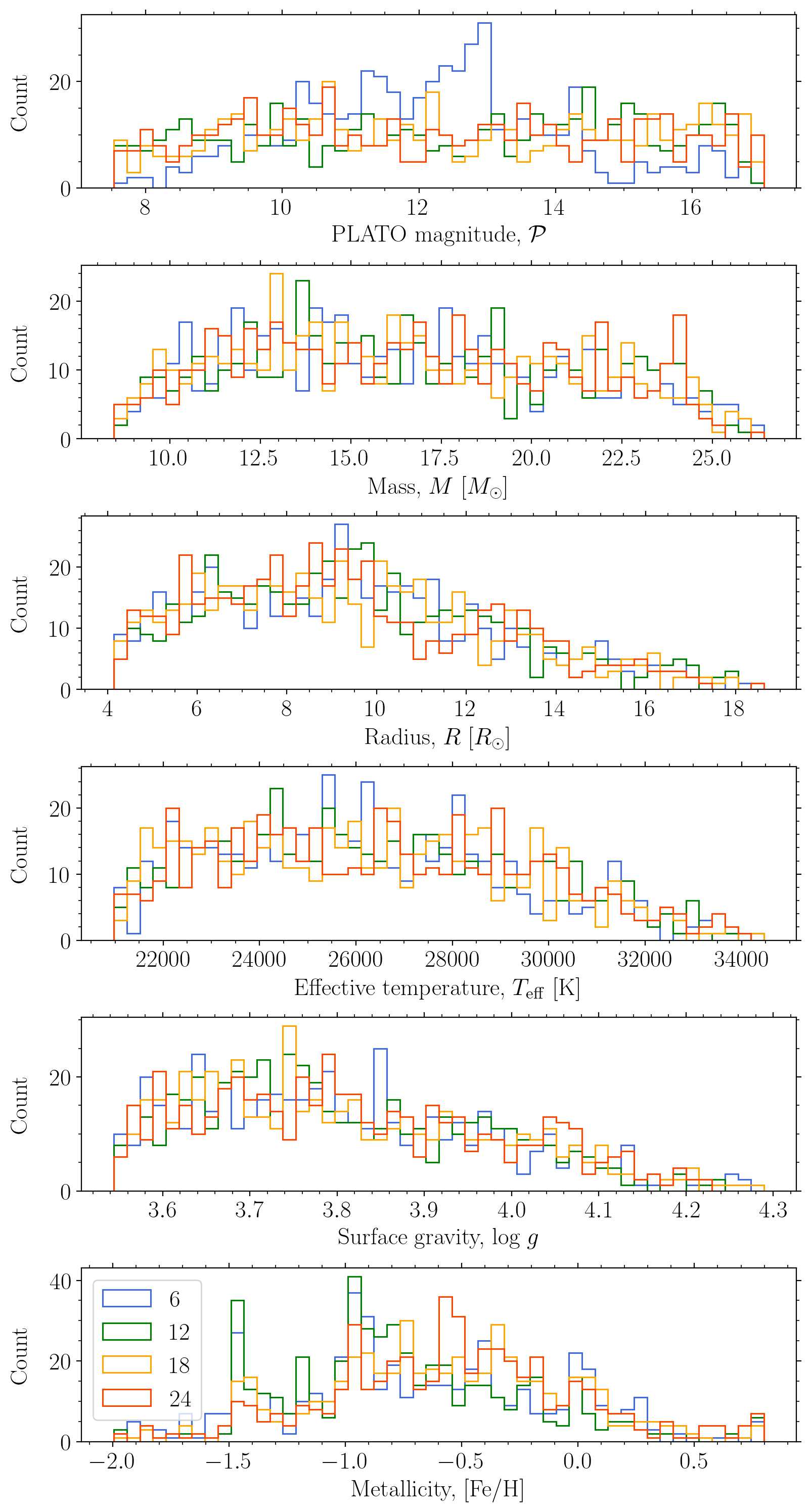}
\includegraphics[width=0.48\columnwidth]{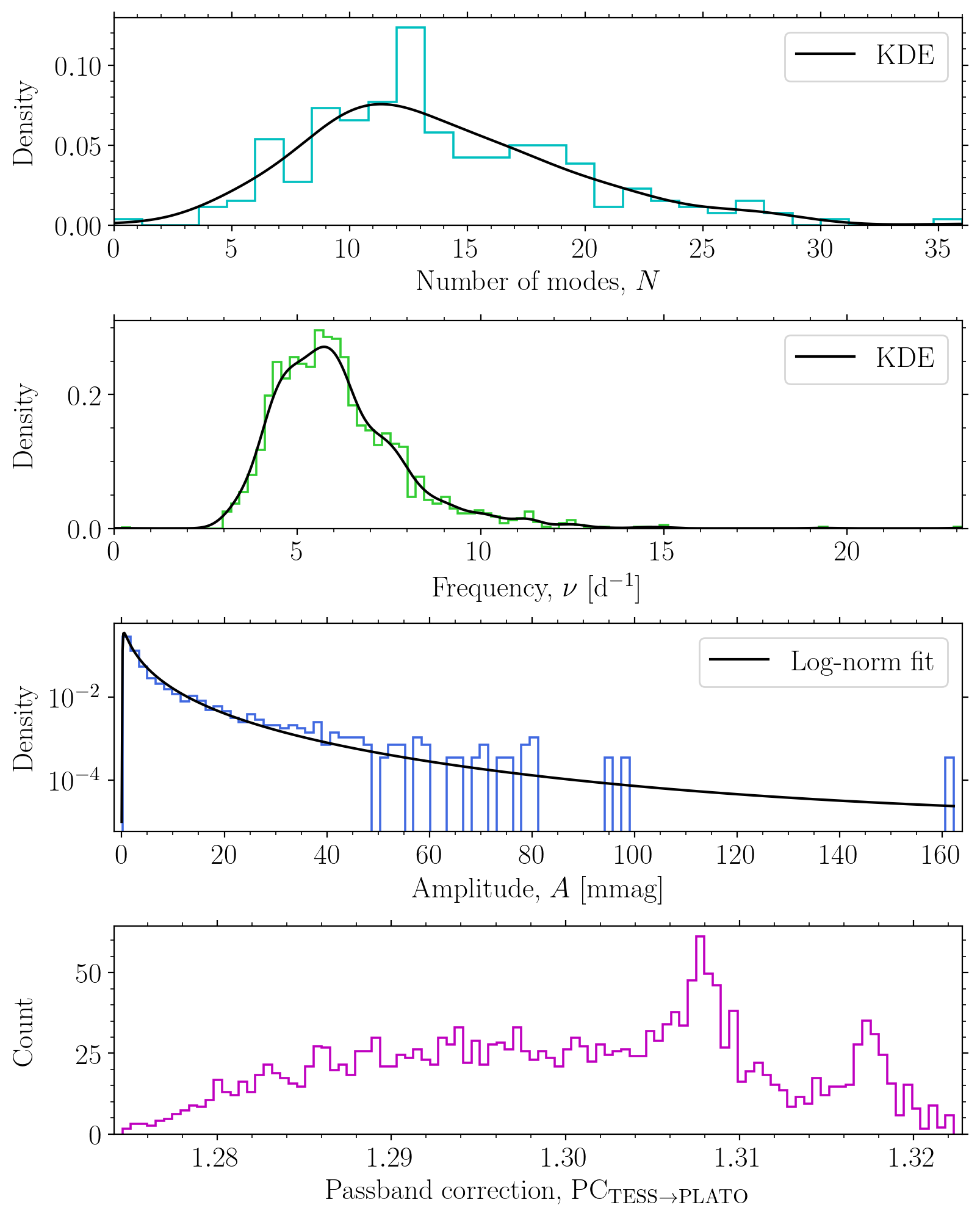}
\caption[]{Model parameters for the \bcep{} sample. See description of Appendix~\ref{app:parameters}.} 
\label{fig:sample_bcep}
\end{figure}

In this appendix we give a general overview of the stellar parameter space for used to generate the simulations of MOCKA. This include the sky catalogue and the variability model of each asteroseismic sample. We present a full-page plot with several subpanels for the \gdor{} (Fig.~\ref{fig:sample_gdor}), SPB (Fig.~\ref{fig:sample_spb}), \dsct{} (Fig.~\ref{fig:sample_dsct}), and \bcep{} (Fig.~\ref{fig:sample_bcep}) sample:
\begin{itemize}
\item In the top left plot, the sky distribution of each stellar sample is shown with respect to the PLATO FOV (similar to Fig.~\ref{fig:LOPS2}), and colour coded by magnitude. As expected for non-evolved stars, the higher the mass, the more concentrated the sample is towards high density regions (e.g. seen from a comparison between the \gdor{} and \dsct{} stars to that of the SPB and \bcep{} stars). 

\item The top right plot is a HRD of each stellar sample, colour coded by surface gravity, together with some MIST evolutionary tracks \citep{choi2016mesa}. Relevant instability strips are also displayed, specifically for \gdor{} stars \citep[red lines from][based on theoretical computations]{dupret2005convection}, \dsct{} \citep[orange lines from][based on observations]{murhpy2019gaia}, SPB and \bcep{} stars \citep[respectively blue and purple lines from][being theoretical p and g mode instability regions]{burssens2020variability}.

\item The lower left plot shows six histograms of the target star parameter space \{$\Pb$, $M$, $R$, $\teff$, $\logg$, \metal{}\} across the four N-CAM visibilities (see legend in bottom panel). 

\item The bottom right plot shows multiple distributions of the pulsation modes used to create each variable signal. Note that for the \gdor{} and SPB stars, the distribution of the period-spacing pattern gradient (red histograms) is only shown for illustrative reasons. Within the same plot, the bottom panel always shows the distribution of the applied passband corrections either from \textit{Kepler} to PLATO or from TESS to PLATO.
\end{itemize}

\subsection{\gdor{} and \dsct{} stars}

We refine the \gdor{} sample by applying the following physical cuts: $1.2 < M/\sole{M} < 2$, $\log g >3.5$, and $L > \SI{2}{\Lsun}$. Note, we exclude star with $\tx{T}{eff}<\SI{6500}{\kelvin}$ to roughly match the theoretical red border of the instability strip \citep[see the solid red lines in the top-right plot of Fig.~\ref{fig:sample_gdor}; adopted from][corresponding to a mixing length of $\alpha_{\rm MLT}>1.5$]{dupret2005convection}, since there is no strong observational evidence for the existence of \gdor{} pulsators beyond it. On the other hand, many \gdor{} stars have been observed hotter than the blue border of the instability region \citep[e.g.][]{hey2024confronting}. Hence, we include stars up to $\tx{T}{eff}<\SI{9000}{\kelvin}$. 

We refine the \dsct{} sample by applying the following physical cuts: $1.5 < M/\sole{M} < 2.5$, $\log g >3.5$, and $L > \SI{2}{\Lsun}$. After these cuts, the \dsct{} sample was compiled by querying stars within the observational \dsct{} instability strip from \cite{murhpy2019gaia} (see the solid orange lines in the top-right plot of Fig.~\ref{fig:sample_dsct}).  

\subsection{SPB and \bcep{} stars}

As the LOPS2 features the Large Magellanic Cloud (LMC), special care was taken to avoid querying faint but numerous OB-type members as part of our SPB star and \bcep{} variable samples. Due to a strong metallicity dependence of the oscillation excitation mechanism of these pulsators, a dearth of massive pulsators is observed in the low metallicity environment of the LMC \citep[e.g.][Bowman et al. subm.]{salmon2012testing}. 


The parameter space of the SPB and \bcep{} sample was artificially generated. Although these pulsators occasionally are observed beyond the instability strips \citep[e.g. see][for \bcep{} stars]{fritzewski2024mode}, we stick to theoretical predictions for simplicity. First, $\tx{T}{eff}$ and $L$ were queried randomly within a HRD polygon defined by the instability strips from \cite{burssens2020variability} together with (or in addition to) the zero-age main-sequence (ZAMS) and the terminal-age main sequence (TAMS). For this exercise, we used the MESA isochrones and stellar tracks \citep[MIST v.1.2;][]{choi2016mesa} library, computed with the stellar evolution code MESA \citep{paxton2011mesa, paxton2013mesa, paxton2015mesa}. The mass borders of the SPB star polygon were defined by a \SI{2.5}{\Msun} and \SI{8}{\Msun} track as the lower and upper boundary, respectively. The mass borders of the \bcep{} polygon was defined by a \SI{8}{\Msun} and \SI{26}{\Msun} track as the lower and upper boundary, respectively. For simplicity, we only include main-sequence \bcep{} stars (i.e. up to the TAMS). 

Having $L$, $M$ was calculated from the mass-luminosity relation $L/\sole{L} \approx 1.4 (M/\sole{M})^{3.5}$ valid for main sequence stars in the range $\SI{2}{\Msun} < M < \SI{55}{\Msun}$ \citep{kuiper1938emperical}. With $L$ and $\tx{T}{eff}$, $R$ was calculated from the Stefan-Boltzmann equation for a blackbody. Next the $\log g$ was estimated using Newton's law of gravity. Lastly, \metal{} was used directly from the stellar information of \textit{Gaia}, however, values below $\metal{} < -2$ were recalculated by drawing from a normal distribution of the underlying \metal{} distribution.


\subsection{RR Lyrae and Cepheid stars}

\begin{figure}[h!]
\center
\includegraphics[width=0.49\columnwidth]{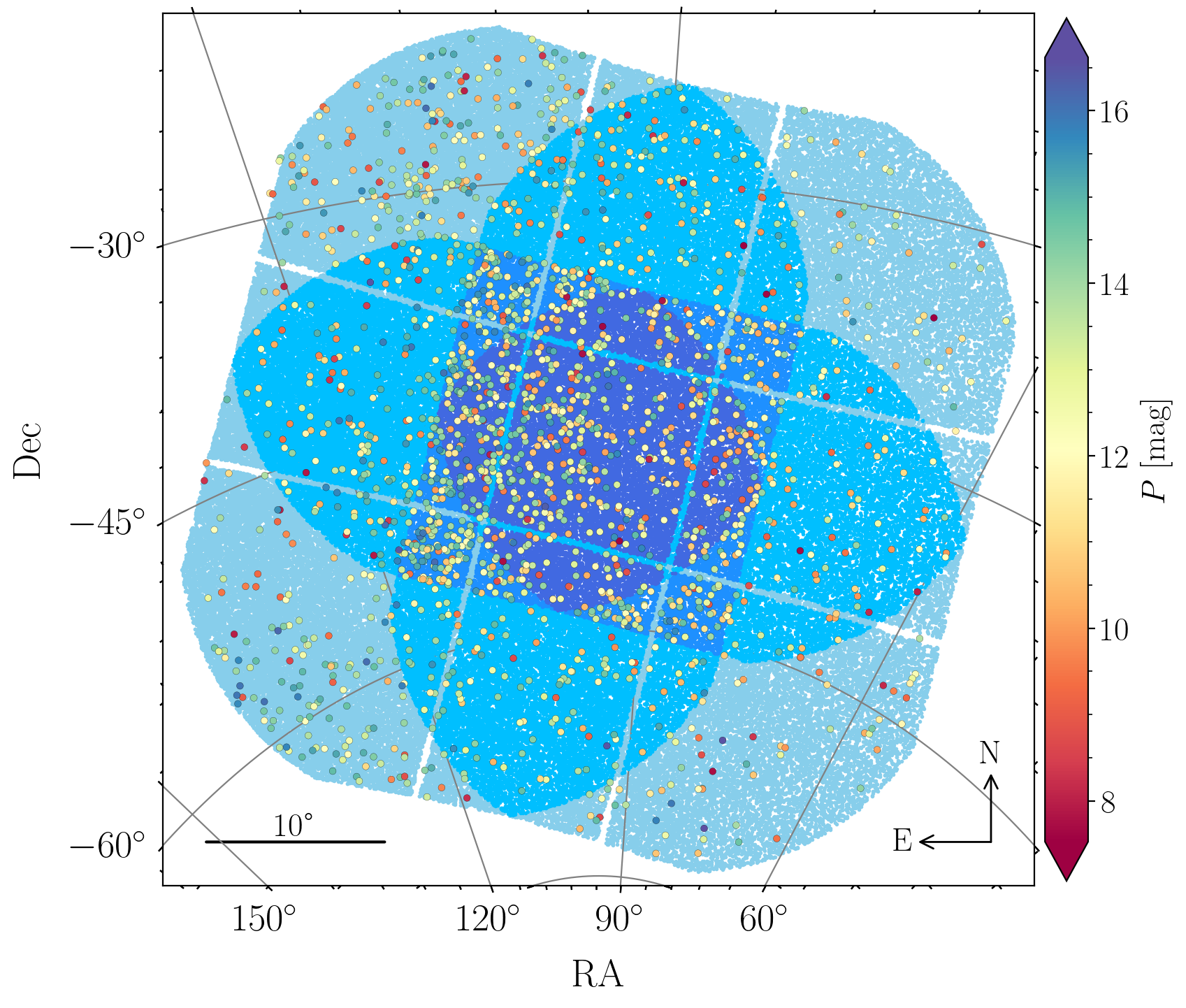}
\includegraphics[width=0.49\columnwidth]{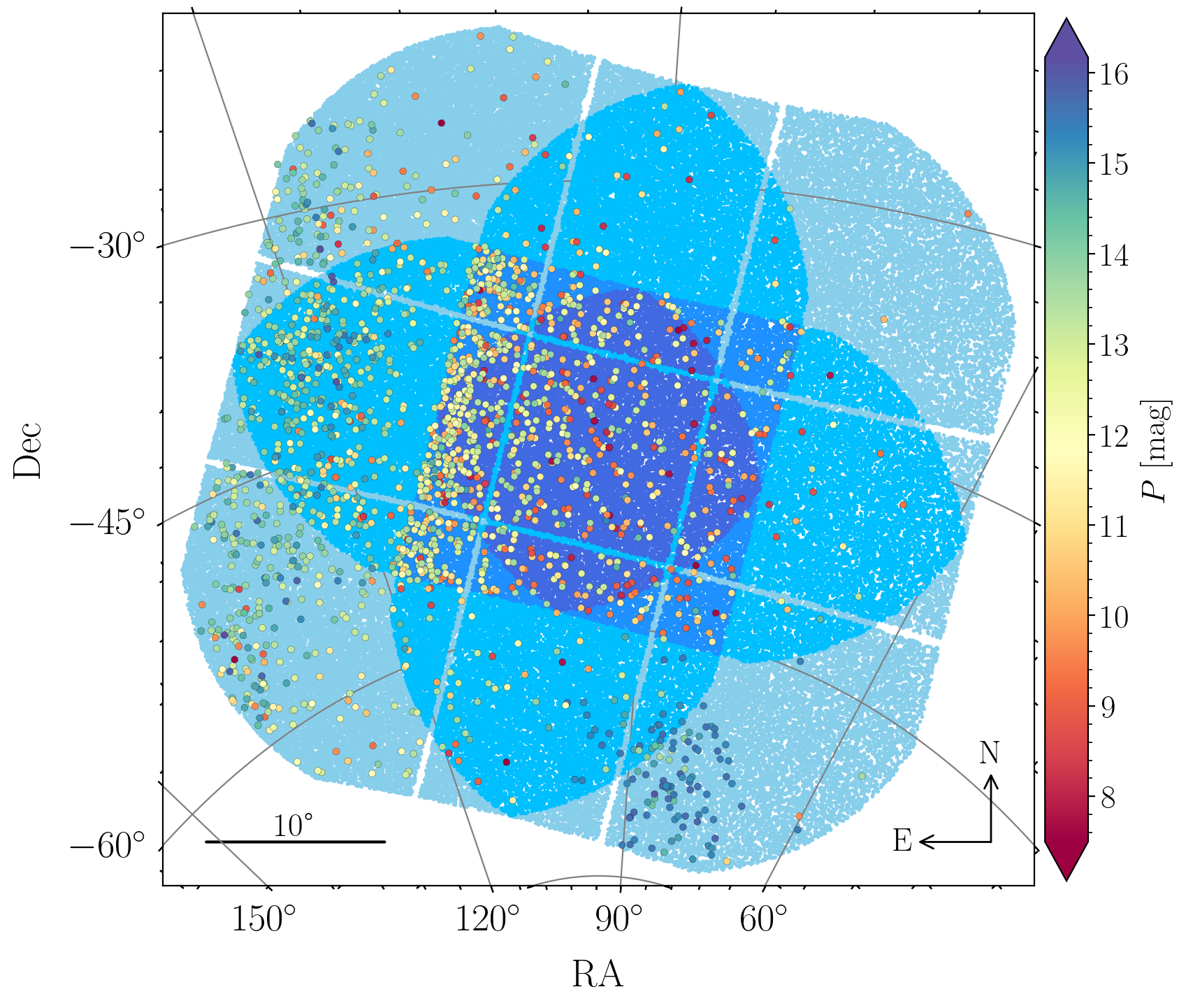}
\includegraphics[width=0.48\columnwidth]{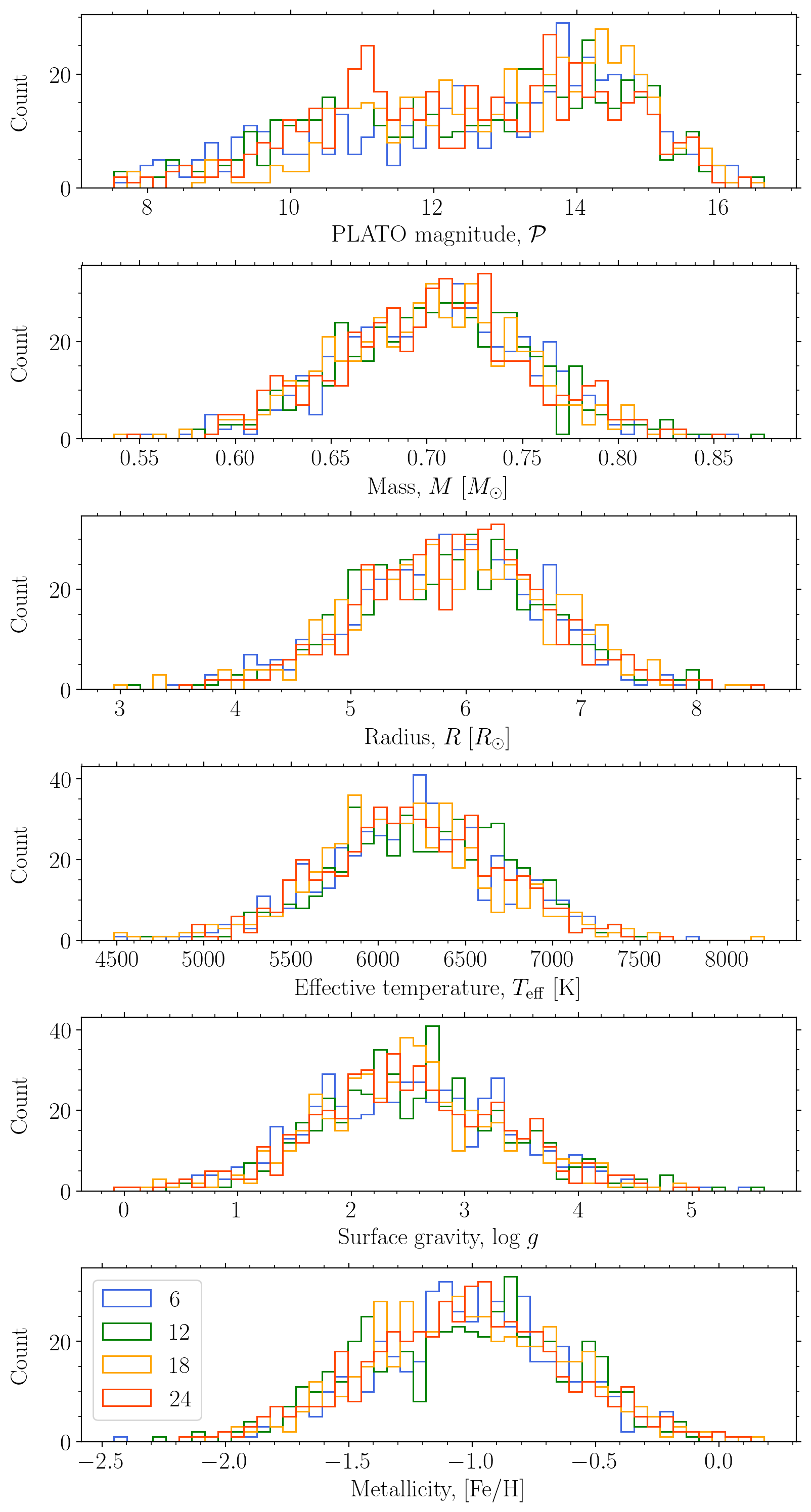}
\includegraphics[width=0.48\columnwidth]{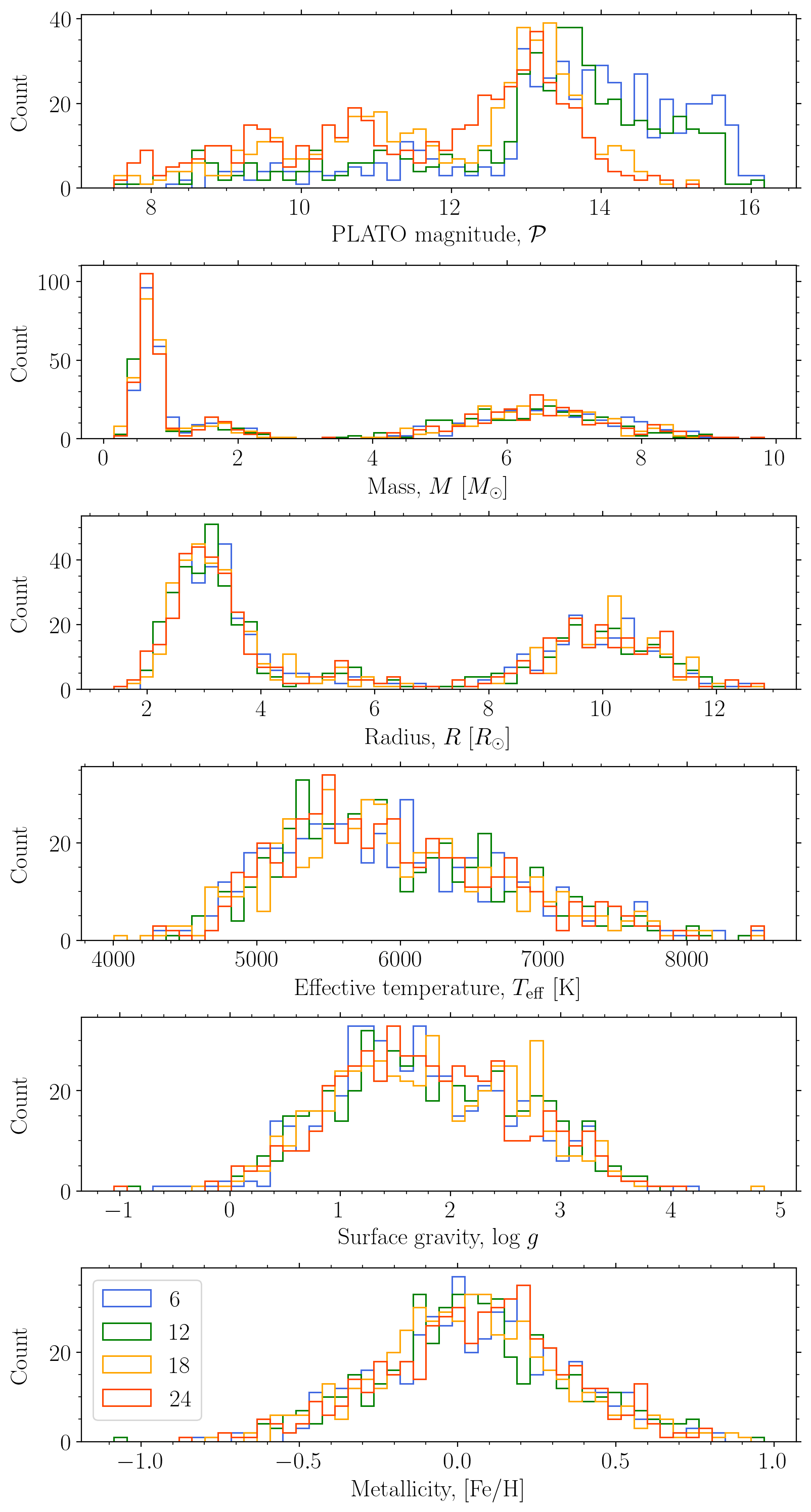}
\caption[]{Model parameters for the RR Lyr sample (left panels) and Cepheid sample (right panels). See description of Appendix~\ref{app:parameters}.} 
\label{fig:sample_rrlyr_ceph}
\end{figure}

For the RR Lyrae and Cepheid samples, we show the sky distribution and parameter distributions in Fig.~\ref{fig:sample_rrlyr_ceph} (left and right panels, respectively). Cepheid pulsators are, apart from the thin disc, typically found in the LMC. With the LOPS2 only partially covering a small portion of the galactic thin disc, effort was made to include more stars from the LMC into the Cepheid sample (see upper right-hand panel of Fig.~\ref{fig:sample_rrlyr_ceph}). 

The parameter space of each sample was generated using a normal distribution, except for the stellar radius of Cepheids being computed with an appropriate period-radius relation (with the radius distribution plot represents the minimum radius). We note that the distributions are the cumulative distributions of all the sub-types of pulsators within each class. 

\begin{table}[h!]
\caption[]{Reference table of RR Lyrae parameters.}
\begin{center}
\begin{tabular}{llll}
\hline\hline
Parameter & Sub-class & Distribution & Reference \\
\hline
$M$ [\si{\Rsun}]	& All		& $\mathcal{N}(0.7,\,0.05)$ 	& \cite{netzel2023asteroseismology} \\
$R$	[\si{\Msun}]	& RRab(d) 	& $\mathcal{N}(6.0,\,0.8)$ 		& \cite{marconi2005predicted} \\
					& RRc		& $\mathcal{N}(5.5,\,0.8)$ 		& \cite{marconi2005predicted} \\
$\tx{T}{eff}$ [K]	& All 		& $\mathcal{N}(6200,\,500)$		& \textit{Gaia} \texttt{FLAME} pipeline \\
$\log g$			& RRab(d) 	& $\mathcal{N}(2.25,\,0.75)$	& \cite{molnar2023grow} \\
					& RRc 		& $\mathcal{N}(2.25,\,0.75)$ 	& \cite{molnar2023grow} \\
\metal{}			& All		& $\mathcal{N}(-1.07,\,0.5)$ 	& \cite{clementini2023gaia} \\
\hline
\end{tabular}
\end{center}
\label{tab:rrlyrae}
\end{table}

\begin{table}
\caption[]{Reference table of Cepheid parameters. During a pulsation cycle Cepheids drastically change in radius, hence we model the radius following a radius-period relation. Typical radii changes are $\sim$\SIrange{27}{381}{\Rsun} for Classical Cepheids, $\sim$\SIrange{6}{100}{\Rsun} for Anomalously Cepheids, and $\sim$\SIrange{6}{100}{\Rsun} for Double-mode Cepheids.}
\begin{center}
\begin{tabular}{llll}
\hline\hline
Parameter & Sub-class & Distribution & Reference \\
\hline
$M$ [\si{\Msun}]	& Classical Cepheids 	& $\mathcal{N}(6.50,\,1.00)$				& \cite{aerts2010asteroseismology} \\
					& Anomalously Cepheids 	& $\mathcal{N}(1.65,\,0.35)$				& \cite{aerts2010asteroseismology} \\
					& Double-mode Cepheids 	& $\mathcal{N}(0.65,\,0.15)$ 				& \cite{bono2020evolutionary} \\
$R$ [\si{\Rsun}]		& Classical Cepheids 	& $\log R = 1.763 + 0.653 (\log P - 0.9)$	& \cite{trahin2021inspecting} \\
					& Anomalously Cepheids 	& $\log R = 0.87 + 0.54 \log P$				& \cite{groenewegen2017period} \\
					& Double-mode Cepheids  & $\log R = 0.87 + 0.54 \log P$  			& \cite{groenewegen2017period} \\
$\tx{T}{eff}$ [K]	& Classical Cepheids 	& $\mathcal{N}(5400,\,450)$ 				& \cite{espinoza2024empirical} \\
					& Anomalously Cepheids 	& $\mathcal{N}(7100,\,650)$ 				& \cite{ripepi2024first} \\
					& Double-mode Cepheids 	& $\mathcal{N}(6350,\,575)$ 				& \cite{schmidt2011spectra} \\
$\log g$			& Classical Cepheids 	& $\mathcal{N}(1.25,\,0.625)$ 				& \cite{lemasle2020atmospheric} \\
					& Anomalously Cepheids 	& $\mathcal{N}(1.65,\,0.675)$ 				& \cite{ripepi2024first} \\
					& Double-mode Cepheids 	& $\mathcal{N}(2.50,\,0.600)$ 				& \cite{schmidt2011spectra} \\
\metal{} 			& All types				& $\mathcal{N}(0.05,\,0.3)$					& \cite{ripepi2023gaia} \\
\hline
\end{tabular}
\end{center}
\label{tab:cepheids}
\end{table}

\subsection{sdBV and WD stars}

\begin{figure}[h!]
\center
\includegraphics[width=0.495\columnwidth]{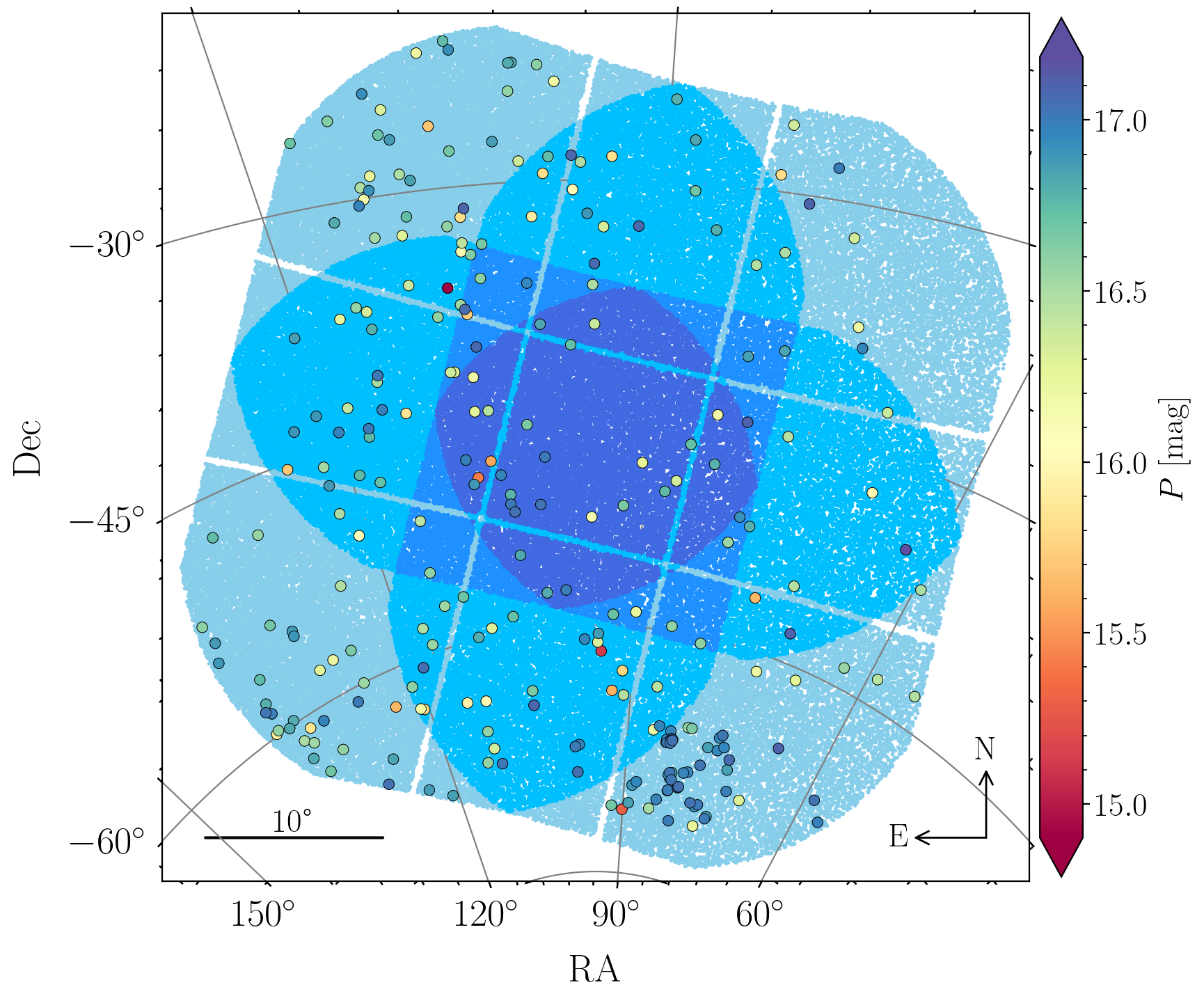}
\includegraphics[width=0.495\columnwidth]{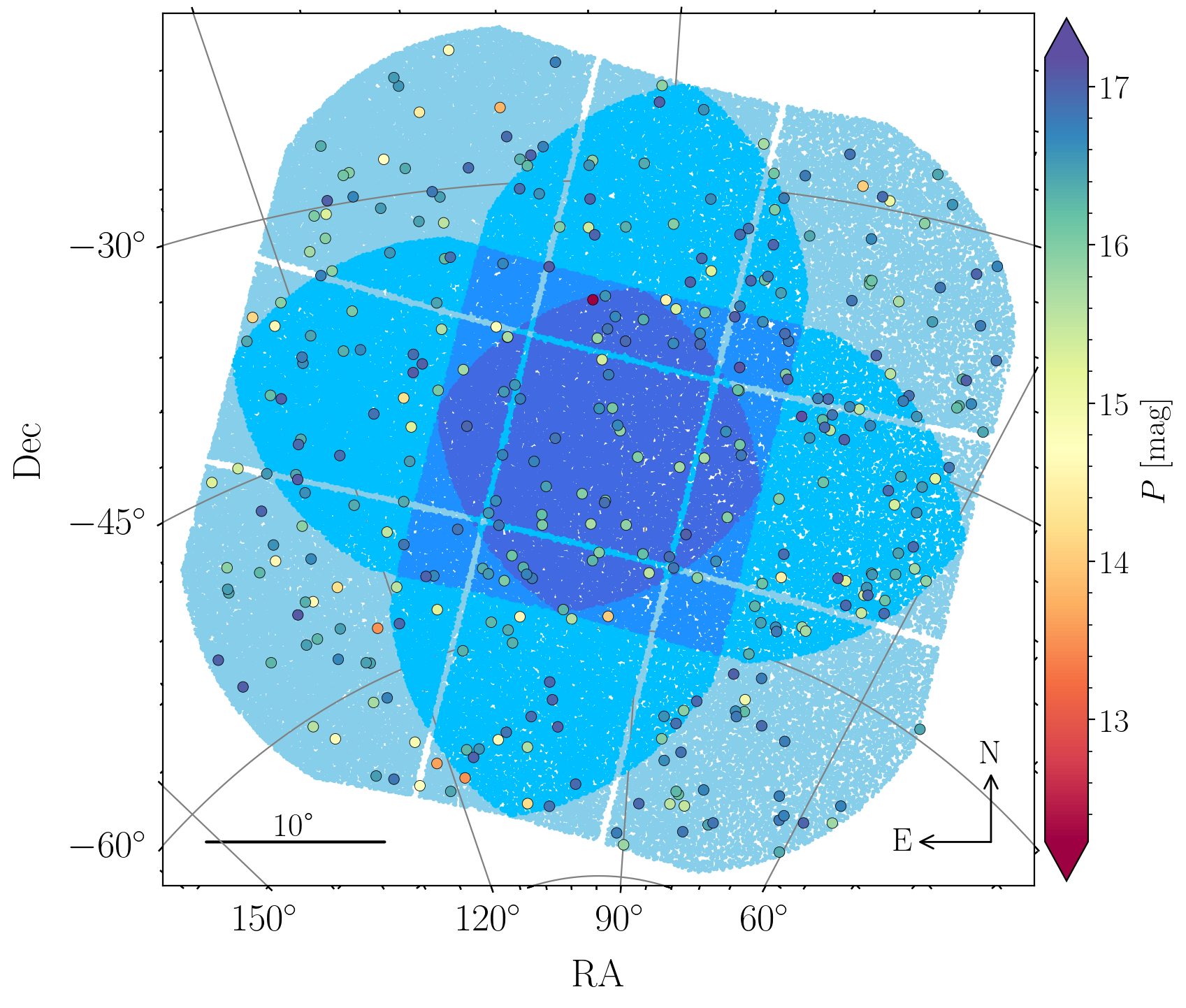}
\caption[]{Sky distribution for the sdBV sample (left) and the WD sample (right). See description of Appendix~\ref{app:parameters}.} 
\label{fig:sample_compact}
\end{figure}

For the compact pulsators, being the samples of sdBV and WD stars, we show the sky distribution in Fig.~\ref{fig:sample_compact} (left and right, respectively) of all stars that may be used as target stars during the ten random draws for each benchmark pulsator. Note that due to their isolated location in the CaMD, the sdBV and WD samples are drawn (highly) potential candidate of each pulsation class, hence, they are much fainter. 

For the compact pulsators we exceptionally use benchmark stars (c.f. Sect.~\ref{sec:var_tar_compact}): 17 sdBs and 30 WDs. As shown in Fig.~\ref{fig:CaMD}, we select target candidates directly from the CaMD by confining compact pulsators below the blue dotted line. We then define sdBs having $M_G < 6$ and WDs having $M_G > 7$ and $\varpi > 2$. These limits were chosen to match the variable star legacy of \textit{Gaia} \citep[][Fig.~3]{gaia2018dr3varability}. In total, 268 sdBs candidates with $N(\tx{n}{CAM}) \in \{137,\,86,\,20,\,25\}$ and 347 WD candidates with $N(\tx{n}{CAM}) \in \{143,\,134,\,27,\,43\}$ were found. For each benchmark star, ten random stars were iteratively drawn from these candidates, thus enforcing ten different magnitude and noise budget realisations per star simulated in line with the future expected observations of compact objects in the LOPS2.

\section{Spot modulation software: \texttt{pyspot}}\label{app:spot}

We here elaborate on the functionality of the spot modulation code \texttt{pyspot} developed by S. Aigrain. The code is only valid for main-sequence stars, and as input it takes $\tx{T}{eff}$, the desired time-sampling (cadence and duration), and optionally the stellar inclination in degrees. All other parameters are derived or selected at random from distributions based on \cite{meunier2019activity} (hereafter M19) and references therein, as we will discuss in the following.

The rotation period is derived from $B-V$ and activity level ($\log\tx{R'}{HK}$) following \cite{noyes1984rotation} and \cite{mamajek2008improved}. The prescription for differential rotation and the range of spot latitudes closely follows Sect.~2.5 M19. Specifically, the differential rotation parameter, $\alpha$, is derived from $\tx{T}{eff}$ and rotational period, $\tx{P}{rot}$, following Eq.~3 and 4 of M19, which are based on fits to the \textit{Kepler} sample of \cite{reinhold2015rotation}. The minimum and maximum observed periods (corresponding to the maximum and minimum latitudes at which spots appear) are related to $\alpha$ and $\tx{P}{rot}$ according to Eq.~1 and 2 of M19. The maximum spot latitude is drawn from a uniform distribution, $\mathcal{U}(32, 52)^{\circ}$ and the minimum is always set to zero (i.e. the equator). Note that while M19 set the maximum spot latitude between \SIrange{22}{42}{\degree}, this refers to the mean latitude at the start of the cycle, rather than the absolute maximum. For the employed code in this work the maximum latitude represents the absolute maximum over the entire cycle, allowing for the dispersion about the mean at each point in time, hence, \SI{10}{\degree} is added to account for that.

Next, $\log\tx{R'}{HK}$ is drawn from a uniform distribution within bounds which depend on $B-V$ (cf. Fig.~3 of M19). $B-V$ is obtained from $\tx{T}{eff}$ by interpolating Table~1 of M19. If the input $\tx{T}{eff}$ is outside the range spanned by this table, the $B-V$ corresponding to the nearest extreme of the table is returned. The equations defining the limits as a function of $B-V$ are given in Table~E.3 of M19, and are based on \cite{noyes1984rotation} and \cite{mamajek2008improved}. The activity cycle period is drawn from a uniform distribution in log, whose limits follow a power-law dependence on rotation period (as illustrated in Fig.~6 of M19, with parameters from Table~E.3). The activity cycle amplitude is drawn from a uniform distribution whose limits depend on both $B-V$ and $\log\tx{R'}{HK}$ (see Fig.~7 and Table~E.3 of M19). The cycle amplitude is used to set the rate at which spots appear, which controls the overall variability level. 
 
Following \cite{borgniet2015using}, M19 simulate two populations of spots (isolated spots, and those belonging to large groups) both of which are based on statistics of active regions in the Sun \citep{baumann2005size, pillet1993distribution}. Both populations emerge instantaneously with a log-normal distribution of sizes and follow a linear decay law, also with a log-normal distribution of decay rates (with different parameters for the two groups, see Table~E.1). However, for the employed code, only one population of spots is modelled. Hence, the spot sizes are matched to a log-normal distribution with parameters that are intermediate between those used for the two populations in M19. Moreover, instead of a linear growth and decay law as in M19, a squared exponential law is used since the growth and decay pattern at the time sampling of PLATO seems excessively discontinuous (compared e.g. to \textit{Kepler} light curves). Thus, the the growth- and decay times are set to the half-life of the squared exponential and a log-normal distribution is used for the determination of the spot decay rates. 

The times and locations at which spots emerge, as well as their peak magnetic flux density, are simulated using the active region emergence code developed by \cite{llama2012using}. The code models an activity cycle with the specified period, $\tx{P}{cyc}$, a butterfly pattern with a fixed width (solar-like), and the specified minimum and maximum latitudes. If not specified by the user an overlap value between consecutive activity cycles is (arbitrarily) drawn from a uniform distribution, $\mathcal{U}(0,\,0.1)\,\tx{P}{cyc}$. The active region emergence code computes a peak magnetic flux density, $\tx{B}{em}$, for each active region. The peak size of each spot is then set to 300 micro-hemispheres%
\footnote{A micro-hemispheres is a unit of area equivalent to one millionth of half the surface of the Sun.} %
times $\tx{B}{em}^2$. This relationship was defined by trial and error in order to produce a distribution of spot sizes that is approximately log-normal with a median and width intermediate between the parameters used by M19 for their two types of spots. 

Each spot is assigned a decay rate drawn from a log-normal distribution with median 10 and mean 15 micro-hemispheres per day. As noted above, these decay rates are somewhat lower than those used in \cite{borgniet2015using} and M19, because the latter gave rise to very incoherent light curves where recovering the rotation period signal would be very hard. The decay is linear, meaning that a spot with initial size 100 micro-hemispheres and with decay rate 10 micro-hemispheres per day disappears 10 days after reaching its peak size. Unlike M19, the spots do not emerge with their maximum sizes but grow linearly from zero. However, the growth rate is 10 times faster than the decay rate -- matching what is been observed in the Sun \citep{howard1992growth}.

Given its latitude, longitude, and the evolution of its size over time, the photometric signature of each spot is simulated using the analytical model of \cite{aigrain2012simple, aigrain2023simple}, which only accounts for foreshortening%
\footnote{Foreshortening describes the decreasing apparent size of a spot when situated at a given distance away from the centre of the stellar disc, as compared to the centre itself.}. %
Like M19, this model does not attempt to take into account the finite size of the spot when computing the foreshortening or modelling the emergence and disappearance of the spots over the stellar rim. Furthermore the spots are treated as completely dark (contrast of 1) and limb darkening is ignored. Lastly, we set the stellar inclination to the cosine of the inclination as drawn from a uniform distribution, $\mathcal{U}(0, 1)\,\text{rad}$, which corresponds to a random orientation of the stellar rotation axis in 3D.

\section{Custom photometric pipeline}\label{app:pipeline}

\begin{figure}[h!]
\center
\includegraphics[width=\columnwidth]{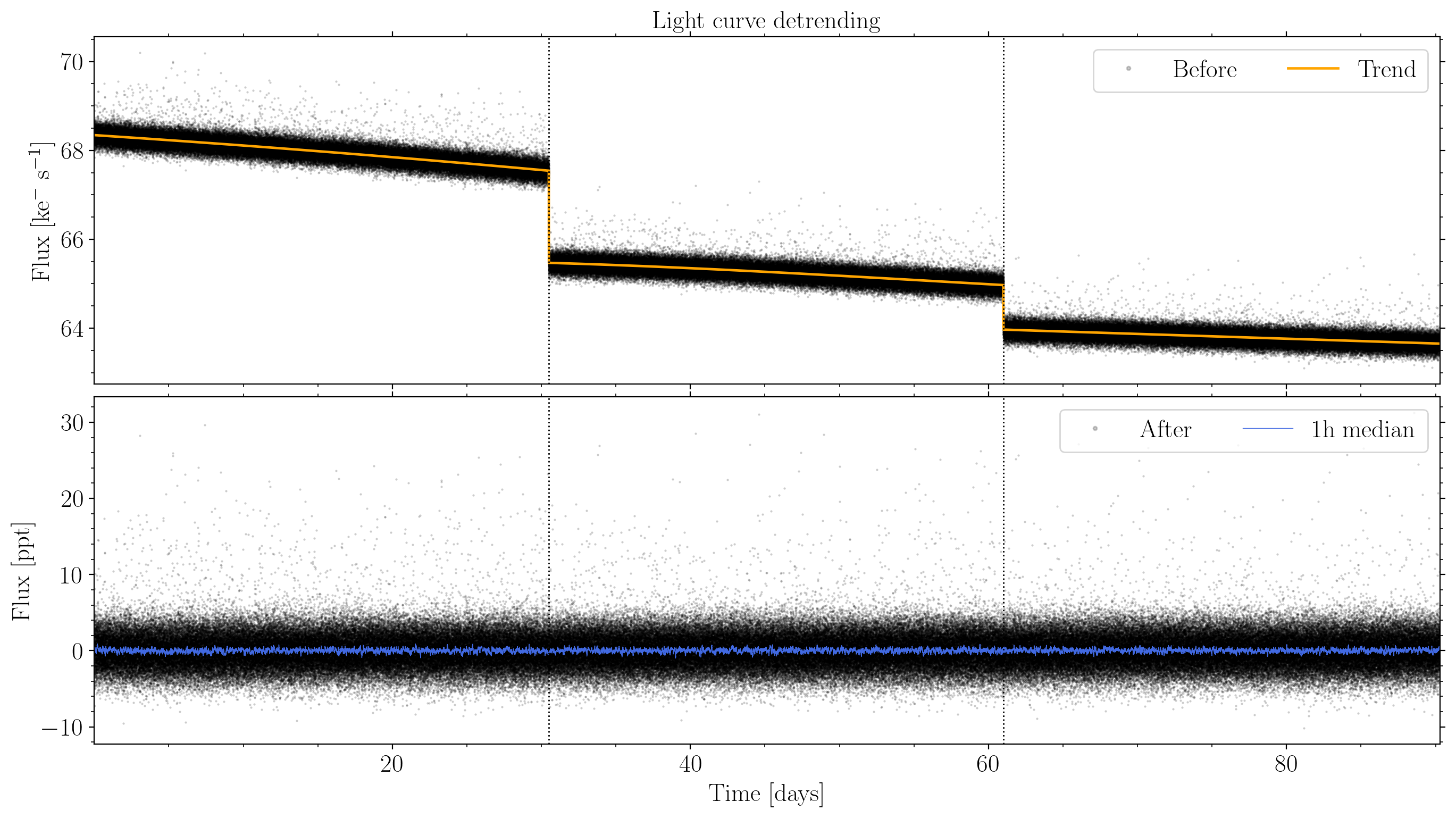}
\caption[]{Performance example of the light curve detrending algorithm for a single N-CAM and mission quarter. The upper panel shows the light curve before detrending (black points) together with the piecewise polynomial model trend fitted to the data (solid orange line). The bottom panel shows the detrended light curve with a \SI{1}{\hour} median filter plotted on top.} 
\label{fig:pipeline_detrending}
\end{figure}

\begin{figure}[]
\center
\includegraphics[width=\columnwidth]{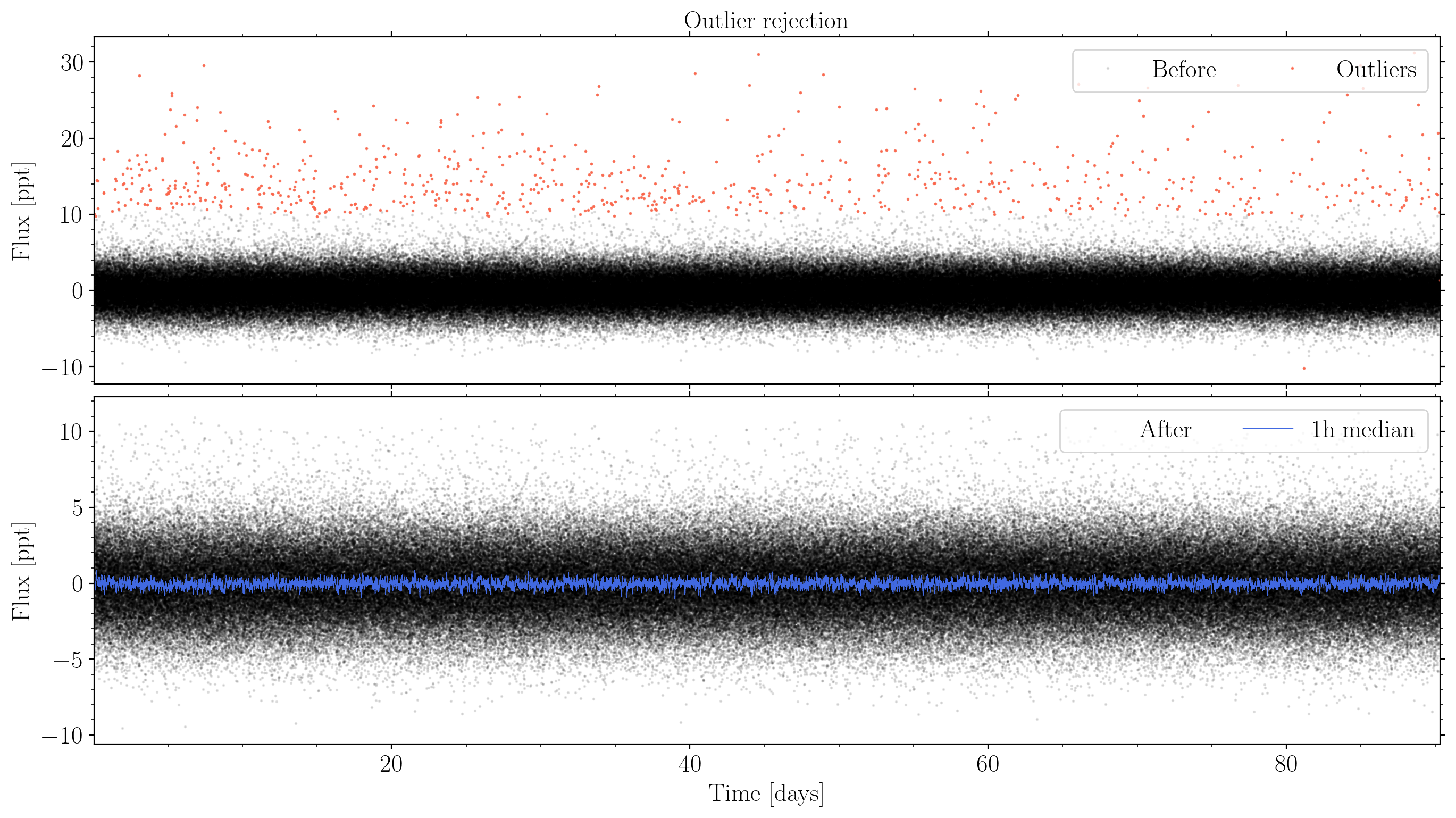}
\caption[]{Illustration of the outlier rejection algorithm for a single N-CAM and mission quarter. The upper panel shows the light curve before outlier rejection (black points) together with outliers identified by the sigma-clipping algorithm (red points). The bottom panel shows the light curve without outliers and a \SI{1}{\hour} median filter plotted on top.} 
\label{fig:pipeline_clip}
\end{figure}

\begin{figure}[]
\center
\includegraphics[width=\columnwidth]{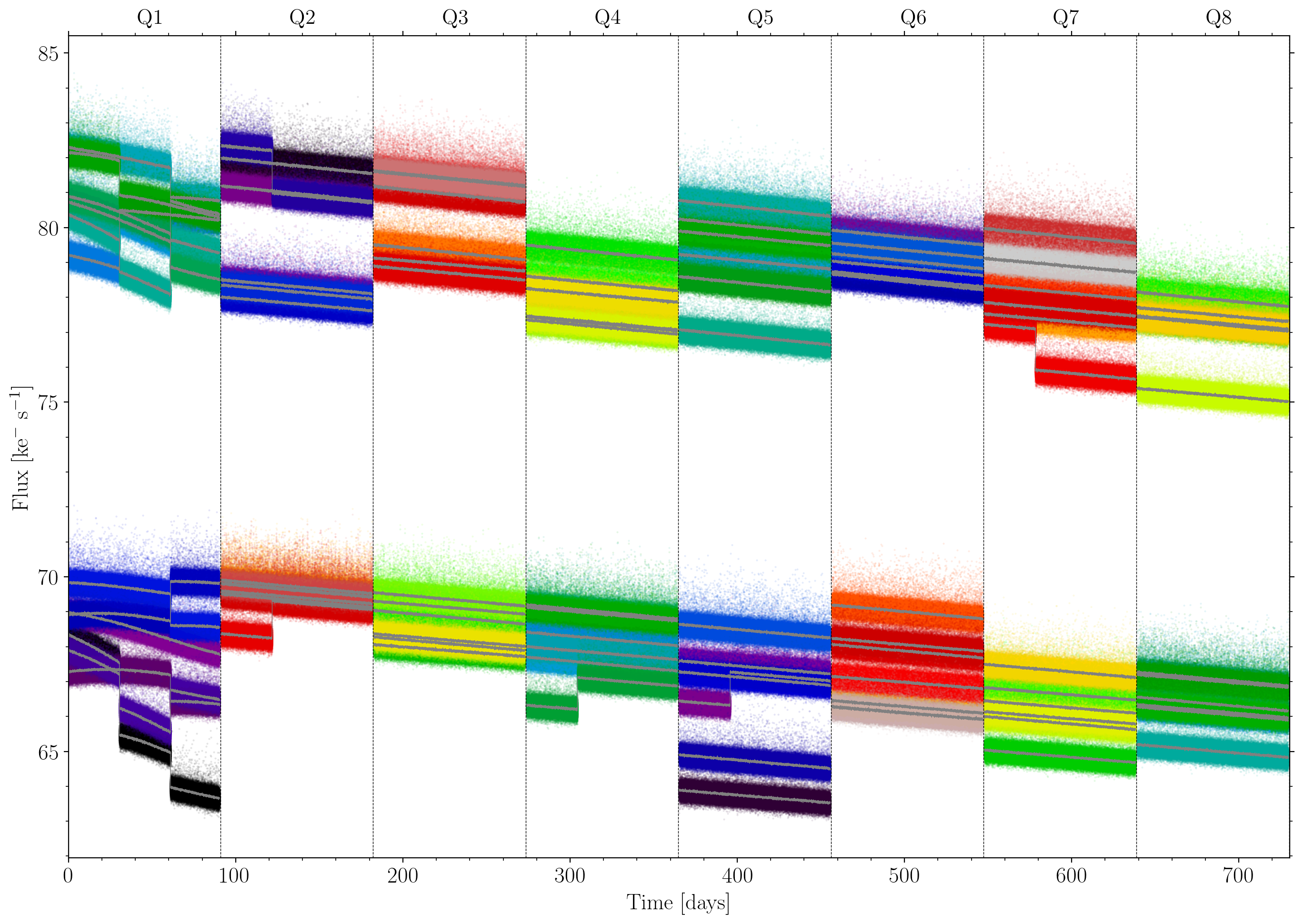}
\caption[]{Illustration of the multi-camera and multi-quarter light curves produced for a single target star observed with 12 N-CAMs. Each colour represents an independent mission quarter segment simulation, and the grey line is the corresponding \SI{1}{\hour} median filter of each segment. The main division between the light curves accumulated in the top and bottom half of the figure corresponds to a difference in optical throughput (since the star is continuously being observed at two specific radial distances).} 
\label{fig:pipeline_multi_camera}
\end{figure}

\begin{figure}[]
\center
\includegraphics[width=\columnwidth]{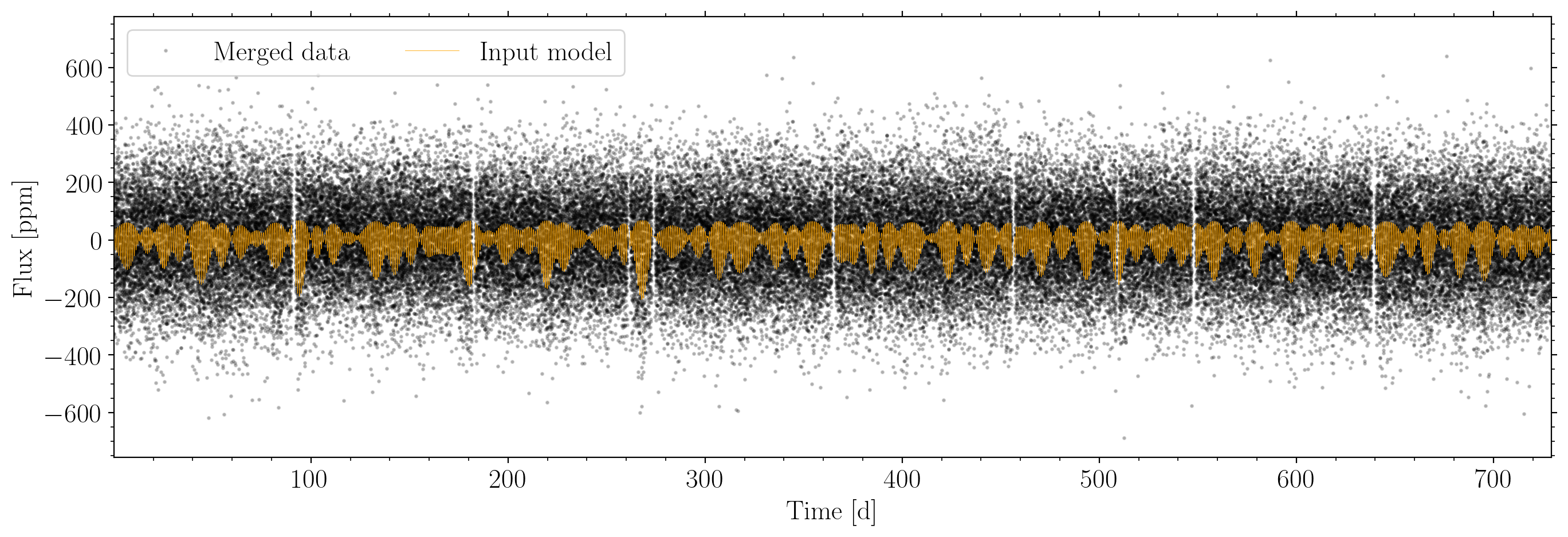}
\caption[]{Illustration of a fully reduced final light curve from the 12 N-CAM observations of Fig.~\ref{fig:pipeline_multi_camera}. This light curve has been reduced to a cadence of \SI{10}{\minute} and data gaps have been applied. The orange solid line is the \gdor{} input model.} 
\label{fig:pipeline_multi_final}
\end{figure}

In this appendix we provide an show case example of the end-to-end reduction pipeline for a $\Pb=9.95$ \gdor{} star of the simulation batch \affogato{}. As explained in Sect.~\ref{sec:sim_photometry}, each mission quarter light curve segment is detrended as shown in Fig.~\ref{fig:pipeline_detrending}, where three aperture mask updates were triggered for this specific observation. Next outliers are removed as shown in Fig.~\ref{fig:pipeline_clip}. At a multi-camera level we show how a light curve of 12 N-CAMs looks like raw before the reduction pipeline in Fig.~\ref{fig:pipeline_multi_camera} and how the final reduced single light curve looks like in Fig.~\ref{fig:pipeline_multi_final}. 

\begin{figure}[h!]
\center
\includegraphics[width=\columnwidth]{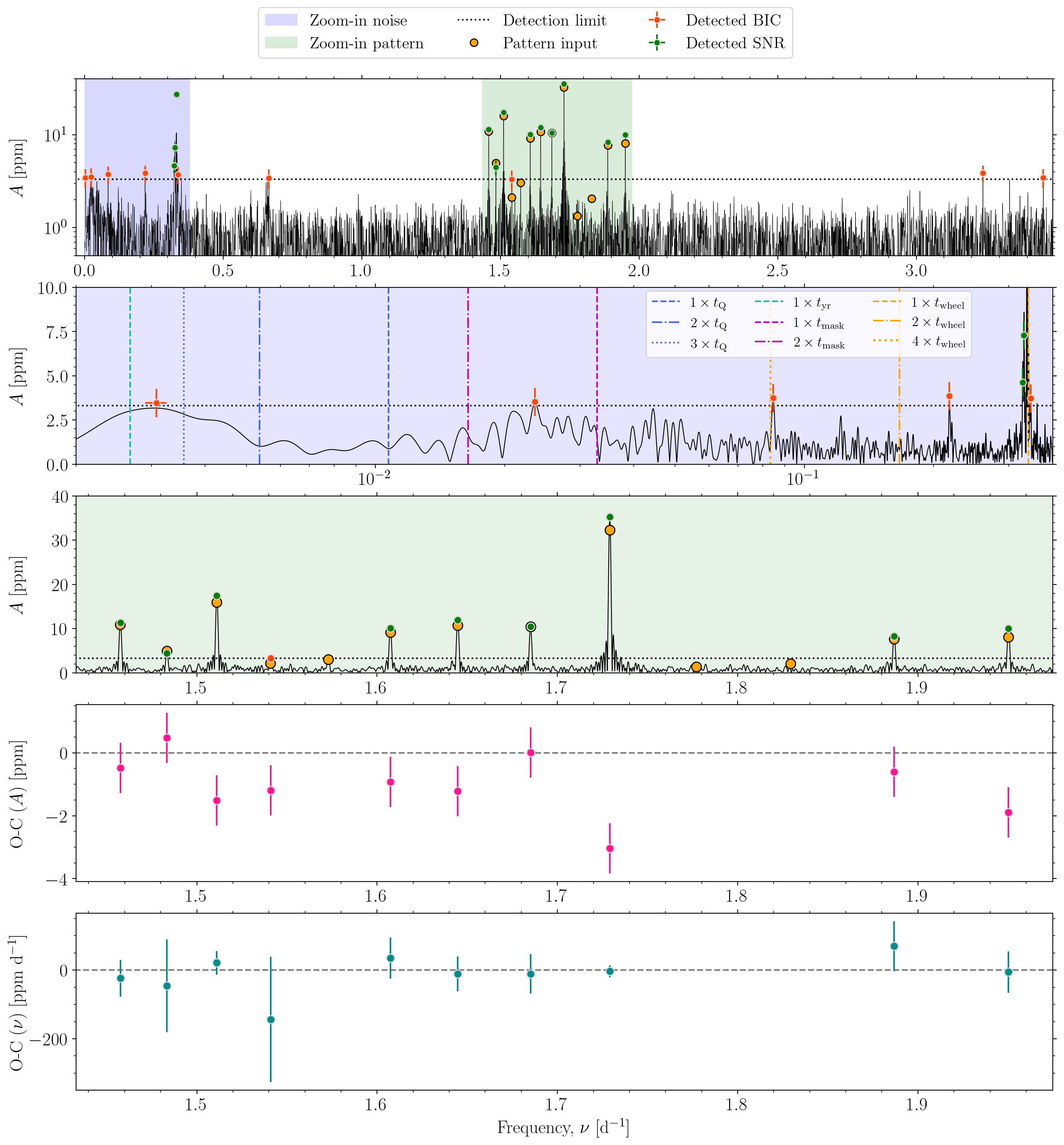}
\caption[]{Illustrative example of the pulsation mode extraction from the Lomb-Scargle periodogram of the light curve from Fig.~\ref{fig:pipeline_multi_final}. In the top three panels, the solid black line represents the amplitude spectrum and the dotted black line is the prewhitening BIC detection limit. The top panel shows a model comparison between injected (orange circles) and extracted pulsation modes using either the BIC (red markers) or the SNR (green markers) significance criterion. Furthermore, two frequency regions are highlighted: the dominant region of potential instrumental systematics (blue shade, second panel) and the region for which the period-spacing pattern resides (green shade, third panel). The vertical lines of the second panel display from right to left the typical frequency of: single, twice, and quadruple the duration of the reaction wheel offloading events (dashed, dashed-dotted, and dotted orange lines, respectively), single and twice the duration of mask-update events (dashed and dotted-dashed purple lines, respectively), single, twice, and triple the duration of quarterly rotation events (dashed, dashed-dotted, and dotted blue lines, respectively), and the yearly harmonic (cyan line). The fourth panel shows the residual amplitude diagram, and the fifth panel shows the residual frequency diagram.} 
\label{fig:pipeline_model_comparison}
\end{figure}

As an extension to the final post-processing step described in Sect.~\ref{sec:sim_frequency}, Fig.~\ref{fig:pipeline_model_comparison} illustrates the final post-processing step of finding which modes are detectable with the SNR and BIC prewhitening stopping criterion, respectively. The top panel shows the amplitude spectrum with injected modes (orange circle), the modes detected using the BIC criterion (red markers), and the modes detected using the SNR criterion (green markers). The plot also illustrates the presence of instrumental systematics (shown in the second panel) and the region of the injected period-spacing pattern (shown in the third panel). The fourth and fifth panels show the O-C diagram for the frequencies and amplitudes of the extracted pulsation modes. These two diagrams are used to compute the error propagated root mean square (rms) frequency and amplitude precision per star, respectively.

We emphasise that the residual systematics that changes on a time scale less than a few days are not corrected for in our pipeline. That means residual noise, especially from wheel-offloading events (with a duration of around three days), is present in the final light curves. The last mission quarter of Fig.~\ref{fig:pipeline_multi_final} shows this systematic, however, the effect is more pronounced for the simulation batch \cortado{} due to the larger injected TED amplitudes. As seen in Fig.~\ref{fig:pipeline_model_comparison}, in the frequency domain the long-term varying systematics rarely overlap in frequency with g- or p-mode pulsators on the main-sequence.

%


\end{appendix}

\end{document}